\documentclass[lettersize,journal]{IEEEtran}

\usepackage{graphicx}
\usepackage{epsfig} 
\usepackage{adjustbox}
\usepackage{siunitx}
\usepackage{booktabs}

\usepackage{hyperref}
\usepackage{empheq}
\usepackage{amsmath}
\usepackage{amssymb}
\usepackage{amsthm}

\usepackage{algorithm}
\usepackage{algorithmic}

\usepackage{array}
\usepackage{multirow}
 \usepackage{tabularx}
\usepackage{stfloats}
\usepackage{url}
\usepackage{verbatim}

\usepackage[caption=false,font=normalsize,labelfont=sf,textfont=sf]{subfig}

\usepackage{url}
\usepackage{fancyhdr}
\usepackage{color}
\usepackage{xcolor, soul}
\sethlcolor{cyan}
\usepackage[noadjust]{cite}

\begin{document}

\title{Efficient Rate-Splitting Multiple Access for the Internet of Vehicles: Federated Edge Learning and Latency Minimization}

\author{Shengyu~Zhang, Shiyao~Zhang,~\IEEEmembership{Member,~IEEE}, Weijie~Yuan,~\IEEEmembership{Member,~IEEE}, Yonghui~Li,~\IEEEmembership{Fellow,~IEEE}, and Lajos~Hanzo,~\IEEEmembership{Life~Fellow,~IEEE} 
\thanks{This work is supported in part by the General Program of Guangdong Basic and Applied Basic Research Foundation No. 2021KQNCX078. This work is supported in part by National Natural Science Foundation of China under Grant 62101232, and in part by the Guangdong Provincial Natural Science Foundation under Grant 2022A1515011257. L. Hanzo would like to acknowledge the financial support of the Engineering and Physical Sciences Research Council projects EP/W016605/1 and EP/P003990/1 (COALESCE) as well as of the European Research Council's Advanced Fellow Grant QuantCom (Grant No. 789028).(Corresponding author: Shiyao Zhang)}
\thanks{Shengyu Zhang is with the Department of EEE, The University of Hong Kong, HKSAR, China (e-mail: zhangsy@eee.hku.hk)}
\thanks{Shiyao Zhang is with the Research Institute for Trustworthy Autonomous Systems, Southern University of Science and Technology, Shenzhen 518055, China (e-mail: zhangsy@sustech.edu.cn)}
\thanks{Weijie Yuan is with the Department of EEE, Southern University of Science and Technology, Shenzhen, China (e-mail: yuanwj@sustech.edu.cn).}
\thanks{Yonghui Li is with the School of EIE, University of Sydney, Sydney, NSW 2006, Australia (e-mail: yonghui.li@sydney.edu.au).}
\thanks{Lajos Hanzo is with the School of ECS, University of Southampton, SO17 1BJ, U.K. (e-mail: lh@ecs.soton.ac.uk).}
}

\maketitle

\begin{abstract}

Rate-Splitting Multiple Access (RSMA) has recently found favour in the multi-antenna-aided wireless downlink, as a benefit of relaxing the accuracy of Channel State Information at the Transmitter (CSIT), while in achieving high spectral efficiency and providing security guarantees. These benefits are particularly important in high-velocity vehicular platoons since their high Doppler affects the estimation accuracy of the CSIT. To tackle this challenge, we propose an RSMA-based Internet of Vehicles (IoV) solution that jointly considers platoon control and FEderated Edge Learning (FEEL) in the downlink. Specifically, the proposed framework is designed for transmitting the unicast control messages within the IoV platoon, as well as for privacy-preserving FEEL-aided downlink Non-Orthogonal Unicasting and Multicasting (NOUM). Given this sophisticated framework, a multi-objective optimization problem is formulated to minimize both the latency of the FEEL downlink and the deviation of the vehicles within the platoon. To efficiently solve this problem, a Block Coordinate Descent (BCD) framework is developed for decoupling the main multi-objective problem into two sub-problems. Then, for solving these non-convex sub-problems, a Successive Convex Approximation (SCA) and Model Predictive Control (MPC) method is developed for solving the FEEL-based downlink problem and platoon control problem, respectively. Our simulation results show that the proposed RSMA-based IoV system outperforms both the popular Multi-User Linear Precoding (MU–LP) and the conventional Non-Orthogonal Multiple Access (NOMA) system. Finally, the BCD framework is shown to generate near-optimal solutions at reduced complexity.

\end{abstract}


\section{Introduction}
\IEEEPARstart{I}{n} recent years, the smart city concept stimulated many innovative ideas and applications for enhancing the quality of life in urban areas. Through incorporating innovative Information and Communication Technologies (ICTs), the quality and efficacy of various urban operations and services have been improved with the aid of the urban Internet of Things (IoT) \cite{6740844}. In this context, the Internet of Vehicles (IoV) and Intelligent Transportation Systems (ITSs) have played a conducive role in harnessing the Vehicle-to-Everything (V2X) concept \cite{7562526,8232587}.

Autonomous driving has brought about great challenges for organizing and managing large-scale Autonomous Vehicles (AVs). Vehicular platooning has been proposed as a promising solution to address this problem \cite{8944077}. The vehicular platoon enables several AVs to be controlled and managed in a cooperative manner relying on the IoV concept \cite{6702516}. Since the vehicular platoons are operated in a semi-autonomous control mode, these vehicles are capable of supporting additional services via their spare computing power, such as distributed learning \cite{9525183}. These services further impose challenging demands on communication. Some services require multicasting transmission, while others require unicasting transmission e.g. video streaming. To support this hybrid communication scenario, an efficient IoV system is desired.

Several existing studies have investigated the design of low-latency V2X communication systems \cite{7990497, 8673568, 8490729}. The deployment of such techniques in Vehicular Ad-hoc NETworks (VANETs) can help improve traffic efficiency and road safety in urban areas. When considering the development of communication systems, 5G V2X communication systems are capable of enhancing the vehicular performance, while improving the traffic efficiency in smart cities \cite{9536953, 7974737, 8678396, 9448149}, relying for example on Non-Orthogonal Multiple Access (NOMA). However, since NOMA was shown to be outperformed by Rate-Splitting Multiple Access (RSMA), we investigate its performance in V2X communication systems \cite{rsma_2018}.

The application and impact of RSMA in wireless communication systems have been extensively studied in the literature of \cite{9445019, 9234747, 9226406, 8756076}. However, most existing works are developed based on the users engaged in downlink communications, hence they are not applicable to vehicular networks due to the stochastic nature of vehicular movements. To tackle this issue, the authors of \cite{9491092, 9519666} analyzed the feasibility of RSMA-based communication systems in conjunction with diverse vehicle mobility patterns. In practice, reliable wireless communications play a crucial role in facilitating vehicular operations. For instance, a vehicular platoon constitutes a promising vehicular management strategy, which requires efficient wireless transmission in support of safe driving by keeping the inter-vehicle distance in each lane. This increases the road capacity of each lane and reduces the fuel consumption of each vehicle. Existing studies focus on the implementation of efficient wireless communication protocols conceived for vehicular platoons \cite{8764451, 8836641, 8485763, 9562539}. Although these treatises have evaluated the feasibility of several wireless communication protocols, the benefits of RSMA in complex driving conditions have not as yet been quantified.

To fill this research gap in the context of platooning schemes, we propose an RSMA-based low latency Internet of Vehicles (IoV) system relying on FEderated Edge Learning (FEEL) aided downlink broadcasting for autonomous driving. In particular, we focus our attention on quantifying the benefits of RSMA-based platoon control under well-defined constraints. The main contributions of this work can be summarized as follows:

\begin{itemize}
\item We design an efficient RSMA-based communication and platoon control system for the IoV, which supports autonomous driving. More explicitly, we jointly consider the wireless communication and platoon motion control system. The motion control of the platoon enhances the efficiency of the RSMA-based communication system via lower CSIT estimation error and larger channel gain. And, RSMA provides reliable and robust wireless communication for the vehicular platoon, which improve the safety of vehicular platooning.
\item We propose a NOUM-enabled RSMA scheme for FEEL downlink in the designed platoon system, where the unicast control messages of the vehicular platoon and for the FEEL-based autonomous driving messages are transmitted simultaneously, in support of Non-Orthogonal Unicasting and Multicasting (NOUM) transmission in the downlink, where the global model of FEEL would be broadcast to all the vehicles.
\item We formulate a joint optimization problem for minimizing the latency of the FEEL DL and platoon control. To efficiently solve this problem, we devise a Block Coordinate Descent (BCD) based framework. This framework decouples the original problem into a pair of sub-problems for minimizing the latency of the FEEL downlink and platoon control, respectively. 
\item Due to the non-convexity of the problems formulated, a Successive Convex Approximation (SCA) approach is developed for optimizing our RSMA-based communication problem, while a Model Predictive Control (MPC) regime is conceived for solving our IoV platoon control problem. We then evaluate the performance of the proposed RSMA-based IoV system through a comprehensive series of case studies.
\end{itemize}

The rest of this paper is organized as follows. Section II presents the prior art and identifies the open challenges motivating this work. In Section III, we develop an RSMA-based IoV platoon system. Section IV formulates our optimization problem to be solved by a block coordinate descent framework in a parallel manner. Finally, Section V details our case studies, while Section VI concludes this treatise.

\section{Related Work}

\subsection{Federate Edge Learning}

The FEEL has been seen as a promising solution for supporting ITS in the smart city \cite{9454587,9690142,9829308}. It provides a privacy-preserving approach to realize a large-scale learning system in an efficient and scalable manner \cite{9318241}. In FEEL, without sacrificing the model accuracy, the metric of interest is to minimize the total training time \cite{Fedavg,8664630,9155494,9725259,9562538}. Many well-known FEEL mechanisms have been proposed since the first work in \cite{Fedavg}, where the FEEL is synchronous. Specifically, in each iteration, the edge server has to receive the local models from all workers before the next communication round \cite{8664630,9155494}. To achieve lower latency in each iteration, Wang \textit{et al.} \cite{9725259} proposed an asynchronous federated learning over wireless communication. Though asynchronous FL can well tackle the edge heterogeneity, it requires frequent model transfers, resulting in massive communication resource consumption. Additionally, Ma \textit{et al.} \cite{9562538}, propose a semi-asynchronous federated learning mechanism (FedSA), where the parameter server updates the global model relying on only a certain number of local models by their arrival order in each round. 

Based on the distributed learning framework, the FEEL has recently been considered in vehicular networks for supporting autonomous driving \cite{9457207}. In this context, Zhou \textit{et al.} \cite{9424984} proposed a two-layer FEEL model in a 6G environment for reducing communication overheads. Moreover, Shinde \textit{et al.} \cite{9650754} propose a FEEL framework, which jointly considers the latency and the energy consumption in FEEL downlink. To avoid the straggler in each communication round, a flexible FEEL framework called FedCPF has been proposed in \cite{9505307}. However, most of these frameworks improve the communication efficiency of the FEEL via the perspective of user scheduling, which neglects the impact of wireless communication schemes.

\subsection{Internet of Vehicles}
Given the gradual roll-out of the IoV in urban areas, it is becoming an essential component in ITS, by sensing their surrounding environment, without human involvement. The V2X communication systems enable efficient wireless transmission under the IoVs. In this context, Lee \textit{et al.} \cite{7990497} verified the feasibility of V2X services, and then analyzed their latency, when relying on shortened transmission time intervals. Additionally, Ma \textit{et al.}\cite{8673568} reviewed a suite of application scenarios in the context of High-Reliability and Low-Latency (HRLL) wireless IoT networks, while a hybrid architecture was designed in \cite{8490729} for mitigating the contention latency, thus facilitating long-distance communications at a given delay. By relying on 5G systems, Kose \textit{et al.}, \cite{9536953} surveyed the ultra-reliable and low-latency communication requirements related to V2X applications. A dense vehicular communication network was conceived by Di \textit{et al.} \cite{7974737} in support of reliable broadcasting among vehicles during each transmission period. Additionally, Liu \textit{et al.} \cite{8678396} developed a pair of relay-assisted NOMA transmission schemes for mitigating traffic congestion and for reducing the latency of 5G V2X communications. In \cite{9448149}, a self-adaptive fuzzy logic-based strategy was proposed by Zhang \textit{et al.} for V2X communications.

The authors of \cite{9234747, 9226406} conducted a comprehensive survey of the key multiple access techniques conceived for aerial networks, advocating RSMA. A scalable and robust RS scheme was proposed in Cloud Radio Access Networks (C-RAN) based on statistical Channel State Information (CSI) \cite{9445019}. In \cite{8756076}, an efficient RSMA scheme was designed for the downlink of a C-RAN. However, these contributions only considered static users in the downlink. As a future advance, considering the mobility of users, the performance of RSMA was studied in \cite{9491092}, under realistic imperfect CSI at the Transmitter (CSIT), which always becomes outdated due to the user mobility and latency of the wireless network. Additionally, a new RSMA architecture was proposed in \cite{9519666} for Unmanned Aerial Vehicles (UAV). In a nutshell, all these contributions have shown the benefits of RSMA schemes in V2X communication systems.

\subsection{Vehicular Platoon}
The integration of efficient wireless communication protocols into vehicular platoon systems is also expected to improve the vehicular management strategy, by guaranteeing low-latency critical information exchange in the face of high mobility. In this context, it is of vital importance to provide each vehicle with the real-time movement information of nearby vehicles equipped with Vehicle-to-Vehicle (V2V) transmission techniques within the communication range. Numerous researches have studied the IoV supported by V2V communications. For instance, Thunberg \textit{et al.} \cite{8764451} proposed an analytical framework for integrating the characteristics of V2V communication into the physical mobility characteristics of IoVs. In \cite{8836641}, an integrated platoon control framework was developed by Xu \textit{et al.}, for heterogeneous vehicles traveling around realistic curved roads by considering the wireless communication delays. Additionally, Mei \textit{et al.} \cite{8485763} jointly optimized the radio resource allocation in the LTE-V2V network and the associated control parameters of IoVs. The objective of this work was to minimize the tracking error of vehicles and guarantee the reliability of V2V communication. An AV motion planning strategy was designed in \cite{9774893}, which was based on motion prediction and V2V communication. Since the NOMA principle is capable of increasing the capacity, the associated uplink transmission power allocation problem was formulated in \cite{9562539}, in the context of a UAV platoon system. Each UAV acted as a relay assisting in inter-vehicle communications. The problem of NOMA-based resource allocation was formulated and analyzed in V2X systems in \cite{9685664}. However, the aforementioned studies do not consider the utilization of RSMA in their platoon. Therefore, due to the inevitable channel estimation error and the limited CSI feedback \cite{4641946}, IoV platoons tend to suffer from packet loss during their information sharing since the position and velocity information become outdated and may lead to accidents.

To summarize, the RSMA-based IoV system provides reliable and robust wireless communication for the IoV system of the vehicular platoon. By relying on the RSMA, the vehicular platoon has less probability to suffer from the outage problem in emergency motion change, which enhances the safety of the vehicular platooning. On the other hand, the smooth and well-organized platoon control provides the IoV system with lower CSIT estimation error and higher channel gain, which improves the spectral and energy efficiency of the V2X communication. Finally, the FEEL downlink benefits from this efficient RSMA-based IoV platoon system in terms of latency.

\section{System Model}

\subsection{IoV Platoon System}
In this study, we design an IoV system to support the FEEL in vehicular platooning. In contrast to the conventional wireless communication systems, two major challenges should be met in the vehicular platoons: 1) the motion of the vehicles is highly dynamic, which leads to imperfect CSIT \cite{9491092}, 2) the FEEL and platoon control messages require both multicasting and unicasting. To tackle these problems, NOUM-enabled RSMA is considered \cite{8846706}. This RSMA scheme is specifically designed for the NOUM transmission scenario and is resilient to the imperfect CSIT in vehicular platooning. By relying on the RSMA, the platoon has less probability to suffer from the outage problem in emergency motion change \cite{9519666}, which enhances the safety of the vehicular platooning.

\begin{figure*}
\centering 
\includegraphics[width = 0.7\linewidth]{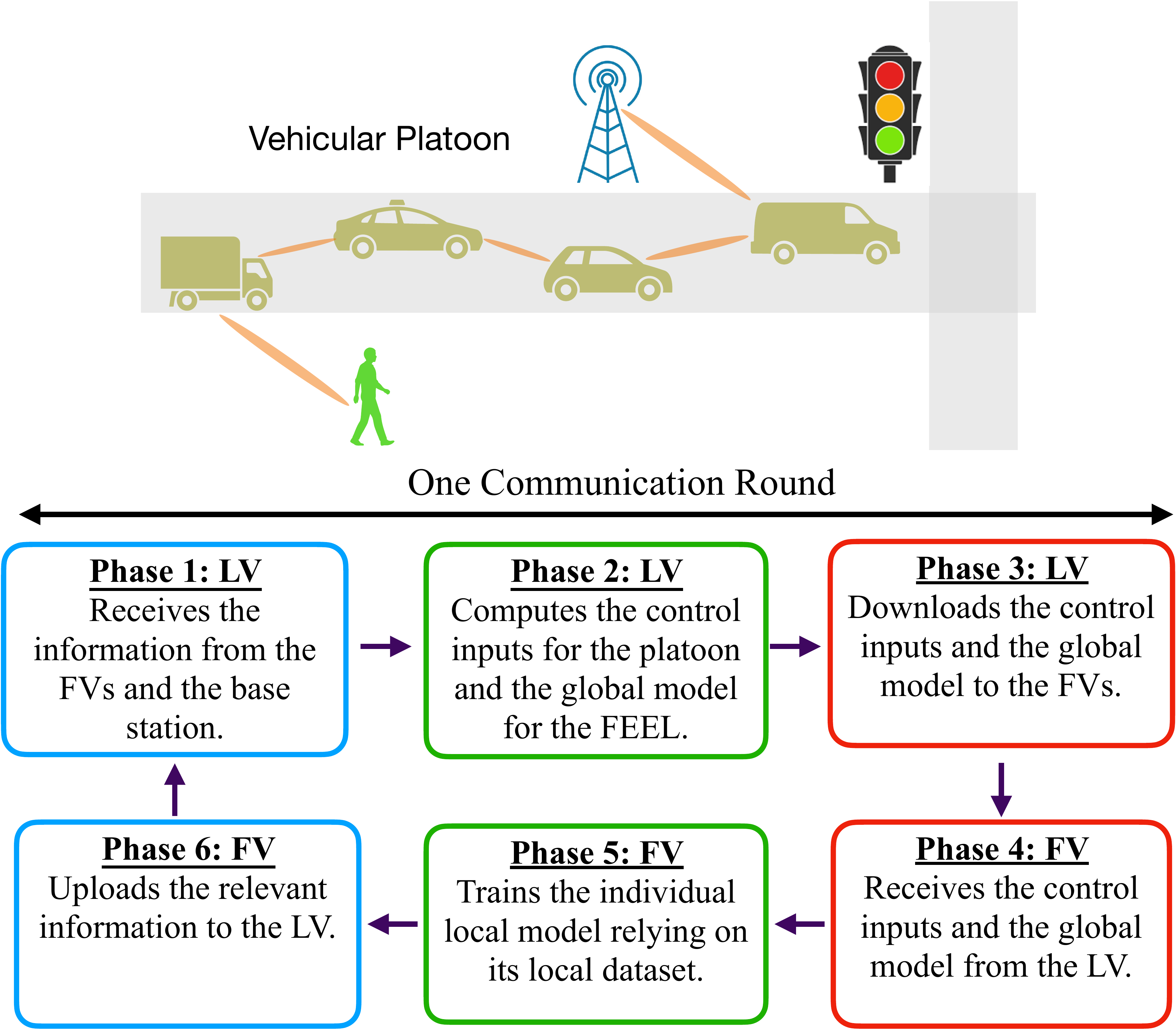} \\
\vspace{-0.4cm} 
\caption{Proposed system model.}
\label{figure_1}

\end{figure*}

The proposed RSMA-based platoon system relies on jointly solving the FEEL problem and platoon control problem  illustrated in Fig. \ref{figure_1}. It is comprised of three major components, namely the FEEL system, the RSMA communication system, and the platoon control system.  Each platoon is composed of a Lead Vehicle (LV) and a number of Follower Vehicles (FVs). Considering the Quality of Service (QoS) constraints, e.g., the outage problem, the scale of a single platoon is less than 10 vehicles. All these vehicles shall support RSMA for V2X communication, i.e. equipped with 1-layer SIC and multiple antennas. The LV is responsible for the entire platoon's behavioral decisions controlled by broadcasting its timely status and control information to the FVs via V2V communication. The FVs in the platoon are controlled by the LV for ensuring that they are traveling at the same speed, while maintaining a safe inter-vehicle distance. Since the FVs follow the LV and operate in a semi-autonomous control mode, these vehicles are capable of supporting additional services via their spare computing power, such as distributed learning \cite{9525183}. However, it is unlikely that all vehicles are willing to upload their local data to the LV due to privacy concerns. To tackle this problem, the FEEL system is proposed as a promising solution. In the system considered, each FV trains the individual local model relying on their local dataset, and then uploads the relevant information to the LV. Then, the LV updates the global model accordingly. After each iteration, the LV can download the global model to all the FVs. However, it is inefficient to transmit this global model via the unicast stream supporting the individual uploads to the LV. To address this issue, the NOUM-enabled RSMA concept is proposed. In this system, the FEEL model will be sent from the LV to the FVs via the super common message, and the platoon control results will be contained in private messages as well as common messages, which will be detailed as follows.

\subsection{Federated Edge Learning System}

According to \cite{flav}, the adoption of the FEEL system is capable of preserving privacy while reducing both the communication and annotation costs, rather than using the conventional centralized learning methods relying on data aggregation. Hence, we consider a Stochastic Gradient Descent (SGD) FL system comprising an LV, which acts as an edge server, and $K$ FVs. In each communication round, say the $n$-th round, the LV broadcasts its current model $\boldsymbol{w}^n$ to all the FVs. Each FV collects its own private data via interacting with the individual adjacent FV so as to constitute its local dataset. Let $\mathcal{D}_k$ denote the local dataset collected at the $k$-th FV. The local loss function of the model vector $\boldsymbol{w}$ on $\mathcal{D}_k$ is given by:
\vspace{-0.2cm} 
\begin{equation}\label{eqn_1}
\begin{aligned}
F_k(\boldsymbol{w})=\frac{1}{|\mathcal{D}_k|}\sum_{d_i\in \mathcal{D}_k}{f(\boldsymbol{w},d_i)},
\end{aligned}
\end{equation}
where $f(\boldsymbol{w},d_i)$ is the sample-wise loss function that quantifies the prediction errors of the model $\boldsymbol{w}$ w.r.t the $i$-th part of the training sample, namely $d_i$, $d_i\in \mathcal{D}_k$. In each communication round of the FEEL, say the $n$-th round, FV $k$ computes a local estimate of the gradient of the loss function in \eqref{eqn_1} using its local dataset $\mathcal{D}_k$ and the global model $\boldsymbol{w}^n$. Let $\boldsymbol{g}^n_k$ denote the local gradient estimate at the $k$-th FV in the $n$-th communication round. This local gradient estimate would then be uploaded to the LV, and the global estimate of the gradient of the loss function in \eqref{eqn_1} can be calculated as: 
\vspace{-0.2cm} 
\begin{equation}\label{eqn_2}
\begin{aligned}
\bar{\boldsymbol{g}}^{n}=\frac{1}{K}\sum_{k=1}^{K}\boldsymbol{g}^{n}_k.
\end{aligned}
\end{equation}
Then, the global model would be updated as:
\vspace{-0.2cm} 
\begin{equation}\label{eqn_3}
\begin{aligned}
\boldsymbol{w}^{n+1}=\boldsymbol{w}^{n}-\eta \bar{\boldsymbol{g}}^{n}, 
\end{aligned}
\end{equation}
where $\eta$ is the step size. Next, the global model $\boldsymbol{w}^{n+1}$ would be broadcast back to the FVs for the next iteration. The learning process involves iterations between \eqref{eqn_1} and \eqref{eqn_3}, until the model converges.

\subsection{RSMA Communication System}

In addition to the FEEL service, the LV can provide both communication and computing resources for other applications (e.g. platoon control). Thus, we consider a NOUM transmission system, where both the unicast and multicast services are activated in the same time-frequency resource blocks. We assume that the multicast service is provided for the FEEL broadcast, while the unicast service is mainly for platoon control. A single-layer RS-assisted NOUM is utilized to communicate with $K$ FVs, where the LV is equipped with $M$ antennas \cite{8846706}. The difference between the RS-assisted NOUM \cite{8846706} and the conventional RMSA \cite{rsma_2018,9491092,9831440} is that, in RS-assisted NOUM, the multicast message is jointly encoded with the unicast messages into a common-super stream, which would be decoded by all FVs.

Specifically, in the $t$-th time slot, FV $k$, $\forall k \in \mathcal{K}$, requires a dedicated unicast message $W_k(t)$ and a multicast message $W_0(t)$. The unicast messages $[W_1(t), ..., W_K(t)]$ are split into a common sub-message $W_k^c(t)$ and a private sub-message $W_k^p(t)$. The private sub-messages $[W^p_1(t),..., W^p_K(t)]$ of the unicast messages are independently encoded into private streams $[s_1(t), ..., s_K(t)]$, while the common sub-messages $[W_1^c(t), . . . , W_K^c(t)]$ of the unicast messages are jointly encoded with the multicast message $W_0(t)$ into a common-super stream $s_c(t)$, which has to be decoded by all FVs. These data streams, given by the set $\boldsymbol{s}(t)\triangleq [s_c(t), s_1(t), ..., s_K(t)]$, are linearly precoded by using the transmit precoding (TPC) matrix $\boldsymbol{P}(t) \triangleq [\boldsymbol{p}_c(t), \boldsymbol{p}_1(t), ...,\boldsymbol{p}_K(t)] \in \mathbb{C}^{M\times(M+1)}$. Here, $\mathbb{C}$ denotes the complex space. Thus, the transmitted signal is described as:
\vspace{-0.2cm} 
\begin{equation}\label{eqn_4}
\begin{aligned}
\boldsymbol{L}^{\text{tr}}(t)=\boldsymbol{P}(t) \boldsymbol{s}(t)=\boldsymbol{p}_c(t) s_c(t)+\sum_{k\in \mathcal{K}}\boldsymbol{p}_k(t) s_k(t). 
\end{aligned}
\end{equation}

The transmit signal vector $\boldsymbol{L}^{\text{tr}} = \boldsymbol{Ps}$ is subject to the power constraint $\mathbb{E}\{||\boldsymbol{L}^{\text{tr}}||^2\} \le P_t$. Here, $\mathbb{E}\{\cdot\}$ means the expectation. At the $t$-th time slot, the signal received at the $k$-th FV can be written as:
\vspace{-0.2cm} 
\begin{equation}\label{eqn_5}
\begin{aligned}
\boldsymbol{L}^{\text{re}}_k(t)=\textbf{h}^H_k(t)\boldsymbol{L}^{\text{tr}}(t)+n_k(t),
\end{aligned}
\end{equation}
where $\boldsymbol{L}^{\text{re}}_k(t)$ is the signal received by FV $k$, $\boldsymbol{h}_k(t) \in \mathbb{C}^{M\times1}$ is the fading channel vector of the link spanning from the LV to FV $k$, and $n_k$ is the circularly symmetric complex Additive White Gaussian Noise (AWGN) with zero-mean and variance $\sigma^2$, where the operator $(\cdot)^H$, represents the Hermitian transpose of a matrix. 

At the receiver side, each FV $k$ decodes both $s_c(t)$ and $s_k(t)$ by using a single layer of SIC. In particular, $s_c(t)$ is decoded first by treating the interference of all other streams as Gaussian noise, where FV $k$ can recover the desired common part $W_k^c(t)$. Then, the effect of $s_c(t)$ is cancelled from the received aggregate signal by SIC. This is achieved by remodulating $W_k^c(t)$ and then subtracting it from $s_c(t)$, leaving behind the desired private part $W_k^p(t)$. Based on $W_k^c(t)$ and $W_k^p(t)$, the desired message $W_k$ is retrieved. The corresponding instantaneous Signal-to-Interference-plus-Noise Ratio (SINR) of $W_k^c(t)$ and $W_k^p(t)$ is given by:
\vspace{-0.2cm} 
\begin{equation}\label{eqn_6}
\begin{aligned}
\gamma^c_k(t)=\frac{|\boldsymbol{h}^H_k(t)\boldsymbol{p}_c(t)|^2}{\sum_{i\in \mathcal{K}}|\boldsymbol{h}^H_k(t)\boldsymbol{p}_i(t)|^2+\sigma^2},
\end{aligned}
\end{equation}
\vspace{-0.2cm} 
\begin{equation}\label{eqn_7}
\begin{aligned}
\gamma^p_k(t)=\frac{|\boldsymbol{h}^H_k(t)\boldsymbol{p}_k(t)|^2}{\sum_{i\in \mathcal{K}  \setminus k}|\boldsymbol{h}^H_k(t)\boldsymbol{p}_i(t)|^2+\sigma^2}.
\end{aligned}
\end{equation}
To guarantee successful decoding by FV $k$, the achievable rates of $W_k^c(t)$ and $W_k^p(t)$ are bounded by:
\vspace{-0.2cm}
\begin{equation}\label{eqn_8}
\begin{aligned}
R^c_k(t)=B\log_2[1+\gamma^c_k(t)],
\end{aligned}
\end{equation}
\vspace{-0.4cm} 
\begin{equation}\label{eqn_9}
\begin{aligned}
R^p_k(t)=B\log_2[1+\gamma^p_k(t)],
\end{aligned}
\end{equation}
where $B$ is the bandwidth of the downlink channel. To guarantee that $W^c$ is successfully decoded by all FVs, its rate $R^c(t)$ must not exceed the value of $\min\{R^c_k(t) | k \in \mathcal{K}\}$. Since $R^c(t)$ is shared by the achievable rates when transmitting the multicast message $W_0(t)$ and the common sub-messages $\{W_1^c(t), . . . , W_K^c(t)\}$ of the unicast messages, we have $C_0(t)+\sum_{k\in \mathcal{K}} C_k(t) = R^c(t)$. Here, $C_0(t)$ represents the specific portion of $R_c(t)$ used for transmitting the broadcast message $W_0(t)$, while $\{C_k(t)| k \in \mathcal{K}\}$ represents the other parts employed for transmitting the unicast message $W_k^c(t)$. The particular portions of the rate, namely $C_0(t)$ and $\{C_k(t)| k \in \mathcal{K}\}$, which are allocated to $W_0(t)$ and $[W_1^c(t), . . . , W_K^c(t)]$ respectively, can be optimized by the associated optimization problem. In the proposed single-layer RS-assisted NOUM transmission, the achievable rate of each unicast message has two constituent parts. The first part is $C_k(t)$ transmitted via $W_k^c(t)$ that is mapped to the supercommon stream $s_c(t)$, while the other part is $R_k(t)$ transmitted via $W_k^p(t)$ that is mapped to the private stream $s_k(t)$. Hence, the achievable rate of transmitting the unicast message $W_k(t)$ of FV $k$ is:
\begin{equation}\label{eqn_10}
\begin{aligned}
R_k^{tot}(t) = C_k(t) + R_k(t).
\end{aligned}
\end{equation}
Due to the mobility of the platoon and the delay of CSI reports, we assume that the CSIT knowledge is imperfect, which can be modeled as \cite{rsma_2018,9491092,9831440,7555358}:
\begin{equation}\label{eqn_11}
\begin{aligned}
\boldsymbol{h}_k=\tilde{\boldsymbol{h}}_k+\hat{\boldsymbol{h}}_k,
\end{aligned}
\end{equation}
where $\tilde{\boldsymbol{h}}_k$ denotes the CSI estimate, and $\hat{\boldsymbol{h}}_k$ is the corresponding CSI estimation error at the transmitter. For each FV $k$, we define $\Gamma_k=|\boldsymbol{p}^H_k(t)\boldsymbol{p}_k(t)|^2\mathbb{E}\{\lVert\boldsymbol{h}_k(t)\lVert ^2 \}$, and assume that $\tilde{\boldsymbol{h}}_k$ and $\hat{\boldsymbol{h}}_k$ have zero means with channel gains of $\Gamma_k=\tilde{\Gamma}_k+\hat{\Gamma}_k$. Here, $\mathbb{E}\{\cdot\}$ means the expectation. Then, we can present $\tilde{\Gamma}_k=(1-\epsilon^2)\Gamma_k$, and $\hat{\Gamma}_k=\epsilon^2\Gamma_k$, where $\epsilon$ denotes a motion-related coefficient, which is calculated as:
\vspace{-0.2cm}
\begin{equation}\label{eqn_12}
\begin{aligned}
\epsilon=\frac{\lVert \boldsymbol{z}_k-\boldsymbol{z}^\text{est}_k \lVert_2}{2\lVert\boldsymbol{z}_\text{max}\lVert_2},
\end{aligned}
\end{equation}
where ${z}_k$ is the state of FV $k$, and $\boldsymbol{z}^\text{est}_k$ is the estimated state of FV $k$. For each $\boldsymbol{h}_k(t)$, the channel coefficient $h^j_k(t)$ $\forall j = \{1,...,M\}$ follows the path loss model of \cite{8647693}, and we have:
\vspace{-0.2cm}
\begin{equation}\label{eqn_13}
h^j_k(t)=d_k(t)^{-\mu},
\end{equation}
where $\mu$ is the path loss exponent and $d_k(t)$ is the distance between the LV and the FV $k$. Table \ref{table.P1} summarizes the parameters used in our system.

\begin{table}[t]
\small
\caption{Table of Notations for RSMA-based FEEL downlink System.}
\centering
\label{table.P1}
\begin{tabularx}{0.95\columnwidth}{cX} 
\toprule
\multicolumn{2}{c}{Parameters} 
\\ \cmidrule(r){1-2}
$\boldsymbol{w}^n$& Global model in $n$-th communication round\\
$\mathcal{D}_k$ & Local dataset collected at $k$-th FV\\
$d_i$ & The $i$-th part of training sample \\
$\boldsymbol{g}^n_k$ & Local gradient estimate at $k$-th FV in the $n$-th round\\
$\eta$ & Step size of the FEEL\\
$B_0$& Data volume of the global model to be broadcast\\
$M$ & Number of antennas in LV\\
$W_k(t)$ & Unicast message of $k$-th vehicle at time $t$\\
$W_0(t)$ & Broadcast message at time $t$\\
$s_k(t)$ & Private stream for $k$-th vehicle at time $t$\\
$s_c(t)$ & Public stream at time $t$\\
$\boldsymbol{L}^{\text{tr}}(t)$ & Transmitted signal\\
$\boldsymbol{L}^{\text{re}}(t)$ & Received signal\\
$P_t$ & Transmit power of LV\\
$\boldsymbol{h}_k(t)$ & Fading channel vector between from LV to FV $k$ at time $t$\\
$n_k(t)$ & Circularly symmetric complex AWGN\\
$\gamma^p_k(t)$& Instantaneous signal-to-interference-plus-noise ratio of $k$-th private message at time $t$\\
$\gamma^c_k(t)$& Instantaneous signal-to-interference-plus-noise ratio of common message received by $k$-th FV at time $t$\\
$B$& Bandwidth of downlink channel\\
$R^p_k(t)$& Achievable rate of transmitting unicast message $W_k(t)$ via private stream $s_k(t)$ at time $t$\\
$R^c_k(t)$& Achievable rate for transmitting common message to $k$-th FV at time $t$\\
$R^\text{tot}_k(t)$& Achievable rate of transmitting unicast message $W_k(t)$ at time $t$\\
$R^\text{th}_k(t)$& Transmission rate requirement for sending control message to $k$-th FV \\
$\tilde{\boldsymbol{h}}_k$ & Estimate of channel vector\\
$\hat{\boldsymbol{h}}_k$ & Corresponding channel estimation error\\
$\Gamma_k$ & Average channel (power) gain\\
$\tilde{\Gamma}_k$& Estimated average channel (power) gain\\
$\hat{\Gamma}_k$& Corresponding estimation error of average channel (power) gain\\
$\epsilon$ & Motion coefficient for estimated average channel (power) gain\\
$d_k(t)$ & Distance between LV and $k$-th FV\\
$\mu$ & Free space path loss factor\\
$Q_t$ and $Q_h$ & Weight factors\\
\midrule
\multicolumn{2}{c}{Optimization variables} 
\\ \cmidrule(r){1-2}
$\boldsymbol{P}$ & Precoding matrix\\
$C_k(t)$& Portion of $R^c(t)$ for transmitting $k$-th unicast message at time $t$\\
$C_0(t)$& Portion of $R^c(t)$ for transmitting broadcast message at time $t$\\
$\psi(t)$ & A bit-mask that $\psi(t)=1$ means $W_0(t)$ is added into common-super messages to transmit the FEEL model to FV; otherwise, $\psi(t)=0$\\
\bottomrule
\end{tabularx}
\label{notations}
\end{table}

\subsection{IoV Platoon Control System}

The platoon considered is modeled based on \cite{DBLP:journals/corr/abs-2003-08595}, which indicates the nonlinear behavior of every vehicle in the system. In this model, each vehicle is regarded as a polytopic set, and the $k$-th vehicle's state vector is denoted by $\boldsymbol{z}_k=[x_k,y_k,\phi_k,v_k]^T$, where $(x_k,y_k)$ represents the two-dimensional (2D) position of the vehicle. Additionally, $\phi_k$ is the heading angle and $v_k$ denotes the velocity of the vehicle. At time $t$, the control input is set to $\boldsymbol{u}_k(t)=[a_k,\delta_k]$, where $a_k$ is the acceleration rate and $\delta_k$ is the steering angle of vehicle $k$. Then, the side slip angle $\beta_k$ is formulated as:
\vspace{-0.2cm}
\begin{equation}\label{eqn_14}
\begin{aligned}
\beta_k=\text{arctan}[\text{tan}\frac{\delta_k l^r_k}{l^f_k+l^r_k}],
\end{aligned}
\end{equation}
where $l^f_k$ and $l^r_k$ represent the distance spanning from the front and rear axles, respectively. Using the vehicular kinematic bicycle model \cite{DBLP:journals/corr/abs-2003-08595}, the vehicle's state evolution model is formulated as:
\vspace{-0.2cm} 
\begin{subequations}\label{eqn_15}
\begin{align}
&x_k(t+1)=x_k(t)+ v_k(t) \cos[\phi_k(t)+\beta_k(t)]\Delta t,\\
&y_k(t+1)=y_k(t)+ v_k(t) \sin[\phi_k(t)+\beta_k(t)]\Delta t,\\
&\phi_k(t+1)=\phi_k(t)+ \frac{v_k(t)\beta_k(t)\tan[\delta_k(t)]}{l^f_k+l^r_k}\Delta t,\\
&v_k(t+1)=v_k(t)+ a_k(t)\Delta t.
\end{align}
\end{subequations}

To avoid any intersection between the vehicles, we define an orthogonal rotation matrix $\boldsymbol{RO}$ and a translation vector $\boldsymbol{J}_r$. This matrix is a function of the vehicle's bearing angle $\phi_k (t)$ at time step $t$, expressed as:
\vspace{-0.2cm}
\begin{equation}\label{eqn_16}
\begin{aligned}
\boldsymbol{RO} [\phi_k (t)]=\begin{bmatrix}
\cos[\phi_k (t)] & -\sin[\phi_k (t)]\\ \sin[\phi_k (t)] & \cos[\phi_k (t)].
\end{bmatrix}
\end{aligned}
\end{equation}

The translation vector $\boldsymbol{J}_r(\cdot)$ is a function of the longitudinal $x_k(t)$ and lateral $y_k(t)$ positions of the vehicle. Hence, the transformed polytope can be represented by $x$ and $y$ coordinates in 2D space. The matrix $\boldsymbol{A}$ and the vector $\boldsymbol{b}$ representing the polytopic sets of the vehicles are defined as:

\begin{equation}\label{eqn_17}
\begin{aligned}
\boldsymbol{A}[\boldsymbol{z}_k(t)]=
\begin{bmatrix}
\boldsymbol{RO} [\phi_k (t)]^T\\-\boldsymbol{RO} [\phi_k (t)]^T
\end{bmatrix},
\end{aligned}
\end{equation}
\vspace{-0.2cm}
\begin{equation}\label{eqn_18}
\begin{aligned}
\boldsymbol{b}(\boldsymbol{z}_k(t))=&
\begin{bmatrix}
\frac{\lambda^{l}}{2},\frac{\lambda^{w}}{2},\frac{\lambda^{l}}{2},\frac{\lambda^{w}}{2}
\end{bmatrix}^T +\boldsymbol{A}(\boldsymbol{z}_k(t))
\begin{bmatrix}
x_k(t),y_k(t)
\end{bmatrix}^T,
\end{aligned}
\end{equation}
where $(\cdot)^{T}$ represents the transpose of the matrix, while $\lambda^{l}$ is the length and $\lambda^{w}$ is the width of the vehicle. Since the area occupied by each vehicle in the system is modeled as a time-varying polytope, at each time step, motion planning is activated for preventing any intersection among the polytopic sets. Table \ref{table.P2} shows the parameters of the platoon control system.

\subsection{Multi-Objective Problem Formulation}
In this section, we formulate our platoon control and vehicular communication problem. The platoon control problem and the vehicular problem are jointly considered because the motion of the platoon would affect the estimation error of the CSIT, and the relative position of the vehicles in the platoon would also determine the interference of the channel. We commence by introducing a bit-mask $\psi(t)$. If $\psi(t)=1$, $W_0(t)$ is mapped into the common-super messages transmitting the FEEL model to the FV; otherwise, it is set to $\psi(t)=0$. We denote the maximum time period of the FEEL downlink process by $T$ and the data volume of the global model to be broadcast from the LV to FVs by $B_0$. Then, for $\forall k\in \mathcal{K}, t\in \mathcal{T}$, the optimization can be formulated as:
\vspace{-0.2cm} 
\begin{subequations}\label{eqn_19}
\allowdisplaybreaks
\begin{align}
&\min \bigg\{Q_t\max_{t}\{\psi(t)t\}+Q_h\sum_{\forall k\in \mathcal{K}}\sum_{t \in \mathcal{T}} \hat{\Gamma}_k+\notag \\
&\qquad \sum_{\forall k\in \mathcal{K}}[\sum_{t \in \mathcal{T}}\lVert Q_z(\boldsymbol{z}_k(t)-\boldsymbol{z}^{\text{Ref}}_k(t)) \lVert ^2_2+\notag \\
&\qquad \sum_{t \in \mathcal{T}}\lVert Q_u\boldsymbol{u}_k(t) \lVert ^2_2+\lVert Q_{\Delta u}\Delta \boldsymbol{u}_k(t) \lVert ^2_2]\bigg\}, \label{19a}\\
&\ \text{s.t.} \quad   C_k(t)+R_k^p(t)\ge R_k^\text{th}(t), \label{19b}\\
&\ \qquad \sum_{t\in \mathcal{T}} \psi(t)C_0(t)\Delta t  \ge B^0, \label{19c}\\
&\ \qquad \psi(t)C_0(t)+\sum_{j\in \mathcal{K}}C_j(t)\le R^c(t), \label{19d}\\
&\ \qquad C_k(t), C_0(t) \ge 0,\label{19e}\\
&\ \qquad \text{trace}(\boldsymbol{P}(t)\boldsymbol{P}^H(t))\le P_t, \label{19f}\\
&\ \qquad \boldsymbol{z}_k(t+1)=f[\boldsymbol{z}_k(t),\boldsymbol{u}_k(t)], \label{19g}\\
&\ \qquad \boldsymbol{z}_\text{min} \le \boldsymbol{z}_k(t) \le \boldsymbol{z}_\text{max}, \label{19h}\\
&\ \qquad \boldsymbol{u}_\text{min} \le \boldsymbol{u}_k(t) \le \boldsymbol{u}_\text{max}, \label{19i}\\
&\ \qquad \Delta\boldsymbol{u}_\text{min} \le \boldsymbol{u}_k(t)- \boldsymbol{u}_k(t-1)\le \Delta\boldsymbol{u}_\text{max}, \label{19j}\\
&\ \qquad \mathcal{P}(\boldsymbol{z}_k(t))\cap \mathcal{P}(\boldsymbol{z}_u(t))=\emptyset,\    \text{if} \ k\ne u,  \label{19k}
\end{align}
\end{subequations}
where \eqref{19a} represents a multi-objective optimization problem. The first term minimizes the latency of the FEEL downlink, where $t$ is the index of the time slot. Thus, $\max_{t}\{\psi(t)t\}$ means that, for any $\psi(t)=1$, the largest time slot index within the maximum delay tolerance $T$, $t \in \mathcal{T}$, would be the latency of a single round of FEEL. The second term represents the penalty owing to the CSI changes and the power consumption. The third term penalizes deviation of the states $\boldsymbol{z}_k(t)$ from the predetermined reference states $\boldsymbol{z}^{\text{Ref}}_k(t)$. The fourth term penalizes the control input effort $u$, which minimizes the change of the vehicular state. In this case, all the vehicles in the platoon would operate in a smooth manner. The fifth term penalizes the input rate (change of control input in two consecutive time steps) $\Delta u$. The parameter $R_k^\text{th}(t)$ in \eqref{19b} represents the transmission rate required for the control message of the platoon, the constraint \eqref{19c} ensures the downlink process of the FEEL, Eq. \eqref{19d} ensures that the common-super stream can be successfully decoded by all FVs, constraint \eqref{19f} ensures that the transmission power constraint of the LV is met, while the operator trace($\cdot$) represents the trace, and finally, the weight factors $Q_t$, $Q_h$, $Q_z$, $Q_u$ and $Q_{\Delta t}$ are positive semidefinite matrices. Furthermore, the function $f(\cdot)$ in \eqref{19g} represents the vehicular kinematic bicycle model \eqref{eqn_15}, for the case of safe driving, \eqref{19h}-\eqref{19j} assist in avoiding heavy changes of the vehicles' states, Eq. \eqref{19k} is developed to avoid collision between the $k$-th vehicle and the others by means of \eqref{eqn_16}-\eqref{eqn_18}.

\begin{table}[t]
\small
\caption{Table of Notations for Platoon Control System.}
\centering
\label{table.P2}
\begin{tabularx}{0.95\columnwidth}{cX} 
\toprule
\multicolumn{2}{c}{Parameters} 
\\ \cmidrule(r){1-2}
$t$ & Index set of time period\\ 
$T$ & Total number of time slot\\
$\Delta t$ & Time slot \\
$\zeta$ & Number of time steps in each horizon \\
$K$ & Number of vehicles \\
$k$ & Index of $k$-th vehicle \\
$\boldsymbol{z}_k$ & States of the $k$-th vehicle\\
$\boldsymbol{z}^\text{Ref}_k$ & Reference states of the $k$-th vehicle\\
$\boldsymbol{u}_k$ & Control input of the $k$-th vehicle\\
$\lambda^l$ & Length of the vehicle  \\
$\lambda^w$ & Width of the vehicle\\
$\beta_k$ & Side slip angle of the $k$-th vehicle\\
$l^f_k$, $l^r_k$ & Distance from the front and rear axles of the $k$-th vehicle\\
$\mathbf{RO}$ & Orthogonal rotation matrix \\
$\textbf{t}_{r}$ & Translation vector \\
$\boldsymbol{A}$ and $\boldsymbol{b}$& Representation of polytopic sets $k$\\
$\mathcal{P}$ & Polytopic representation of the vehicle\\
$f$ & Representation of vehicular kinematic bicycle model\\
$Q_z$, $Q_u$, and $Q_{\Delta t}$& Weight factors\\
\midrule
\multicolumn{2}{c}{Optimization variables} 
\\ \cmidrule(r){1-2}
$v_k$ & Velocity of the vehicle $k$\\
$a_k$ & Acceleration of the vehicle $k$\\
$\phi_k$ & Heading angle of the vehicle $k$\\
$\delta_k$ & Steering angle of the vehicle $k$\\
$x^{k}$ & Longitudinal position of the vehicle $k$\\
$y^{k}$ & Lateral position of the vehicle $k$\\

\bottomrule
\end{tabularx}
\label{notations}
\end{table}

\section{RSMA-Based IoV Operation Framework}

\subsection{Block Coordinate Descent Framework}
In problem \eqref{eqn_19}, the RSMA part is relevant to the platoon part, since they are coupled in the constraint \eqref{eqn_11}-\eqref{eqn_13}. Specifically, in \eqref{eqn_12}, the variable $z_k$ determines the estimation error of the CSIT, which is related to the real-time motion of the platoon. Moreover, in \eqref{eqn_13} the relative distance between vehicles $d_k$ is also determined by the position of the vehicles in the platoon. Therefore, the larger deviation of the platoon control inputs would lead to a larger error in the estimation of the CSIT and to grace interference.

Since the constraint set of \eqref{eqn_19} is decoupled when either the set ${[\psi(t), p_k(t), C_k(t)]}$ or the set ${[u_k(t), z_k(t)]}$ is fixed, the original problem can be decomposed into two sub-problems by utilizing the Block Coordinate Descent (BCD) framework. When ${u_k(t), z_k(t)}$ is fixed, one of the sub-problem of \eqref{eqn_19} for the FEEL downlink is given by:
\vspace{-0.2cm} 
\begin{subequations}\label{eqn_20}
\begin{align}
&\min Q_t\max_{t}\{\psi(t)t\}+Q_h\sum_{\forall k\in \mathcal{K}}\sum_{t \in \mathcal{T}} \hat{\Gamma}_k, \label{20a}\\
&\ \text{s.t.} \quad \eqref{19b}-\eqref{19g}. \label{20b}
\end{align}
\end{subequations}
On the other hand, the platoon control sub-problem is formulated as:
\vspace{-0.2cm}
\begin{subequations}\label{eqn_21}
\begin{align}
&\min Q_h\sum_{\forall k\in \mathcal{K}}\sum_{t \in \mathcal{T}} \hat{\Gamma}_k+ \sum_{\forall k\in \mathcal{K}}[\sum_{t \in \mathcal{T}}\lVert Q_z(\boldsymbol{z}_k(t)-\boldsymbol{z}^{\text{Ref}}_k(t)) \lVert ^2_2 \notag\\
&\qquad + \sum_{t \in \mathcal{T}}\lVert Q_u\boldsymbol{u}_k(t) \lVert ^2_2+\lVert Q_{\Delta u} \Delta \boldsymbol{u}_k(t) \lVert ^2_2], \label{21a}\\
&\ \text{s.t.} \quad   \eqref{19h}-\eqref{19k}. \label{21b}
\end{align}
\end{subequations}

The related BCD framework is illustrated in Algorithm \ref{Algorithm1}. In each iteration, we solve the sub-problem \eqref{eqn_20} by considering that the ${[u_k(t), z_k(t)]}$ is fixed. The set ${[\psi(t), p_k(t), C_k(t)]}$ is updated according to the optimization results. Meanwhile, the set ${[u_k(t), z_k(t)]}$ is updated according to the result of \eqref{eqn_21}. This operation repeats until the stopping criterion is satisfied or the maximum number of iterations is reached.

\begin{algorithm}[ht]
\small
\caption{BCD Framework}
\label{Algorithm1}
\begin{algorithmic}[1]
\STATE Initialize the parameters ${p^{[0]}_k(t), C^{[0]}_k(t), u^{[0]}_k(t), z^{[0]}_k(t)}$.
\STATE Set iteration count as $n=0$ and maximum number of iterations as $N_c$.
\WHILE{$n\le N_c$}
\STATE $n=n+1$.
\STATE Update $[p^{[n]}_k(t), C^{[n]}_k(t)]$ by solving \eqref{eqn_20}.
\STATE Update $[u^{[n]}_k(t), z^{[n]}_k(t)]$ by solving \eqref{eqn_21}.
\IF {the stopping criterion is satisfied}
\STATE \textbf{break}
\ENDIF
\ENDWHILE
\end{algorithmic}
\end{algorithm}

\subsection{SCA-based FEEL downlink algorithm}

Since the partial sub-problem \eqref{eqn_20} is a non-convex problem, we propose a SCA-based FEEL downlink algorithm for efficiently solving this problem. First of all, by introducing two scalar variables $\varphi$ and $\omega(t)$, the FEEL offload latency problem \eqref{eqn_20} is equivalently transformed as:
\vspace{-0.2cm} 
\begin{subequations}\label{eqn_22}
\allowdisplaybreaks
\begin{align}
&\min \varphi+Q_h\sum_{\forall k\in \mathcal{K}}\sum_{t \in \mathcal{T}} \hat{\Gamma}_k, \label{22a}\\
&\ \text{s.t.} \quad   \psi(t)t \le \varphi, \label{22b}\\
&\ \qquad \sum_{t\in \mathcal{T}} \omega(t) \ge B^0/\Delta t,  \label{22c}\\
&\ \qquad \psi(t)C_0(t)\ge\omega(t),  \label{22d}\\
&\ \qquad \eqref{19b}-\eqref{19g}. \label{22e}
\end{align}
\end{subequations}

The equivalence between \eqref{eqn_20} and \eqref{eqn_22} is established based on the fact that the constraint \eqref{22b} must hold with equality at the optimum. The challenge of solving \eqref{eqn_22} results from the non-convex constraints \eqref{19b} and \eqref{19d}. Thus, we introduce the variables $\boldsymbol{\alpha}(t)=[\alpha_1(t),...,\alpha_K(t)]$ to represent the set of private rates. Then, the constraint \eqref{19b} becomes equivalent to:
\vspace{-0.2cm}
\begin{subequations}
  \label{eqn_23}
    \begin{empheq}[left={\eqref{19b} \Leftrightarrow \empheqlbrace\,}]{align}
      & C_k(t)+\alpha_k(t)\ge R_k^{th}(t), \ \forall k\in \mathcal{K}, t\in \mathcal{T},  \label{23a} \\
      &  R_k(t)\ge \alpha_k(t) , \ \forall k\in \mathcal{K}, t\in \mathcal{T}. \label{23b}
    \end{empheq}
\end{subequations}

Moreover, to deal with the non-convex constraint \eqref{23b}, we introduce the variables $\boldsymbol{\vartheta}(t)=[\vartheta_1(t),...,\vartheta_K(t)]$, and then, the constraint \eqref{23b} is derived as
\vspace{-0.2cm} 
\begin{subequations}
  \label{eqn_24}
    \begin{empheq}[left={\eqref{23b} \Leftrightarrow \empheqlbrace\,}]{align}
      & \vartheta_k(t)\ge 2^{\alpha_k(t)}, \ \forall k\in \mathcal{K}, t\in \mathcal{T},  \label{24a} \\
      &  1+\gamma^p_k(t) \ge \vartheta_k(t) , \ \forall k\in \mathcal{K}, t\in \mathcal{T}. \label{24b}
    \end{empheq}
\end{subequations}

Furthermore, by introducing $\boldsymbol{\xi}(t)=[\xi_1(t),...,\xi_K(t)]$ to represent the interference plus noise at each FV to decode its private stream, the constraint \eqref{24b} is transformed as:
\begin{subequations}
  \label{eqn_25}
    \begin{empheq}[left={\eqref{24b} \Leftrightarrow \empheqlbrace\,}]{align}
      & \frac{|\boldsymbol{h}^H_k(t)\boldsymbol{p}_k(t)|^2}{\xi_k(t)} \ge \vartheta_k(t)-1,\notag
      \\ & \forall k\in \mathcal{K}, t\in \mathcal{T}, \label{25a} \\
      &  \xi_k(t) \ge \sum_{i\in \mathcal{K} \setminus k}|\boldsymbol{h}^H_k(t)\boldsymbol{p}_i(t)|^2+\sigma^2, \notag
      \\ &\forall k\in \mathcal{K}, t\in \mathcal{T}. \label{25b}
    \end{empheq}
\end{subequations}

Similarly, we introduce the variables $\boldsymbol{\alpha}^c(t)$, $\boldsymbol{\vartheta}^c(t)$, $\boldsymbol{\xi}^c(t)$ for the common rate constraint \eqref{19d}, and we have:
\begin{subequations}
  \label{eqn_26}
  \allowdisplaybreaks
    \begin{empheq}[left={\eqref{19d}\Leftrightarrow \empheqlbrace\,}]{align}
      & C_0(t)+\sum_{j\in \mathcal{K}}C_j(t)\le \alpha^c_k(t),\ ,\ \forall t\in \mathcal{T}  \label{26a} \\
      &  \vartheta^c_k(t)\ge 2^{\alpha^c_k(t)},\ \forall k\in \mathcal{K}, t\in \mathcal{T}, \label{26b} \\
      &  \frac{|\boldsymbol{h}^H_k(t)\boldsymbol{p}_c(t)|^2}{\xi^c_k(t)} \ge \vartheta^c_k(t)-1,\notag
      \\ & \forall k\in \mathcal{K}, t\in \mathcal{T}, \label{26c}\\
      &\xi^c_k(t) \ge \sum_{i\in \mathcal{K}}|\boldsymbol{h}^H_k(t)\boldsymbol{p}_i(t)|^2+\sigma^2,\notag
      \\ & \forall k\in \mathcal{K}, t\in \mathcal{T}. \label{26d}
    \end{empheq}
\end{subequations}

However, the constraints \eqref{22d}, \eqref{25a} and \eqref{26c} are still non-convex. So the linear approximation method is utilized for approximating the relevant non-convex constraints in each iteration. The left side of constraint \eqref{22d} is approximated by using the first-order lower approximation \cite{8846706}, which is given by:
\vspace{-0.2cm}
\begin{equation}\label{eqn_27}
\begin{aligned}
\psi(t)C_0(t) &\ge \psi^{[n]}(t)C_0(t)+C_0^{[n]}(t)\psi(t)\\&-\psi^{[n]}(t)C_0^{[n]}(t) \triangleq\Omega^{[n]}[\psi^{[n]}(t),C_0^{[n]}(t)],
\end{aligned}
\end{equation}
where $[\psi^{[n]}(t),C_0^{[n]}(t)]$ are the values of the variables $[\psi(t),C_0(t)]$ at the output of the $n$-th iteration. The left side of the constraints \eqref{25a} and \eqref{26c} can be represented by using the linear lower bound approximation at the point $[\boldsymbol{p}_k^{[n]}(t), \xi_k^{[n]}(t)]$ and $[\boldsymbol{p}_c^{[n]}(t), \xi_k^{c[n]}(t)]$, respectively. Then, we have:
\vspace{-0.1cm}
\begin{equation}\label{eqn_28}
\begin{aligned}
\frac{|\boldsymbol{h}^H_k(t)\boldsymbol{p}_k(t)|^2}{\xi_k(t)} &\ge  2\text{Re}[(\boldsymbol{p}_k^{[n]}(t))^H\boldsymbol{h}_k(t)\boldsymbol{h}^H_k(t)\boldsymbol{p}_k(t)]\\
&/\xi_k^{[n]}(t)-[|\boldsymbol{h}^H_k(t)\boldsymbol{p}_k^{[n]}(t)|/\xi_k^{[n]}(t)]^2\xi_k(t)
\\&\triangleq \Psi_k^{[n]}[\boldsymbol{p}_k^{[n]}(t),\xi_k^{[n]}(t)],
\end{aligned}
\end{equation}
\vspace{-0.2cm}
\begin{equation}\label{eqn_29}
\begin{aligned}
\frac{|\boldsymbol{h}^H_k(t)\boldsymbol{p}_c(t)|^2}{\xi^c_k(t)} &\ge  2\text{Re}[(\boldsymbol{p}_c^{[n]}(t))^H\boldsymbol{h}_k(t)\boldsymbol{h}^H_k(t)\boldsymbol{p}_c(t)]\\
&/\xi_k^{c[n]}(t)-[|\boldsymbol{h}^H_k(t)\boldsymbol{p}_c^{[n]}(t)|/\xi_k^{c[n]}(t)]^2\xi^c_k(t)
\\&\triangleq \Psi_k^{c[n]}[\boldsymbol{p}_c^{[n]}(t),\xi_k^{c[n]}(t)].
\end{aligned}
\end{equation}
Here, $\text{Re}\{\cdot\}$ gets the real part of a complex
number. Based on the approximations \eqref{eqn_27}, \eqref{eqn_28} and \eqref{eqn_29}, the problem \eqref{eqn_20} is approximated at the $n$-th iteration as:
\begin{subequations}\label{eqn_30}
\allowdisplaybreaks
\begin{align}
&\min \varphi+Q_h\sum_{\forall k\in \mathcal{K}}\sum_{t \in \mathcal{T}} \hat{\Gamma}_k, \label{30a}\\
&\ \text{s.t.} \quad \Omega^{[n]}[\psi^{[n]}(t),C_0^{[n]}(t)] \ge \omega(t),\ \forall t\in \mathcal{T} \label{30b}
\\&\ \qquad  \Psi_k^{[n]}[\boldsymbol{p}_k^{[n]}(t),\xi_k^{[n]}(t)] \ge \vartheta_k(t)-1,\ \forall k\in \mathcal{K}, t\in \mathcal{T}, \label{30c}\\
&\ \qquad \Psi_k^{c[n]}[\boldsymbol{p}_c^{[n]}(t),\xi_k^{c[n]}(t)] \ge \vartheta^c_k(t)-1,\ \forall k\in \mathcal{K}, t\in \mathcal{T}, \label{30d}\\
&\qquad \eqref{19b}-\eqref{19g},\eqref{22b},\eqref{22c},\eqref{23a},\notag\\
&\qquad \eqref{24a},\eqref{25a},\eqref{26a},\eqref{26b},\eqref{26d}. \label{30e}
\end{align}
\end{subequations}

In this case, the problem formulated in \eqref{eqn_30} is convex. The SCA-based FEEL offload algorithm is outlined in Algorithm \ref{Algorithm2}. In each iteration, \eqref{eqn_30} is solved via the interior-point method and $\psi^{[n]}(t)$, $\ C_0^{[n]}(t)$, $\boldsymbol{P}^{[n]}(t)$, $\boldsymbol{\xi}^{[n]}(t)$, $\boldsymbol{\xi}^{c[n]}(t)$ are updated using the corresponding optimized variables. The beamformer $P^{[0]}$ is initialized by finding the feasible beamformer satisfying the transmit power constraint \eqref{19g}. Initially, we set $C_0(t) = R^c(t)$, $C_k(t)=0$, $\forall k \in \mathcal{K}, \forall t\in \mathcal{T}$. Furthermore, $\psi^{[0]}(t)$, $C_0^{[0]}(t)$, $\boldsymbol{\xi}^{[0]}(t)$, $\boldsymbol{\xi}^{c[0]}(t)$ are initialized by individually replacing the inequalities of \eqref{22d}, \eqref{25b} and \eqref{26d} by the related equities. Besides, $\varphi^{[n]}$ is the minimum FEEL offload latency at the output of the $n$-th iteration and $\varsigma$ is the tolerance bound of the algorithm.

\begin{algorithm}[ht]
\small
\caption{SCA-Based FEEL Offload Algorithm}
\label{Algorithm2}
\begin{algorithmic}[1]
\REQUIRE Initial parameters $\varphi^{[0]}$, $\psi^{[0]}(t)$, $C_0^{[0]}(t)$, $\boldsymbol{P}^{[0]}(t)$, $\boldsymbol{\xi}^{[0]}(t)$, $\boldsymbol{\xi}^{c[0]}(t)$.
\ENSURE Minimum FEEL offload latency $\varphi$.

\STATE Iteration count $n=0$ and maximum number of iterations $N$.
\WHILE{$n\le N$}
\STATE $n=n+1$
\STATE Solve the problem \eqref{eqn_30} by inputting $\psi^{[n-1]}(t)$, $C_0^{[n-1]}(t)$, $\boldsymbol{P}^{[n-1]}(t)$, $\boldsymbol{\xi}^{[n-1]}(t)$, $\boldsymbol{\xi}^{c[n-1]}(t)$. 
\STATE Obtain the optimal values $\varphi^{[n]}=\varphi^*,\ \psi^{[n]}(t)=\psi^*(t),\ C_0^{[n]}(t)=C_0^*(t),\ \boldsymbol{P}^{[n]}(t)=\boldsymbol{P}^*(t),\ \boldsymbol{\xi}^{[n]}(t)=\boldsymbol{\xi}^*(t),\ \boldsymbol{\xi}^{c[n]}(t)=\boldsymbol{\xi}^{c*}(t)$.
\IF {$n>N$ or $|\varphi^{[n]}-\varphi^{[n-1]}|<\varsigma$}
\STATE \textbf{return} the minimum value $\varphi^{[n]}$.
\ENDIF
\ENDWHILE
\end{algorithmic}
\end{algorithm}

\subsubsection{Convergence Analysis}
As \eqref{22d}, \eqref{25a} and \eqref{26c} are relaxed by the first-order lower bounds \eqref{eqn_27}-\eqref{eqn_29}, the solution of the problem \eqref{eqn_30} at iteration [$n$] is also a feasible solution at iteration [$n+1$]. Therefore, the optimized objective is non-increasing as the iteration index increases, and $\varphi^{[n+1]}\le\varphi^{[n]}$ always holds. As the $\varphi^{[n]}$ above is bounded by the transmission power and the maximum FEEL downlik latency, the proposed algorithm is guaranteed to converge. Due to the linear approximation of the constraints \eqref{22d}, \eqref{25a} and \eqref{26c}, the global optimality of the solution cannot be guaranteed.

\subsection{MPC-Based Platoon Control}

Since the first partial sub-problem in \eqref{eqn_30} mainly deals with the FEEL downlink resource allocation problem, we further consider the other partial sub-problem that tackles the proposed IoV platoon control system, which is shown in \eqref{eqn_21}. Note that this sub-problem is also decoupled from the main problem in \eqref{eqn_19}. We harness the MPC approach for efficiently solving the platoon control problem. The proposed MPC-based platoon control problem uses a receding horizon technique. Specifically, instead of solving the platoon control problem in its entirety, the optimization problem is partitioned into slices along the time horizon. In each short time horizon, the optimization problem is solved, and the first time step of the control input is implemented. This facilitates the online control of the platoon. We denote the number of time steps in each horizon by $\zeta$, and then, the sub-problem \eqref{eqn_21} is reformulated as:
\vspace{-0.2cm} 
\begin{subequations}\label{eqn_31}
\begin{align}
&\min \bigg\{ Q_h\sum_{\forall k\in \mathcal{K}}\sum_{t_0\le t\le t_0+\zeta} p_k(t)\Delta h_k(t)+ \notag\\
&\qquad \sum_{\forall k\in K}[\sum_{t_0\le t\le t_0+\zeta}\lVert Q_z(\boldsymbol{z}_k(t)-\boldsymbol{z}^{\text{Ref}}_k(t)) \lVert ^2_2+\notag\\
&\qquad \sum_{t_0\le t\le t_0+\zeta}Q_u\lVert \boldsymbol{u}_k(t) \lVert ^2_2+\lVert Q_{\Delta u} \Delta \boldsymbol{u}_k(t) \lVert ^2_2] \bigg\}, \label{31a}\\
&\ \text{s.t.} \quad   \eqref{19h}-\eqref{19k}. \label{31b}
\end{align}
\end{subequations}

The entire process of the MPC-based platoon control operation is illustrated in Algorithm \ref{Algorithm3}. For every iteration, $m$ of time horizon $[t_0,t_0+\zeta]$, the optimization problem \eqref{eqn_31} is solved via the interior-point method so as to obtain the solution, $\boldsymbol{z}^m$ and $\boldsymbol{u}^m$. Once the convergence conditions are satisfied, the first step of the results, $\boldsymbol{z}^m(t_0)$ and $\boldsymbol{u}^m(t_0)$ is implemented. As shown in Algorithm \ref{Algorithm3}, in order to satisfy the stopping criterion, the maximum number of iterations is set to $M_p$. Moreover, for the convergence of this algorithm, the tolerance is set to $\varepsilon_p=10^{-5}$. For the next time horizon $[t_0+1,t_0+\zeta+1]$, the previous results $\boldsymbol{z}^m(t_0)$ and $\boldsymbol{u}^m(t_0)$ are utilized for the initial states of the platoon.

\begin{algorithm}[ht]
\small
\caption{MPC-Based IoV Platoon Control Algorithm}
\label{Algorithm3}
\begin{algorithmic}[1]
\REQUIRE Initial status of the platoon $\boldsymbol{z}(0)$ and reference states $\boldsymbol{z}^{\text{Ref}}$.
\ENSURE Obtain optimal IoV platoon control results $\boldsymbol{z}^*$ and $\boldsymbol{u}^*$.
\FOR {$t=1:T-\zeta$}
\FOR {$m=1:M_p$}
\STATE Solve the problem \eqref{eqn_31} and obtain results $\boldsymbol{z}^m$ and $\boldsymbol{u}^m$.
\IF {the convergence conditions are satisfied}
\STATE Replace $\boldsymbol{z}^m(t)$ in $\boldsymbol{z}^*$ and $\boldsymbol{u}^m(t)$ in $\boldsymbol{u}^*$.
\STATE \textbf{break} 
\ELSE
\STATE \textbf{continue} the \textbf{for} loop.
\ENDIF
\ENDFOR
\ENDFOR
\STATE \textbf{return} $\boldsymbol{z}^*$ and $\boldsymbol{u}^*$.
\end{algorithmic}
\end{algorithm}

\subsection{Complexity Analysis}
The computational complexity of the SCA-based FEEL downlink procedure formulated in Algorithm 2 is $\mathcal{O}\{(MKT)^{3.5}\}$, and the computational complexity of the MPC-based platoon control of Algorithm 3 is $\mathcal{O}\{(3K\zeta)^{3.5}T\}$. Thus, the computational complexity of the entire BCD framework is $\mathcal{O}\{N_c[(MKT)^{3.5}+(3K\zeta)^{3.5}T]\}$.

\section{Case Studies}
\par In this section, we assess the performance of the proposed RSMA-based IoV system through numerical simulations. Firstly, we detail our experimental setup. Then, we compare the performance of the proposed scheme with the relevant baseline techniques, such as NOMA and MU-LP. Next, we assess the robustness of the proposed model. Then, we investigate the impact of different road conditions and their effect on our proposed system model. Finally, we analyze the convergence trends of the algorithms devised.

\subsection{Simulation Setup}
\par In the simulations, we evaluate the performance of the proposed RSMA-based IoV system. The total number of the operational time periods is set to $T=100$ with $\Delta t$ being equal $0.05$ seconds. The length of each vehicle is $4.5$ meters, and its width is $1.8$ meters. According to the actual highway lane standard in the United States, we set the lane width to $3.7$ meters. The control input limits, $a$ and $\delta$, are chosen as the real physical limits of the actual vehicle \cite{DBLP:journals/corr/abs-2003-08595}. For the IoV driving conditions, the lower and upper bounds of the acceleration rates are set to $-4$ $\text{m}/\text{s}^{2}$ and $4$ $\text{m}/\text{s}^{2}$, respectively, while the acceleration rate changes are $-1$ $\text{m}/\text{s}^{2}$ and $1$ $\text{m}/\text{s}^{2}$. In addition, the lower and upper limits of steering are set to $-0.3$ and $0.3$ radian, and the change rate of the steering is limited to $0.2$ radius per second. According to \cite{DBLP:journals/corr/abs-2003-08595}, the $z_k^\text{Ref}$ is the reference trajectory, which is predetermined by the LV. The estimated trajectory, $z_k^\text{est}(t)$, is obtained through the vehicular motion of the previous time slot $t-1$ via \eqref{eqn_15}, and the $z_k^\text{max}(t)$ is the upper bound of $z_k(t)$. The coefficient of the multi-objective optimization is $Q_t=1$, $Q_h=100$, $Q_z=[1,100,1,0.1]$, $Q_u=[1,1]$. Furthermore, we set the transmission bandwidth, $B$, to $5$ MHz, which follows \cite{9453811}. The vehicles' transmission power $P_t$ is set to $25$ dBm based on \cite{8720192}, and the noise power spectral density is set to $-174$ dBm/Hz through \cite{9484529}. The path-loss coefficient $\mu$ is set to $2$ by following \cite{7572068}. Finally, the simulations are conducted in MATLAB \cite{MATLAB:R2021a} by relying on the solvers of YALMIP \cite{Lofberg2004} and IPOPT \cite{IPOPT}.

\begin{figure}[ht]
\centering  
  \subfloat[]{
      \label{figure_2a}
      \includegraphics[width=0.45\linewidth]{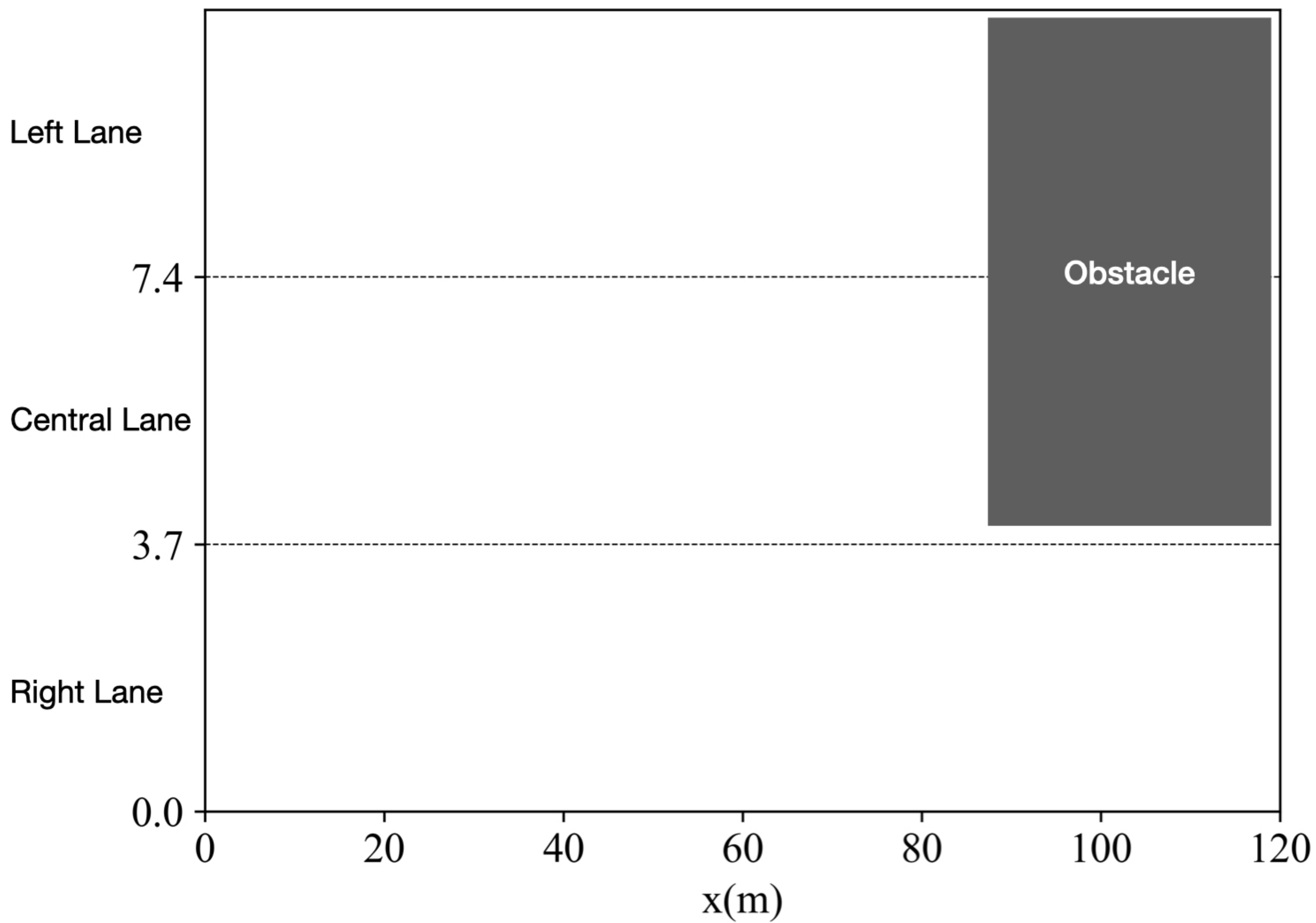}}
 \subfloat[]{
      \label{figure_2b}
      \includegraphics[width=0.45\linewidth]{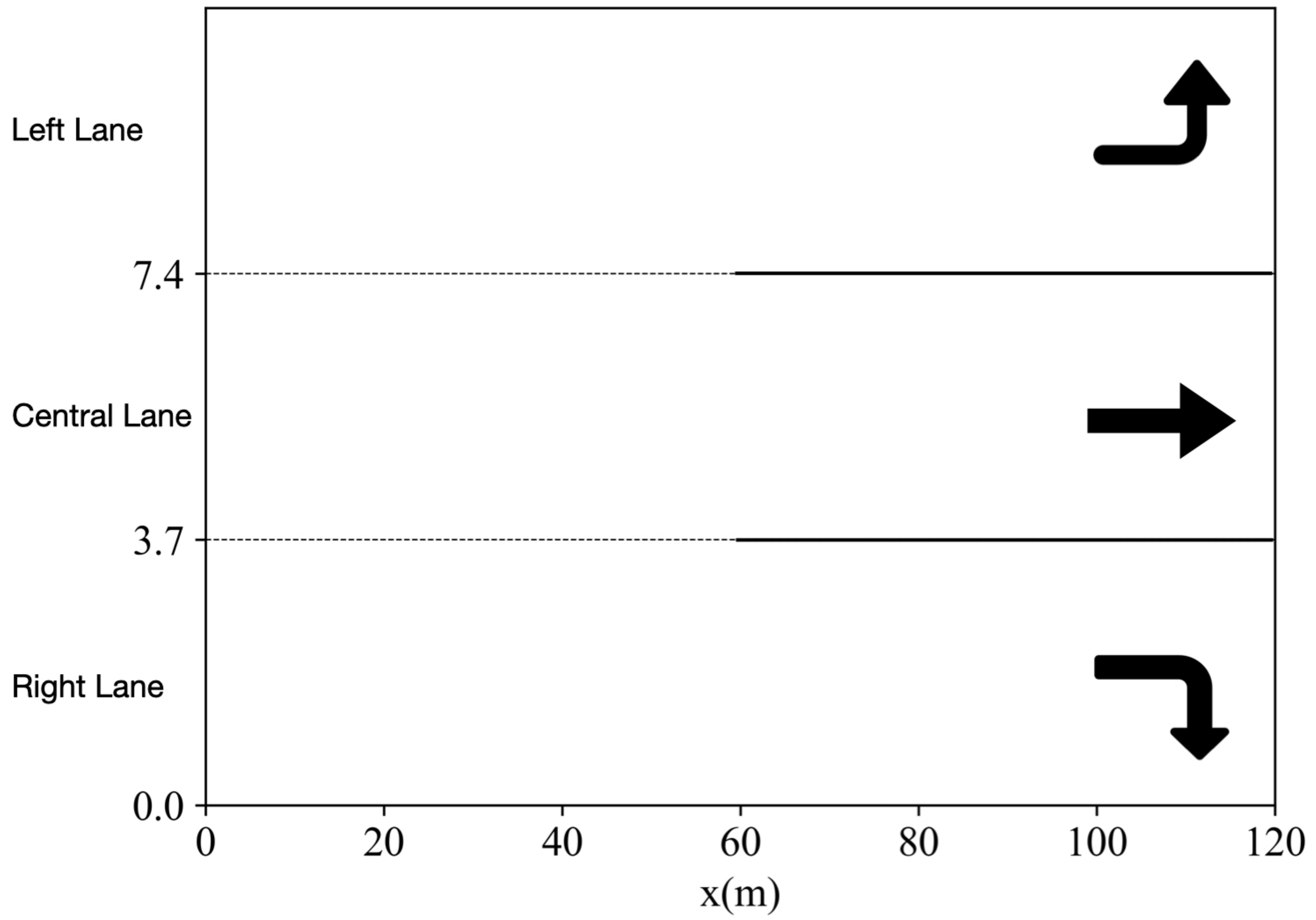}}

\caption{Scenarios for comparison: (a) obstacle avoidance scenario, (b) crossroad zone scenario.}
  \label{figure_2}
\end{figure}

\begin{figure*}[ht]
\centering  
  \subfloat[]{
      \label{figure_3a}
      \includegraphics[width=0.3\linewidth]{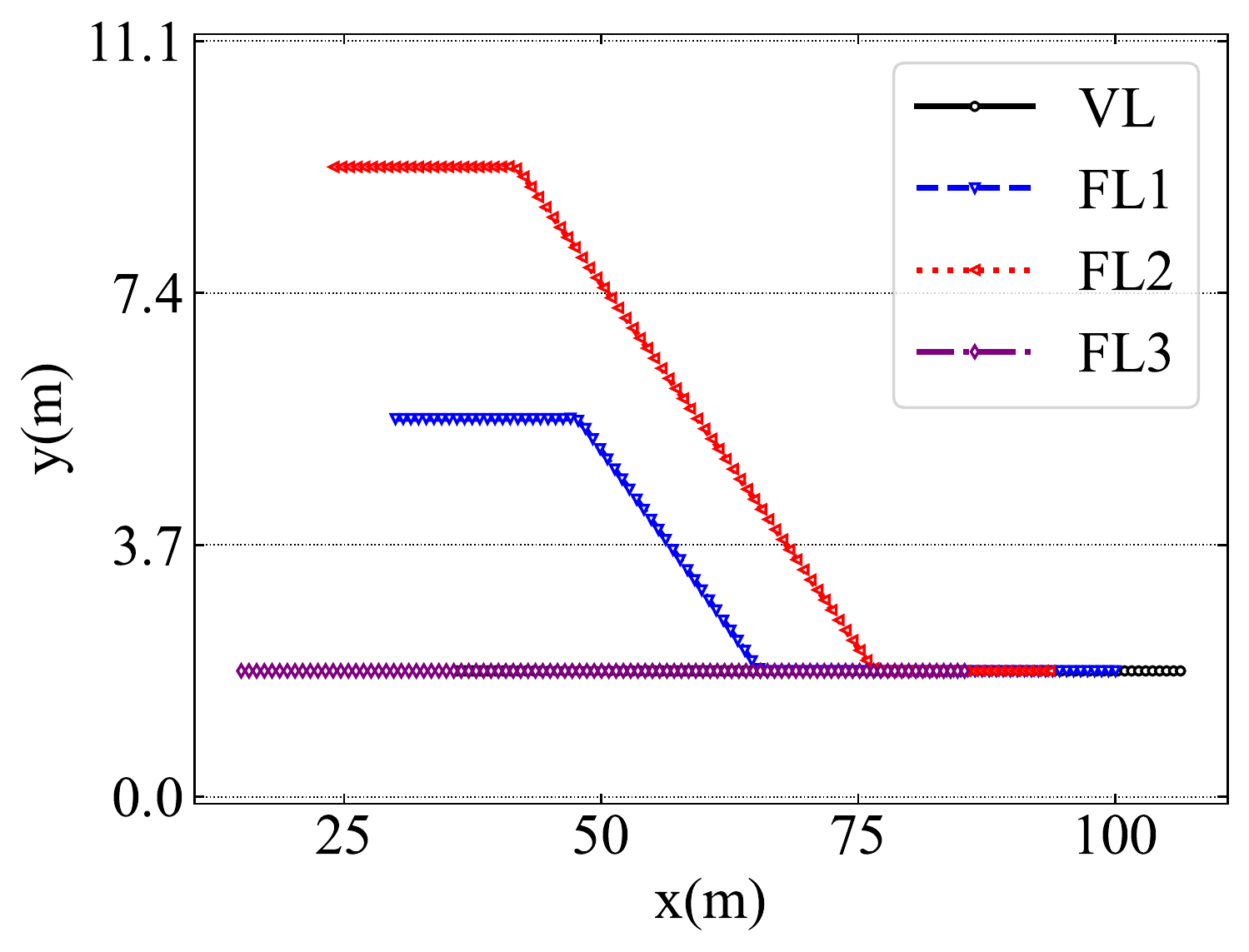}}
  \subfloat[]{
      \label{figure_3b}
      \includegraphics[width=0.3\linewidth]{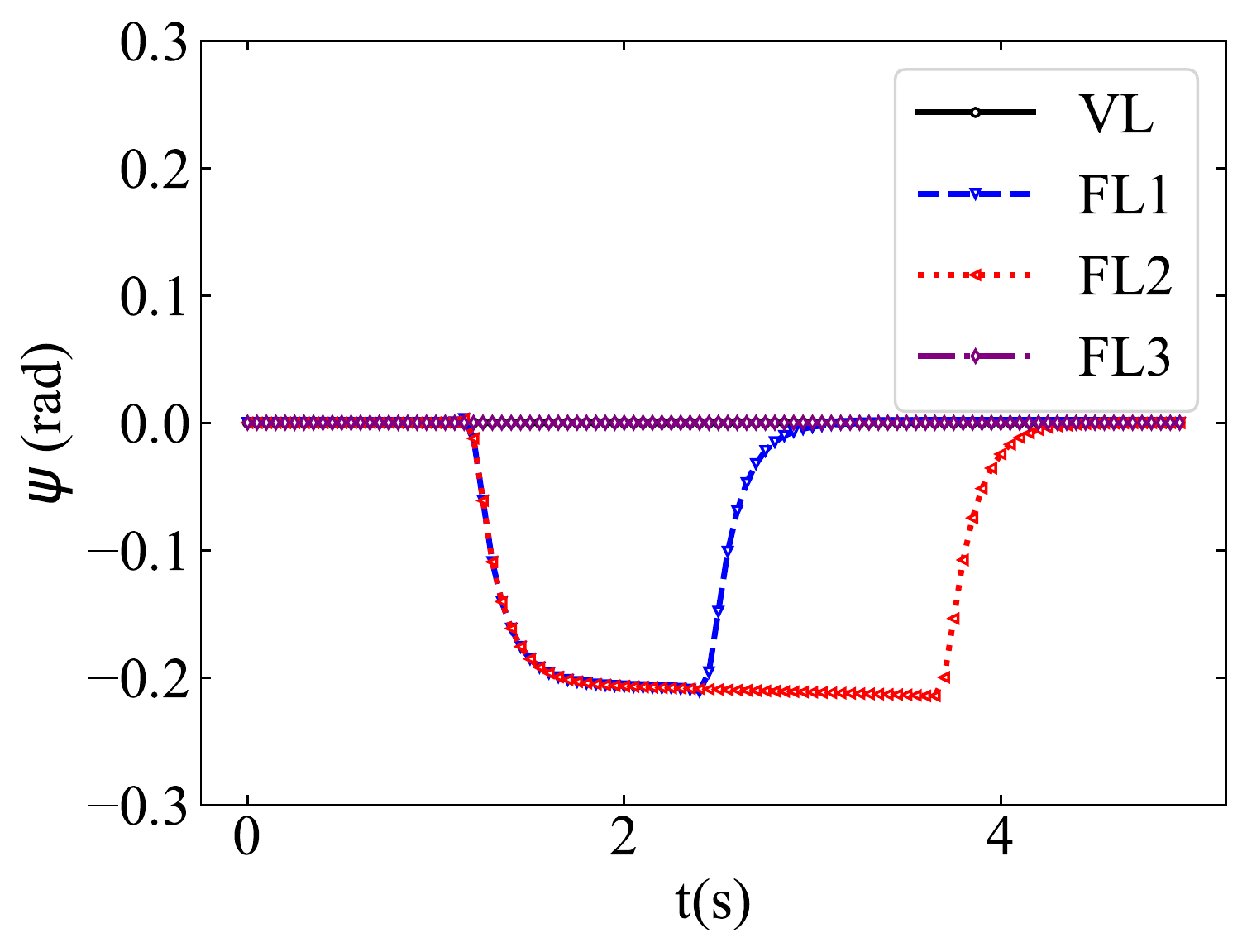}}
  \subfloat[]{
      \label{figure_3c}
      \includegraphics[width=0.3\linewidth]{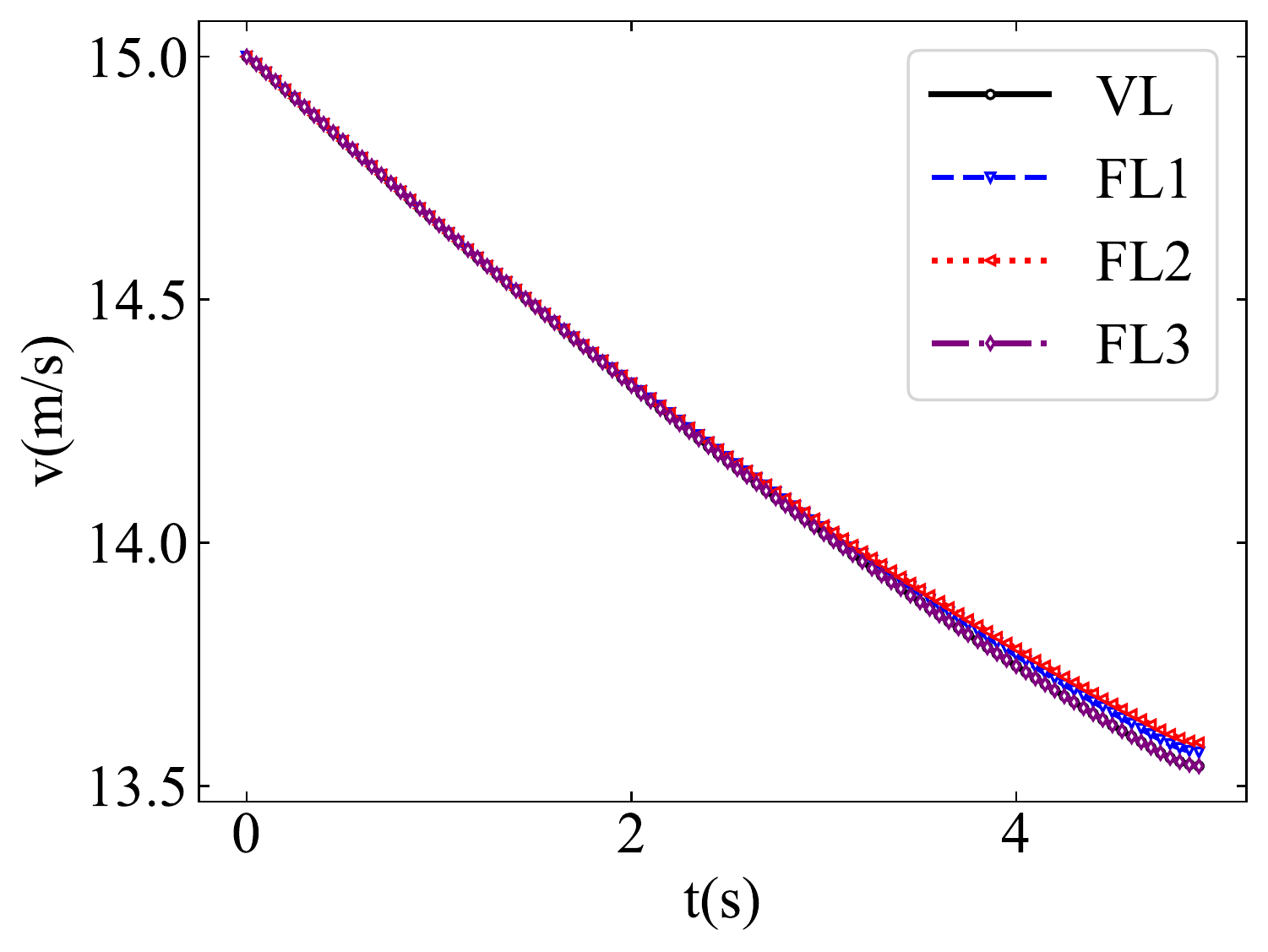}}
      
  \subfloat[]{
      \label{figure_3d}
      \includegraphics[width=0.3\linewidth]{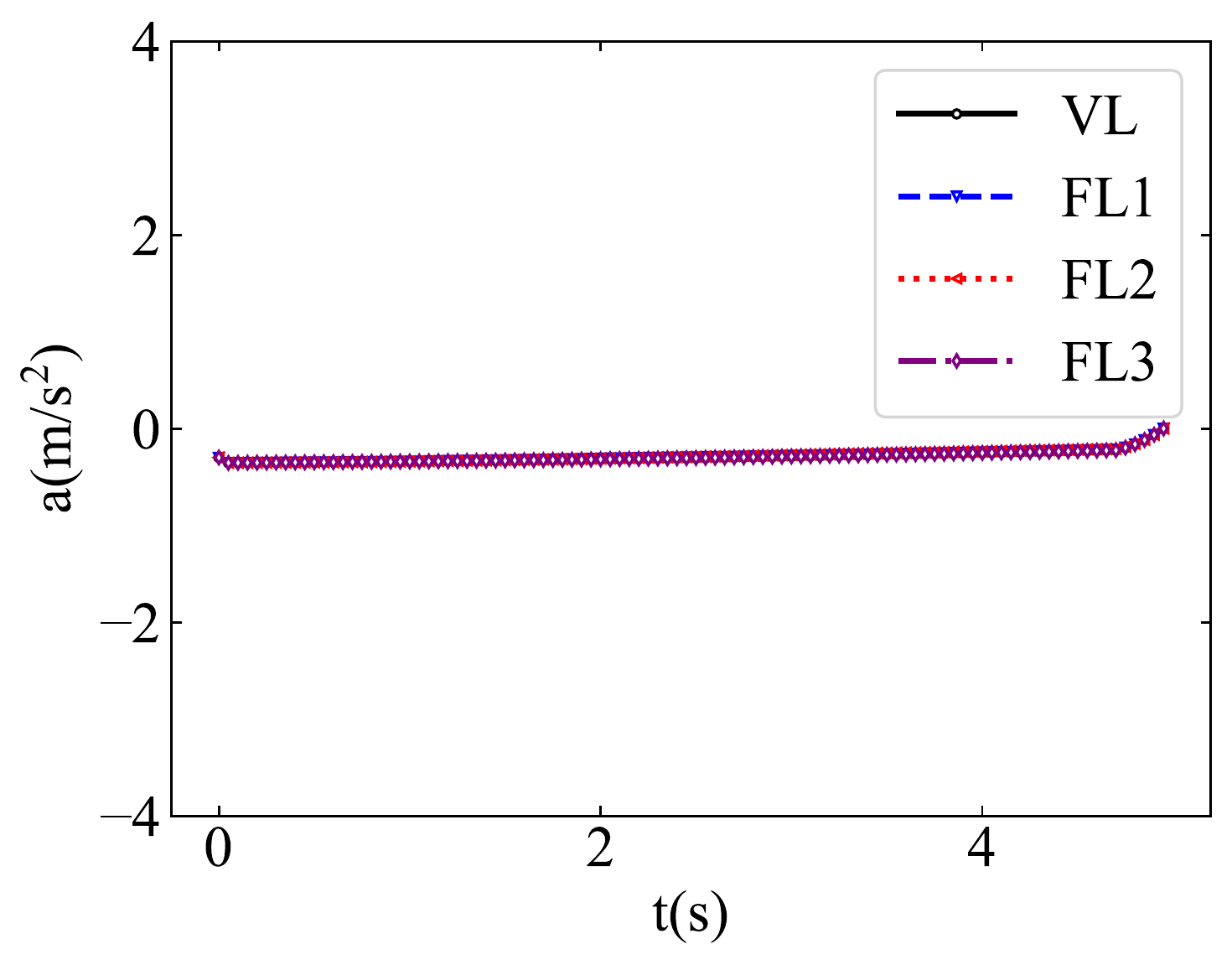}}
  \subfloat[]{
      \label{figure_3e}
      \includegraphics[width=0.3\linewidth]{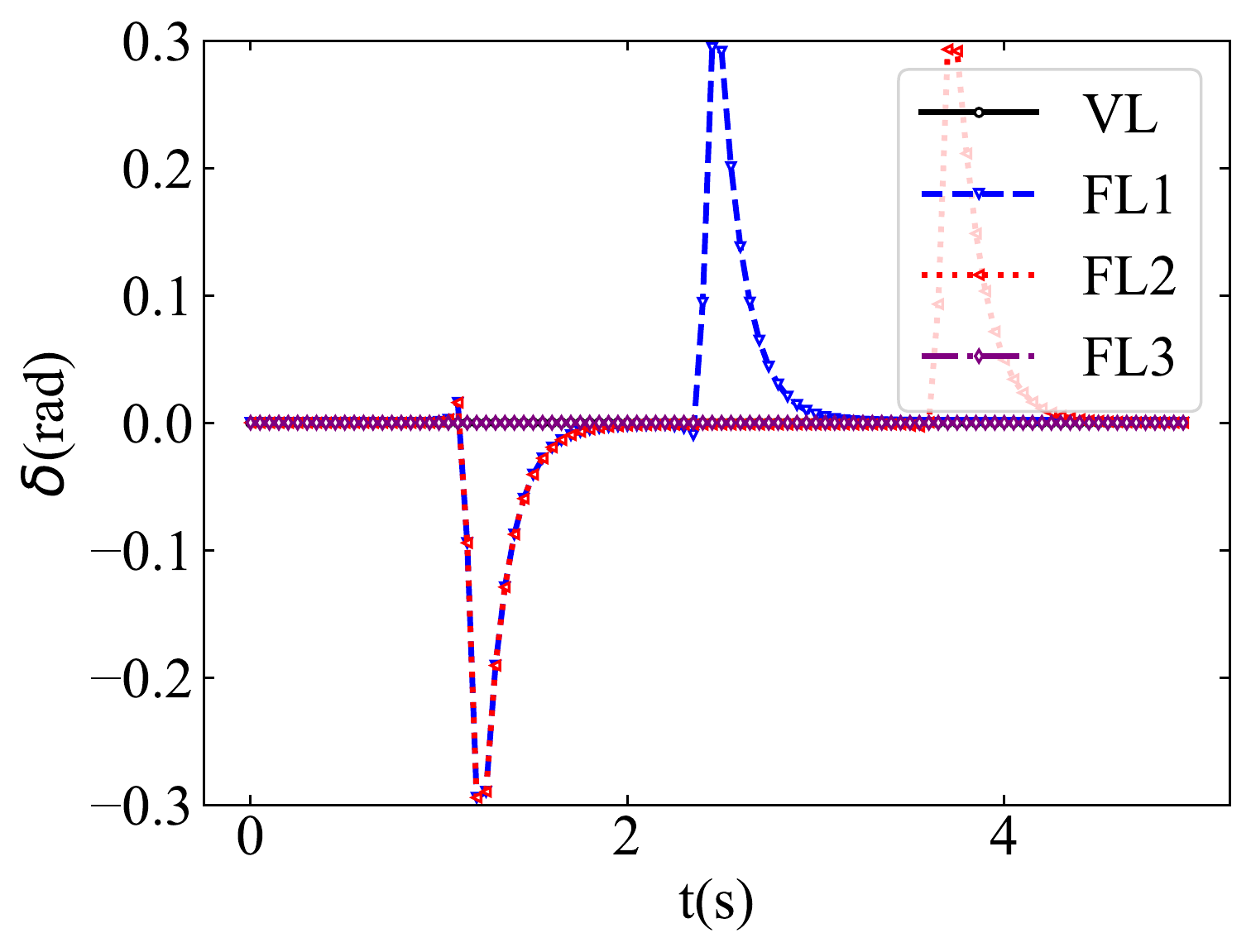}}
  \subfloat[]{
      \label{figure_3f}
      \includegraphics[width=0.3\linewidth]{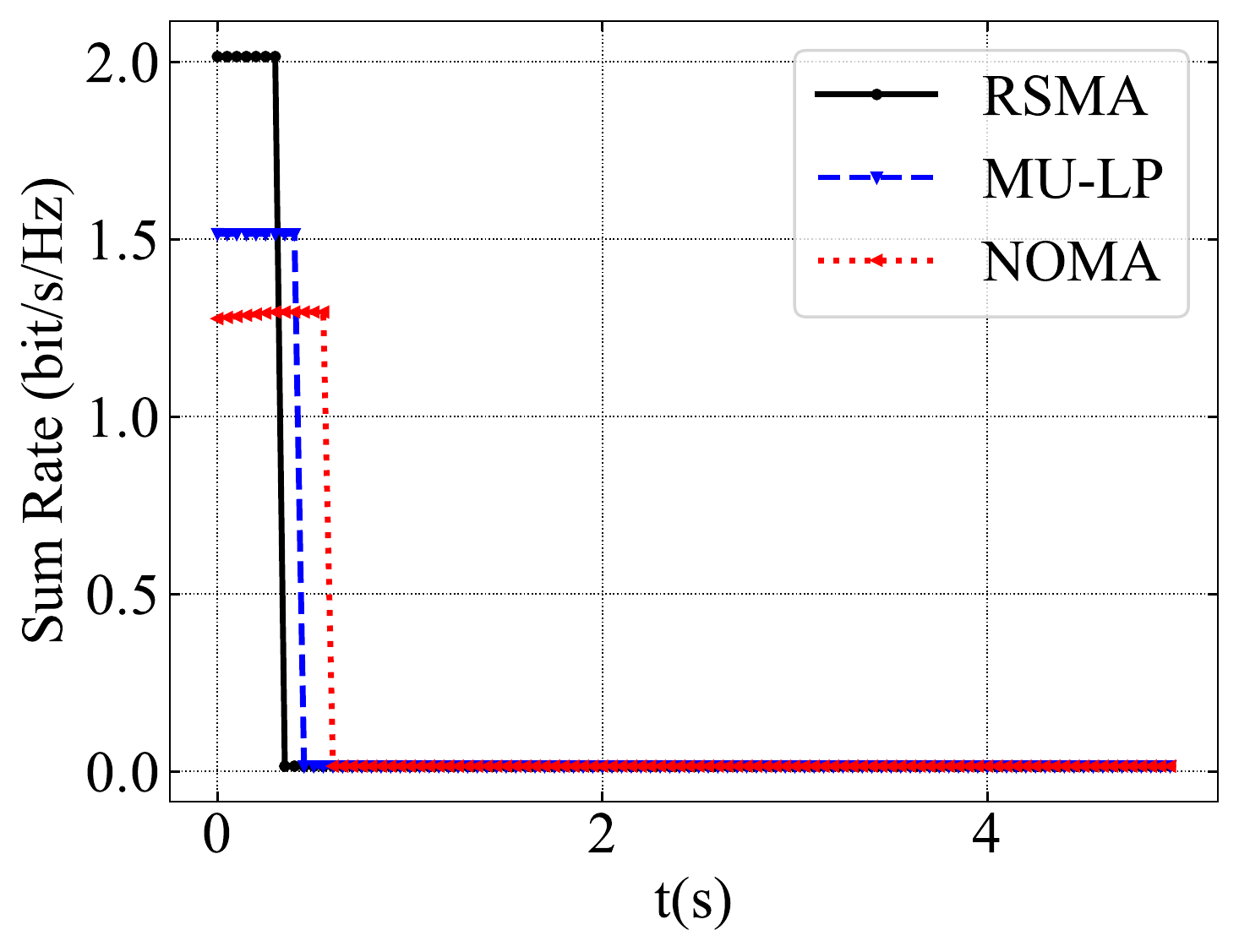}}
  \caption{Vehicles’ states and actions in S1 with RSMA: (a) motion trajectories, (b) heading angle, (c) velocity, (d) acceleration, (e) steering angle, (f) sum rate.}
  \label{figure_3}

\end{figure*}

\begin{figure}[ht]
\centering  
  \subfloat[]{
      \label{figure_4a}
      \includegraphics[width=0.45\linewidth]{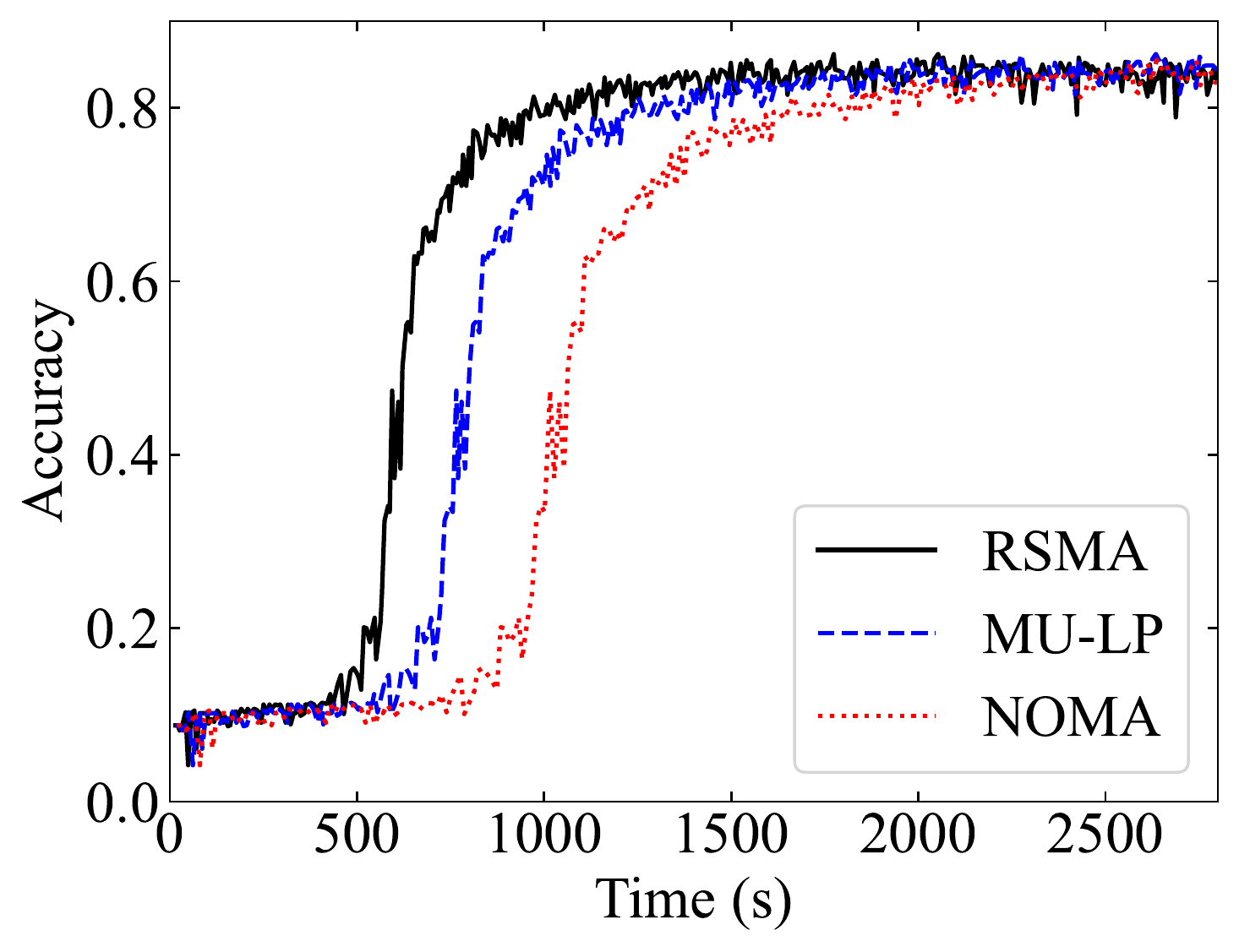}}
 \subfloat[]{
      \label{figure_4b}
      \includegraphics[width=0.45\linewidth]{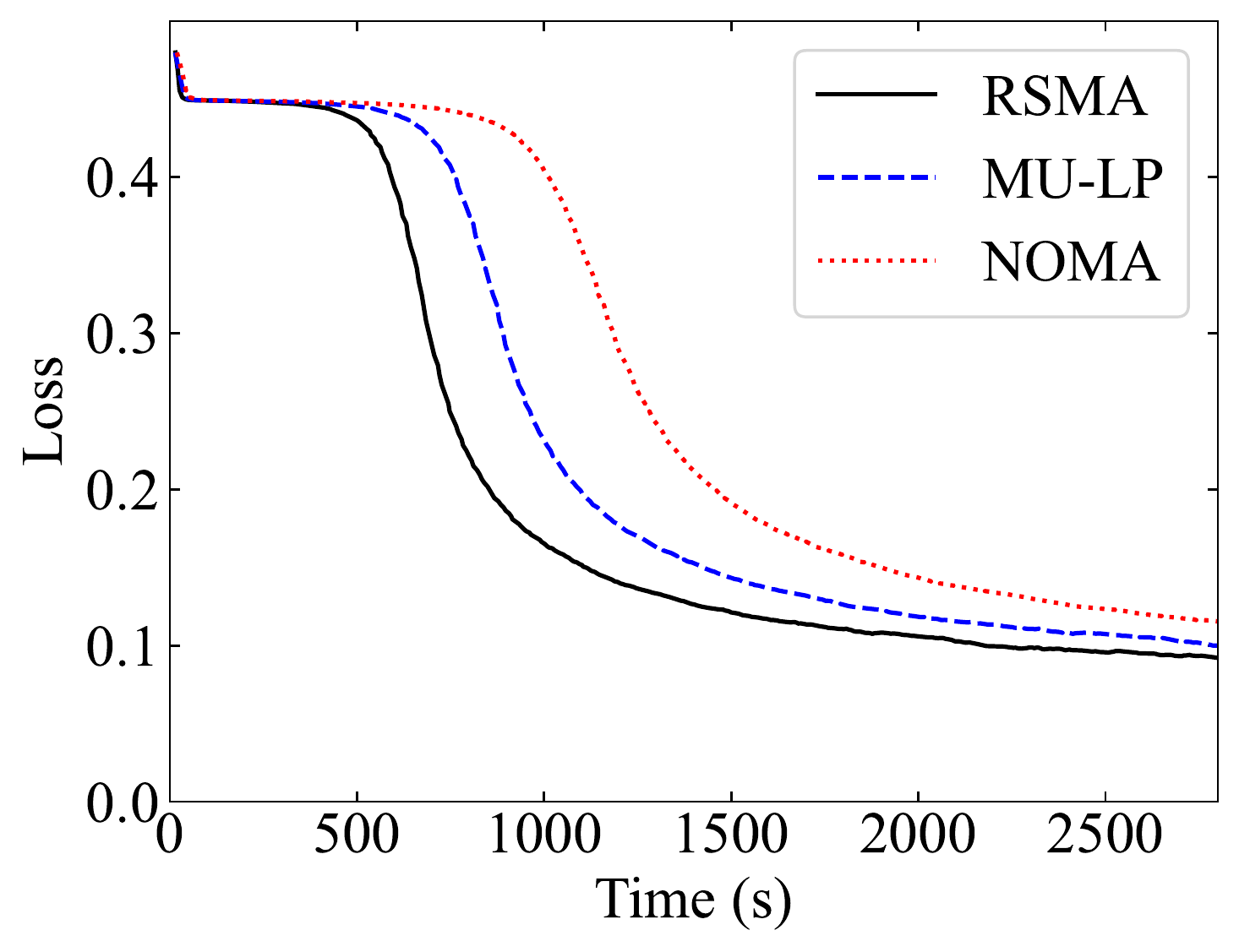}}
  \caption{FEEL performance in S1: (a) accuracy, (b) loss value.}
  \label{figure_4}

\end{figure}

\begin{table*} 
\normalsize
\caption{Comparison of Vehicles' States and Actions in S1.}
\centering
\label{table.S1}
\begin{center}
\begin{tabular}{lcccc} \toprule
\multirow{2}{*}{Metrics of motion states} & \multicolumn{4}{c}{Vehicles’ states and actions under RSMA} \\
\cmidrule(lr){2-5}
 & Heading angle & Velocity & Acceleration & Steering angle \\ \midrule
Mean &  -0.0393 &   14.2120 & -0.2881 & 0.0003 \\ 
Variance &  0.0061 &  0.1843 & 0.0036 & 0.0025 \\ \bottomrule\toprule 
\multirow{2}{*}{Metrics of motion states} & \multicolumn{4}{c}{Vehicles’ states and actions under MU-LP} \\
\cmidrule(lr){2-5}
 & Heading angle & Velocity & Acceleration & Steering angle \\ \midrule
Mean &  -0.0393 &   14.2120 & -0.2881 & 0.0003 \\ 
Variance &  0.0061 &  0.1843 & 0.0036 & 0.0025 \\ 
$\Upsilon$ &  0.0 &  0.0030 & 0.0030 & 0.0 \\\bottomrule\toprule
\multirow{2}{*}{Metrics of motion states} & \multicolumn{4}{c}{Vehicles’ states and actions under NOMA} \\
\cmidrule(lr){2-5}
 & Heading angle & Velocity & Acceleration & Steering angle \\ \midrule
Mean &  -0.0393 &   14.2120 & -0.2881 & 0.0003 \\ 
Variance &  0.0061 & 0.1843 & 0.0036 & 0.0025 \\ 
$\Upsilon$ &  0.0008 & 0.0030 & 0.0409 & 0.0009 \\ \bottomrule
\end{tabular}
\end{center}
\end{table*}

\subsection{Scenarios for Comparison}

\par In this simulation, we consider a pair of different road scenarios for verifying the effectiveness of our proposed IoV platoon control system, namely, the obstacle avoidance scenario (S1) and the crossroad zone scenario (S2). The case study consists of three V2V communication schemes: a) the NOUM-based RSMA, b) the NOUM-aided MU-LP \cite{8846706}, and c) the unicast NOMA \cite{9685664}. 

\par The traffic environment of the obstacle avoidance scenario (S1) is shown in Fig. \ref{figure_2a}. In S1, multiple vehicles are traveling together in a multi-lane platoon formation. In particular, an obstacle blocks the left and the central lanes. Once the obstacle is detected, the LV opts for reconfiguring the platoon to a single-lane formation confined to the right lane. Thus, all the vehicles in the left and the central lanes will move to the right lane. The vehicles in the right lane have to create sufficient space to facilitate the safe and smooth lane changing and merging of the vehicles originally traveling in the lane having the obstacle. The obstacle is modeled as a polytopic set, and some additional obstacle avoidance constraints are also imposed in this scenario.

\par In another typical traffic environment, the crossroad zone scenario (S2) is shown in Fig. \ref{figure_2b}. In S2, multiple vehicles are preparing to enter the crossroad zone. By following the practical traffic laws, these must perform the following actions: 1) all the vehicles should enter the straight lane in the center, and 2) lane changes are not allowed when entering the solid line area. Thus, the LV has to predetermine the distance of the crossroad zone and reconfigure the platoon into a single-lane configuration for shifting to the central lane. In this scenario, the additional traffic law constraints are also entered into the proposed platoon control model.

\subsection{Quality of IoV Platoon System}
\par This simulation characterizes the quality of the proposed IoV platoon model and platoon control operations and the FEEL performances. To quantify the efficacy of the platoon control and the FEEL provided by the proposed system, we compare the model to the following three baseline techniques: 1) NOUM-enabled RSMA, 2) the NOUM-enabled MU-LP, and 3) the unicast NOMA.

\par Based on the road settings in S1, we study the vehicles’ states and actions generated by the three baselines. In this part, we assume that there is one LV and three FVs moving along the road. Fig. \ref{figure_3a} shows the vehicles' motion trajectories generated by NOUM-aided RSMA. As shown in the results, the IoV platoon can help the vehicles to avoid obstacles by reconfiguring the platoon into a single-lane pattern at the right lane. Furthermore, Figs. \ref{figure_3b}-\ref{figure_3e} present the heading angle, velocity, acceleration, and steering angle of the vehicles. In particular, the vehicles perform the lane-change in a smooth manner. Similarly, by considering the same road settings in S1, we evaluate the other two baseline communication techniques in the proposed IoV platoon control system. We also compare the mean and variance of the heading angle, velocity, acceleration, and steering angle of the vehicles. Table \ref{table.S1} presents the related results of characterizing both NOUM-aided MU-LP and unicast NOMA for the V2V communication module. Here, $\Upsilon$ denotes the difference between the variance of the RSMA and MU-LP/NOMA in percentage, which is given by:
\vspace{-0.2cm}
\begin{equation}\label{eqn_32}
\begin{aligned}
\Upsilon=\frac{|V_i-V_0|}{V_0},
\end{aligned}
\end{equation}
where $V_i$ is the variance of the MU-LP/NOMA and $V_0$ is the variance of the RSMA. Observe that, the proposed RSMA-based system generates a relatively modest variance of the profiles for all these four metrics. Then, the dynamic sum rate of the system is shown in Fig. \ref{figure_3f}. The RMSA scheme outperforms the other two conventional communication schemes. Moreover, the unicast NOMA achieves the lowest sum rate due to inefficient unicast communication. At the very beginning of the simulation, the broadcast message of the FEEL downlink occupies the communication channel. Thus, all these three communication schemes achieve a high sum rate until the FEEL downlink finishes. Then, only the unicast message is transmitted from the LV to the FVs, and the communication channel is under-loaded. This means that all these three schemes can meet the transmission rate requirement of the unicast message, and achieve the same sum rate. To summarize, the employment of RSMA can guarantee efficient and smooth IoV platoon control operations and improves the communication throughput in the proposed system.

\par Then, we study the FEEL downlink performance generated by the three baselines in the same settings as previously. Following \cite{9127160,9453811}, we evaluate the proposed system on the MNIST dataset, and the Convolutional Neural Network (CNN) tool is used to train the local model. The test data was the original 10,000 test cases in the MNIST dataset. The training dataset of the MNIST is randomly split into 50,000 training cases and 10,000 validation cases, which would be evenly distributed to the FVs (workers). Fig. \ref{figure_4a} shows the accuracy of the FEEL global model generated by the three baseline techniques. It is apparent that the unicast NOMA is the slowest one to converge, suggesting that it is inefficient to solely utilize the unicasting channel in this circumstance. Meanwhile, the NOUM-enabled RSMA outperforms the MU-LP. The loss value of the FEEL global model, in \ref{figure_4b}, further confirms this result.

\par
\begin{figure}
\centering 
\includegraphics[width = 0.5\linewidth]{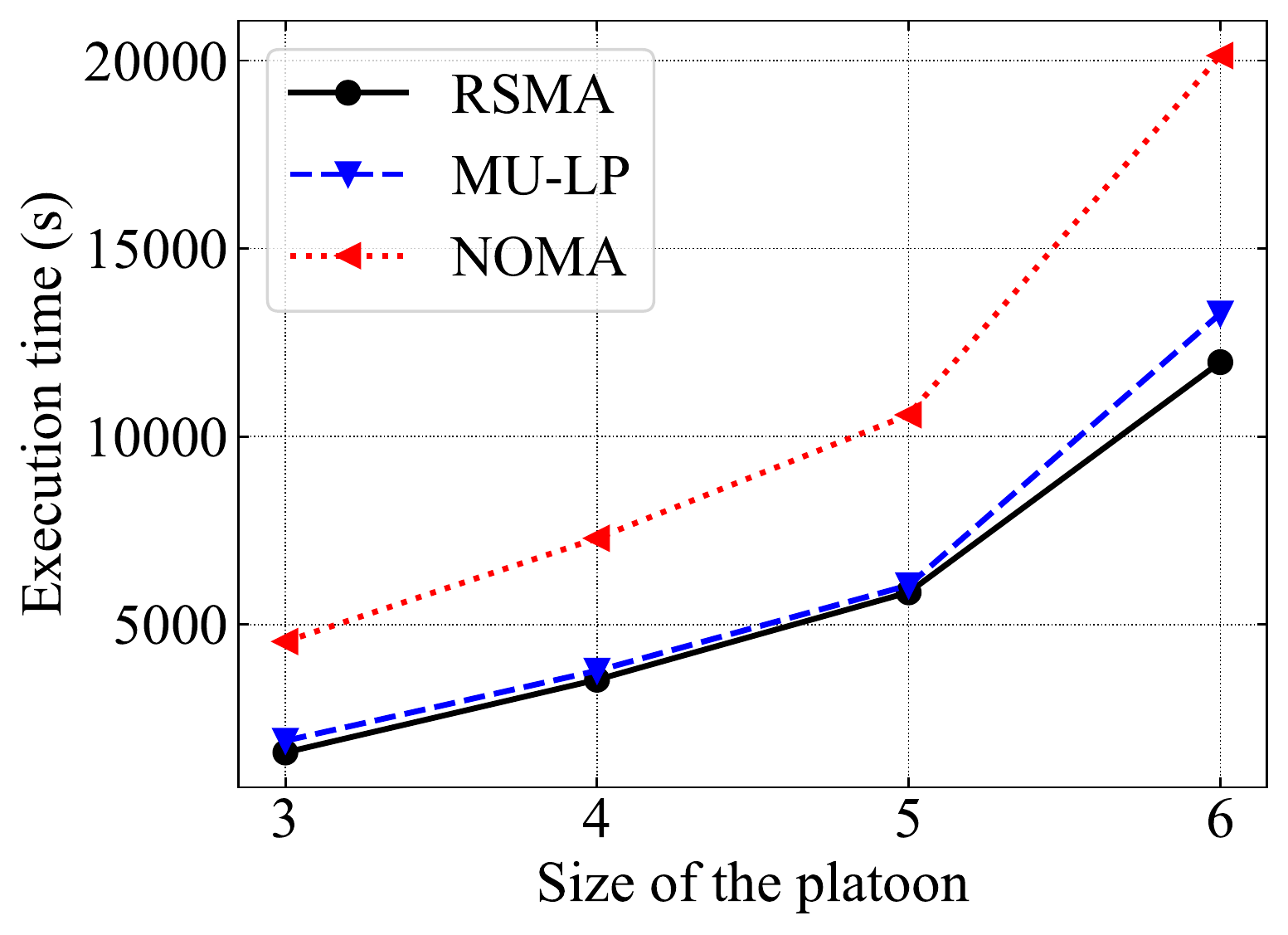} \\
\vspace{-0.4cm} 
\caption{Execution time of the proposed framework.}
\label{figure_5}

\end{figure}

 Finally, we consider the execution time of the three baselines in Fig. \ref{figure_5}. The RSMA schemes cost a minimal execution time to converge, and NOMA is still the worst one. 

\subsection{Robustness of V2V Communication under RSMA}
\par
\begin{figure}[ht]
\centering  
  \subfloat[]{
      \label{figure_6a}
      \includegraphics[width=0.45\linewidth]{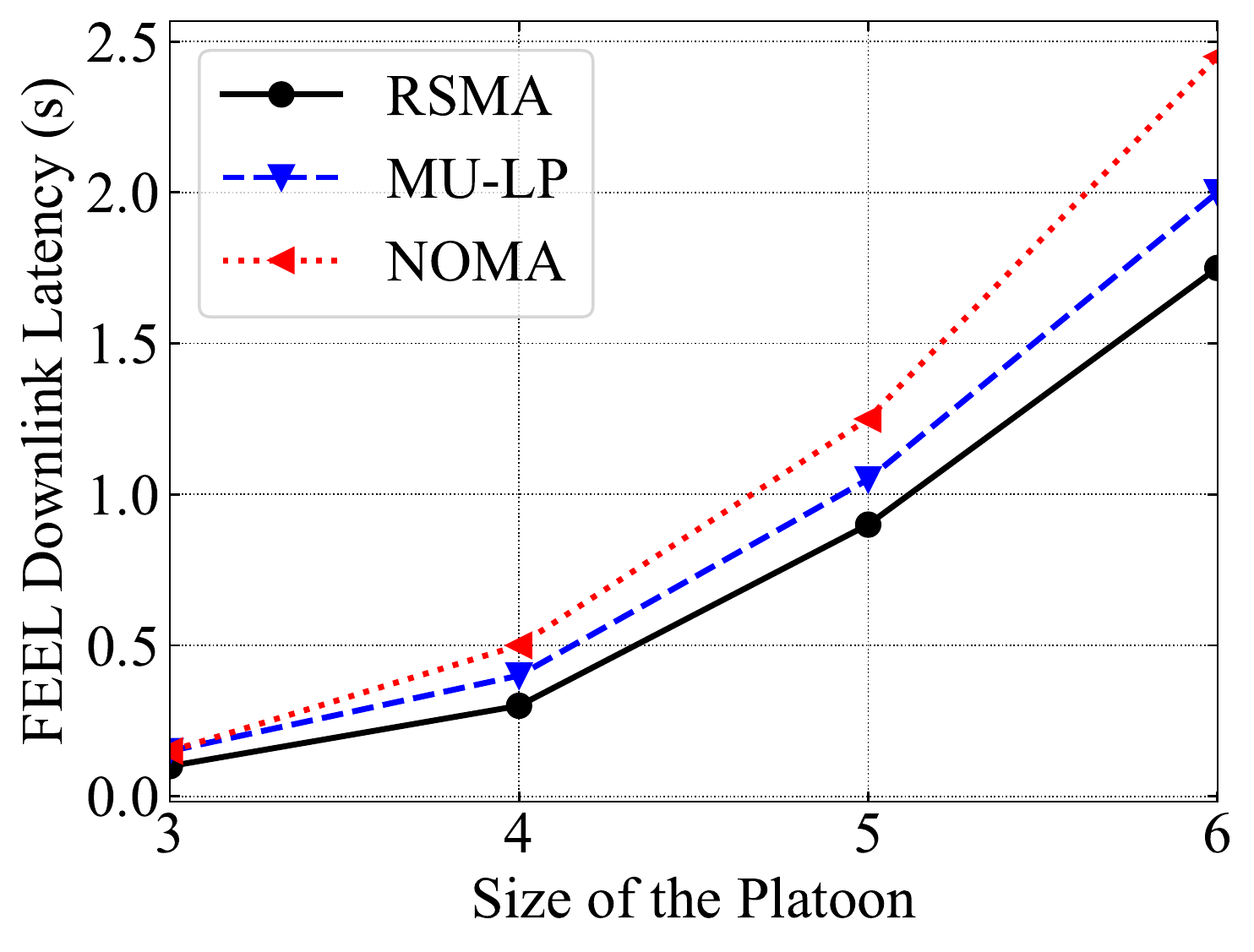}}
  \subfloat[]{
      \label{figure_6b}
      \includegraphics[width=0.45\linewidth]{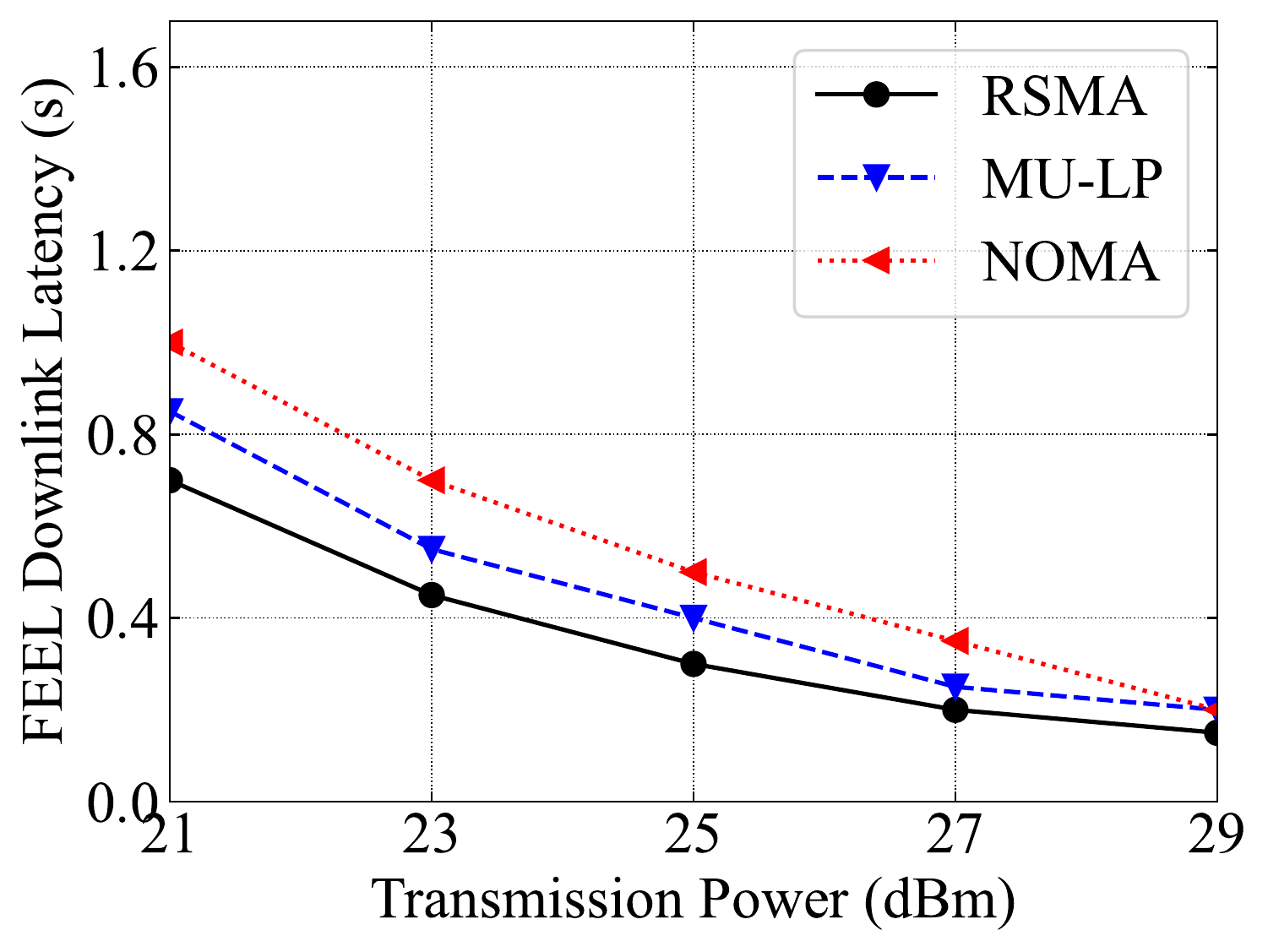}}

  \subfloat[]{
      \label{figure_6c}
      \includegraphics[width=0.45\linewidth]{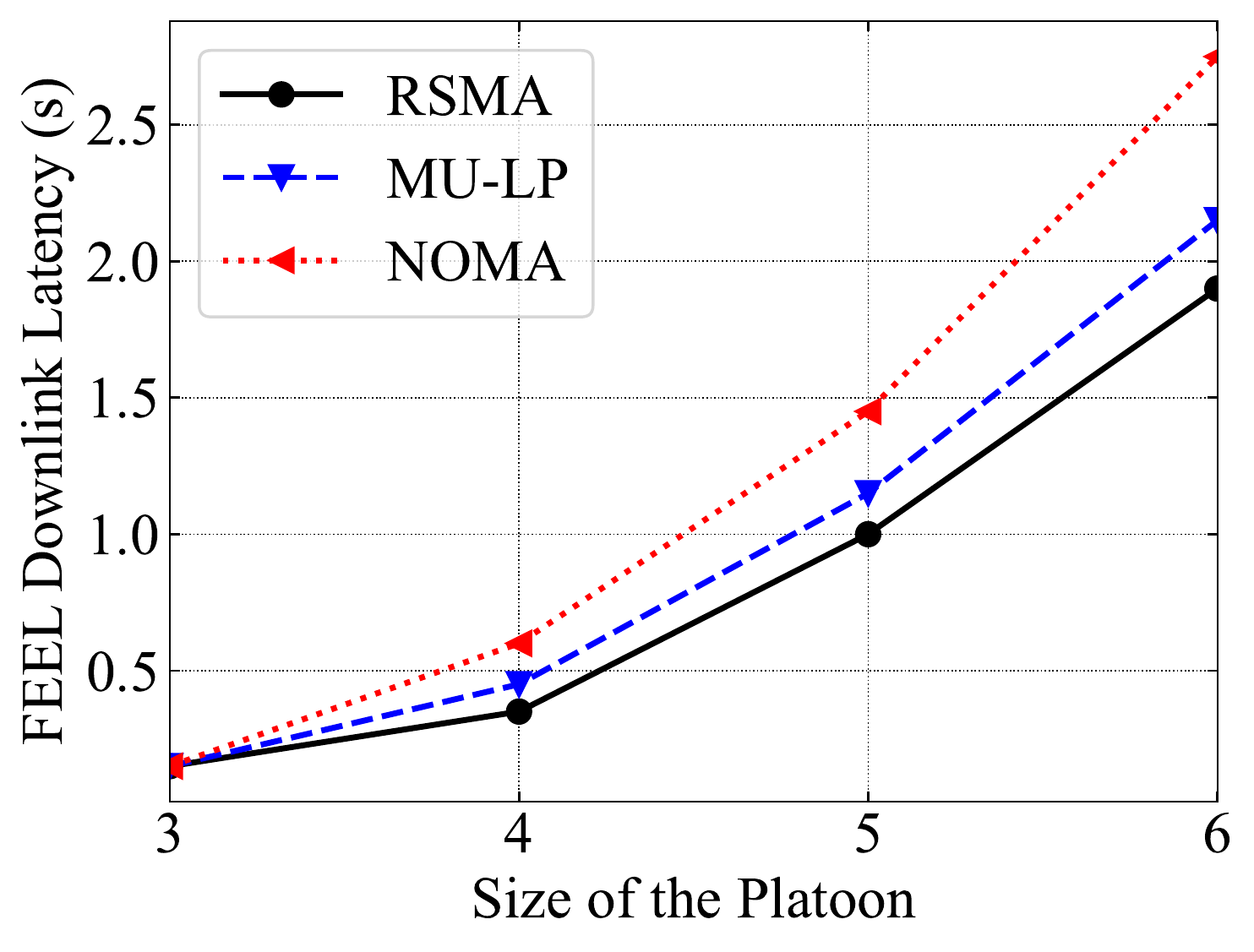}}
  \subfloat[]{
      \label{figure_6d}
      \includegraphics[width=0.45\linewidth]{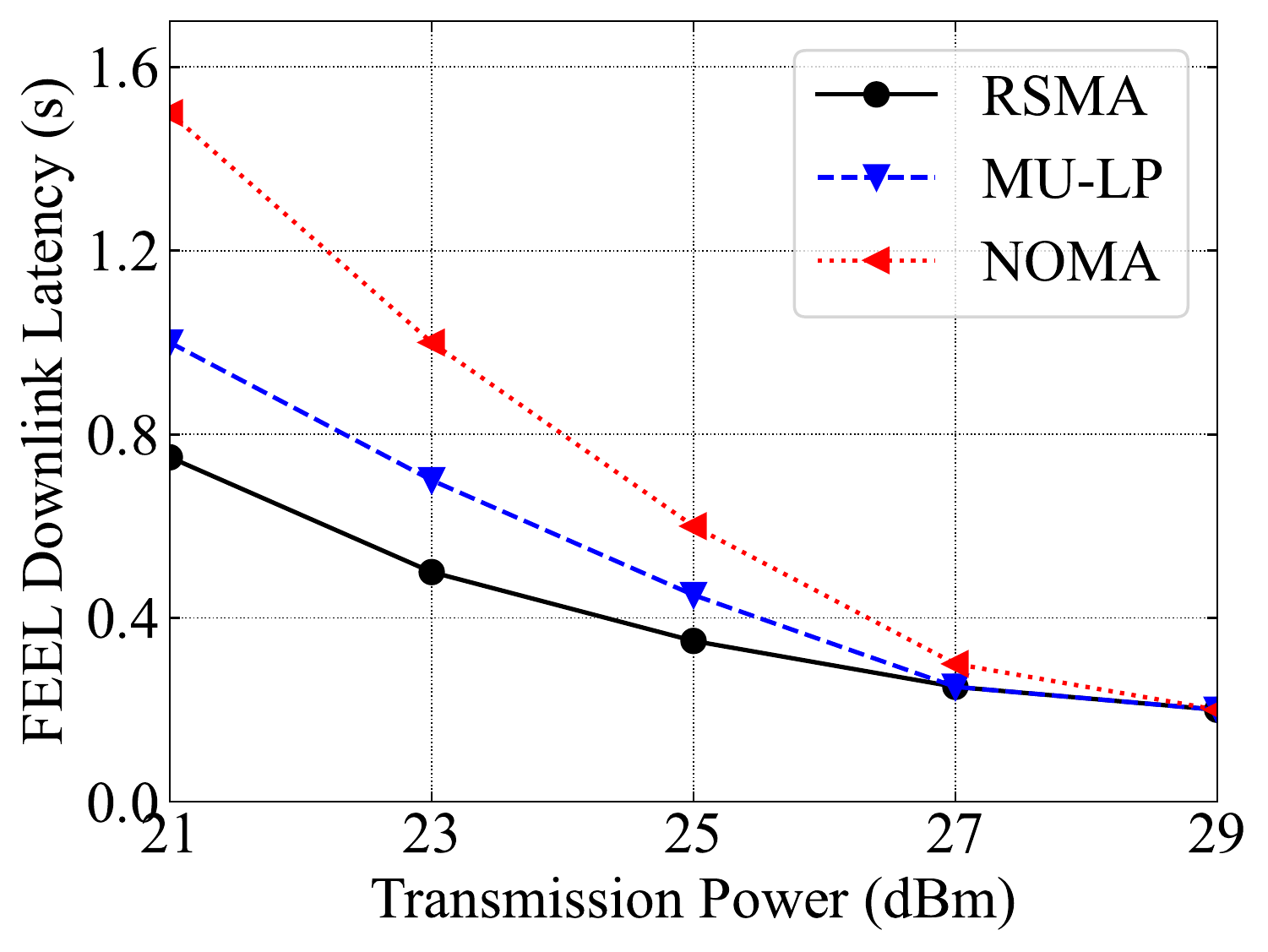}}
  
  \vspace{-0.2cm} 
  \caption{Robustness of V2V communication: (a) size of platoon vs. latency in imperfect CSIT, (b) transmission power vs. latency in imperfect CSIT, (c) size of platoon vs. latency in perfect CSIT, (d) transmission power vs. latency in perfect CSIT.}
  \label{figure_6}

\end{figure}

 Based on the setting of S1, we assess the robustness of the V2V communication module. We first compare the FEEL downlink latency for the different platoon sizes under perfect CSIT in Fig. \ref{figure_6a}. It is noted that unicast NOMA exhibits the highest latency, suggesting that it is inefficient to solely utilize the unicasting channel in this circumstance. Furthermore, the NOUM-enabled RSMA outperforms the MU-LP, because RSMA is more efficient in the over-loaded network. We also compare the latency of the FEEL downlink under different transmission powers in Fig. \ref{figure_6b}. Naturally, a higher transmission power leads to a higher transmission rate, and lower latency. RSMA can outperform the other two baselines. 

Then, we further assess the reliability of the V2V communication module. We also evaluate the efficacy of the RSMA scheme with perfect CSIT in Fig. \ref{figure_6c} and Fig. \ref{figure_6d}. The NOUM-based RSMA still outperforms the conventional MU-LP and NOMA schemes. Moreover, compared to this case in perfect CSIT, both the MU-LP and NOMA schemes consume more energy and have higher latency under imperfect CSIT. In the meantime, the latency gap among these three techniques becomes higher. This indicates that the NOUM-aided RSMA is more reliable than the other two schemes.

\begin{table*}
\small
\caption{Comparison of Vehicles' States and Actions in S2.}
\centering
\label{table.S2}
\begin{center}
\begin{tabular}{lcccc} \toprule
\multirow{2}{*}{Metrics of motion states} & \multicolumn{4}{c}{Vehicles’ states and actions under RSMA} \\
\cmidrule(lr){2-5}
 & Heading angle & Velocity & Acceleration & Steering angle \\ \midrule
Mean &  0.0126 &    14.2191 & -0.2865 & 0.0001 \\ 
Variance &  0.0068 &  0.1823 & 0.0033 & 0.0040 \\ \bottomrule\toprule
\multirow{2}{*}{Metrics of motion states} & \multicolumn{4}{c}{Vehicles’ states and actions under MU-LP} \\
\cmidrule(lr){2-5}
 & Heading angle & Velocity & Acceleration & Steering angle \\ \midrule
Mean &  0.0126 &    14.2192 & -0.2865 & 0.0001 \\ 
Variance &  0.0068 &  0.1823 & 0.0033 & 0.0040 \\ 
$\Upsilon$ &  0.0 &  0.0027 & 0.0034 & 0.0 \\\bottomrule\toprule
\multirow{2}{*}{Metrics of motion states} & \multicolumn{4}{c}{Vehicles’ states and actions under NOMA} \\
\cmidrule(lr){2-5}
 & Heading angle & Velocity & Acceleration & Steering angle \\ \midrule
Mean &  0.01263 &    14.2191 & -0.2865 & 0.0001 \\ 
Variance &  0.0008 & 0.0030 & 0.0409 & 0.0009 \\ 
$\Upsilon$ &  0.0004 &  0.0022 & 0.0332 & 0.0003 \\\bottomrule
\end{tabular}
\end{center}
\end{table*}

\subsection{Impact of Different Road Conditions}

\par
\begin{figure*}[h]
\centering  
  \subfloat[]{
      \label{figure_7a}
      \includegraphics[width=0.3\linewidth]{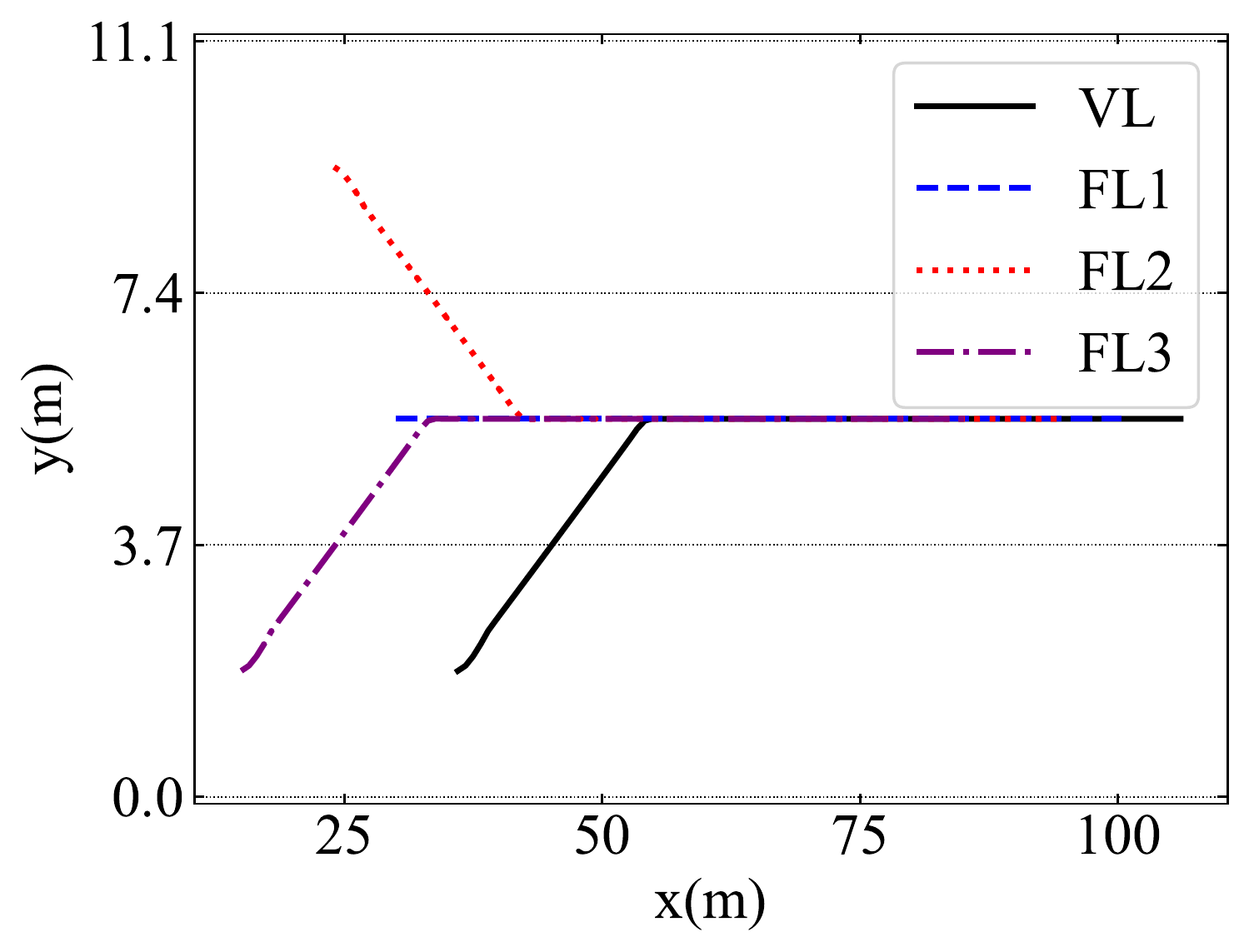}}
  \subfloat[]{
      \label{figure_7b}
      \includegraphics[width=0.3\linewidth]{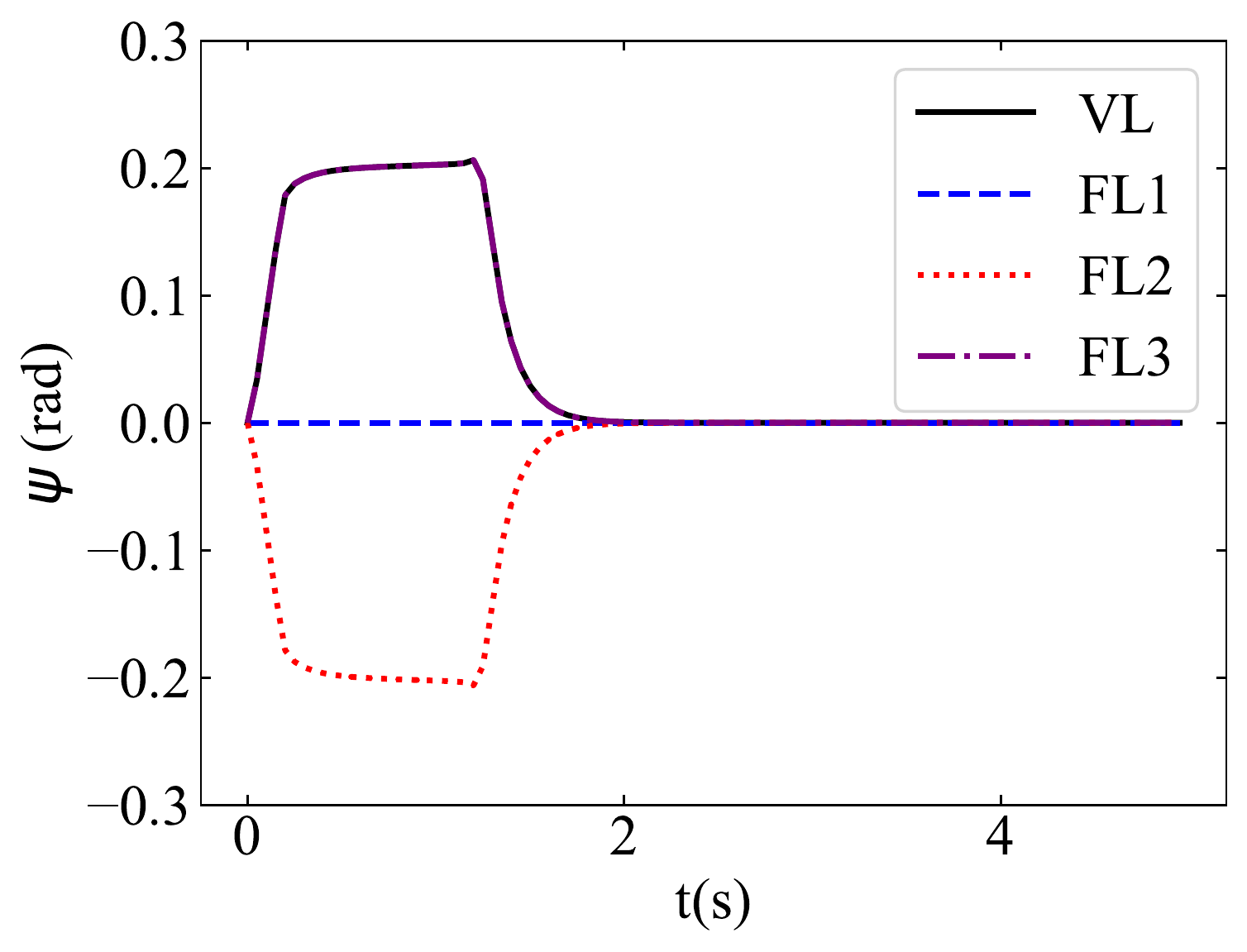}}
  \subfloat[]{
      \label{figure_7c}
      \includegraphics[width=0.3\linewidth]{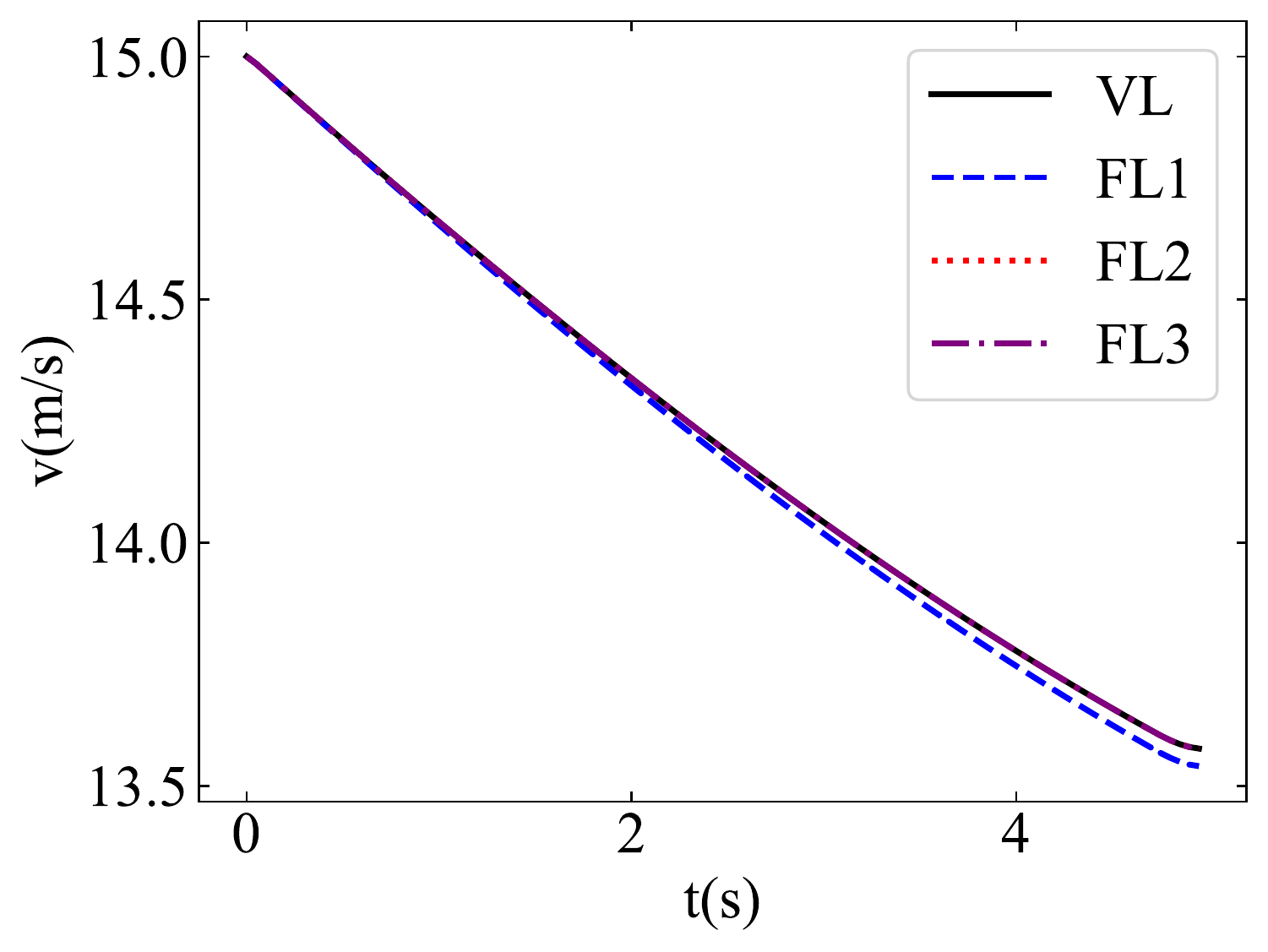}}
      
  \subfloat[]{
      \label{figure_7d}
      \includegraphics[width=0.3\linewidth]{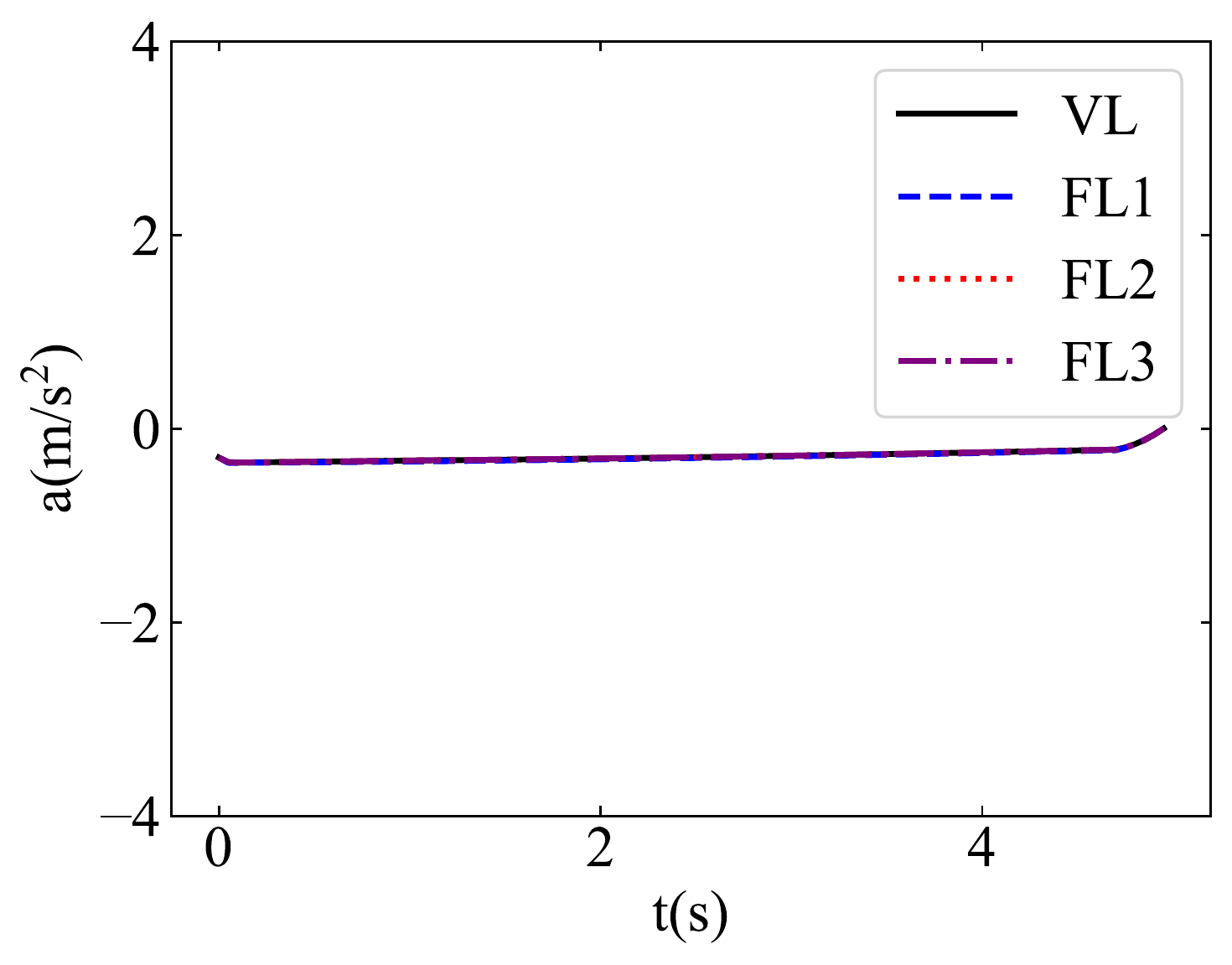}}
  \subfloat[]{
      \label{figure_7e}
      \includegraphics[width=0.3\linewidth]{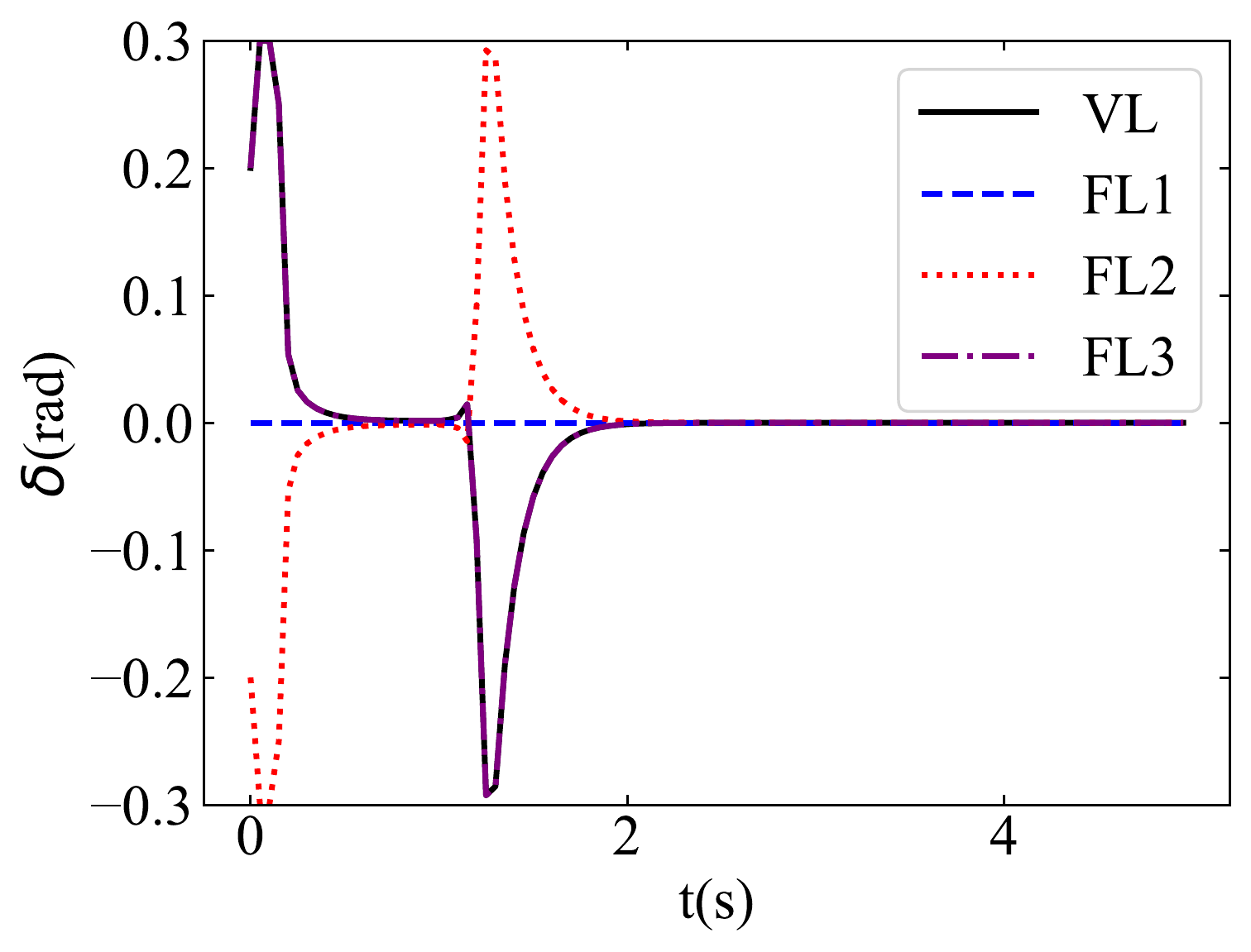}}
  \subfloat[]{
      \label{figure_7f}
      \includegraphics[width=0.3\linewidth]{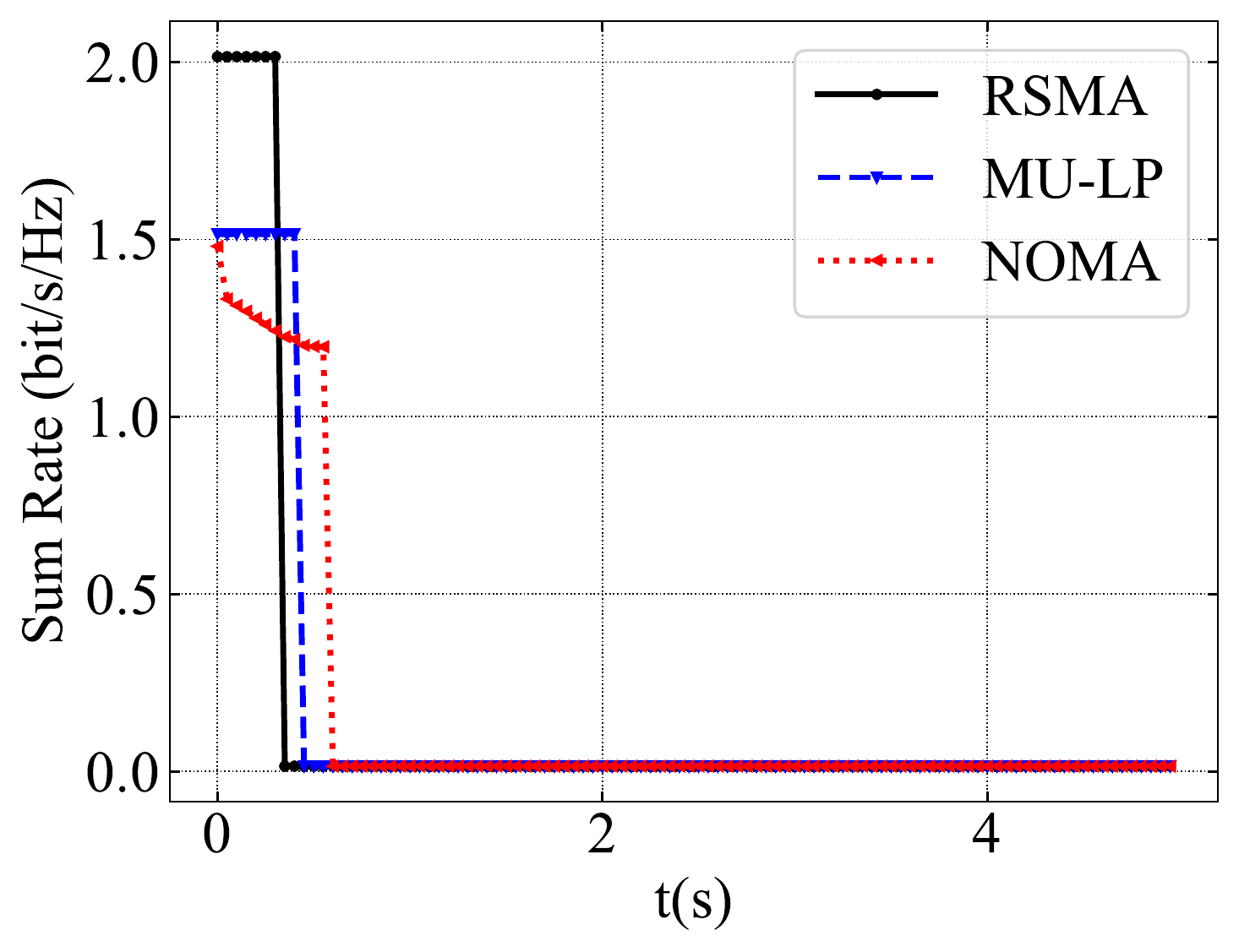}}
\vspace{-0.2cm} 
  \caption{Vehicles’ states and actions in S2 with RSMA: (a) motion trajectories, (b) heading angle, (c) velocity, (d) acceleration, (e) steering angle, (f) sum rate.}
  \label{figure_7}
\end{figure*}

 In Fig \ref{figure_7}, we further investigate the effect of the proposed IoV platoon system under different road conditions. Explicitly, Fig \ref{figure_7} shows the vehicles' states and control inputs: (a) motion trajectories, (b) heading angle, (c) velocity, (d) acceleration, (e) steering angle, and (f) sum rate. Since the platoon system has to follow the traffic regulations, all the vehicles shall enter the straight lane before moving into the solid line area. In this case, the system should obey some additional constraints, in order to smoothly reconfigure the current platoon into a single-lane platoon. Furthermore, compared to the traffic behavior shown in S1, it is more challenging to make safe and smooth lane changes for the crossroad scenario. Nevertheless, our simulation results demonstrate the efficiency of the proposed IoV platoon under different road conditions. Moreover, we also compare the mean and the variance of the heading angle, velocity, acceleration, and steering angle of the vehicles. Table \ref{table.S2} shows the related results characterizing the NOUM-aided MU-LP and the unicast NOMA for the V2V communication module. Compared to S1, the vehicles dedicate more effort to steering control in S2, which leads to a higher vehicular heading and steering angle variance. In this case, the RSMA-aided IoV platoon control scheme can still outperform both baselines in terms of its variance. Hence, these results prove the superiority of the RSMA scheme, as well as the robustness of the proposed IoV platoon system.

Moreover, we also investigate the impact of the proposed IoV platoon system under different weather conditions. We follow similar vehicle dynamics under the severe weather conditions detailed in \cite{Yu:2020vv}. The road adhesion coefficient, $\varrho$, in rainy weather can be calculated as:
\vspace{-0.2cm}
\begin{subequations}\label{eqn_33}
\begin{align}
\varrho=0.9458-0.0057v-0.0108\kappa, \tag{33}
\end{align}
\end{subequations}
where $v$ is the speed of the vehicle and $\kappa$ is the thickness of the water film. We consider $\kappa=20$ in the rainy weather scenario of this study. To guarantee safe driving in rainy weather, the acceleration threshold, $a_\text{thres}$, becomes lower, which can be calculated as:
\vspace{-0.2cm}
\begin{subequations}\label{eqn_34}
\begin{align}
a_\text{thres}=\varrho g,\tag{34}
\end{align}
\end{subequations}
where $g$ is the gravitational acceleration.

\begin{figure*}[h]
\centering  
  \subfloat[]{
      \label{figure_8a}
      \includegraphics[width=0.3\linewidth]{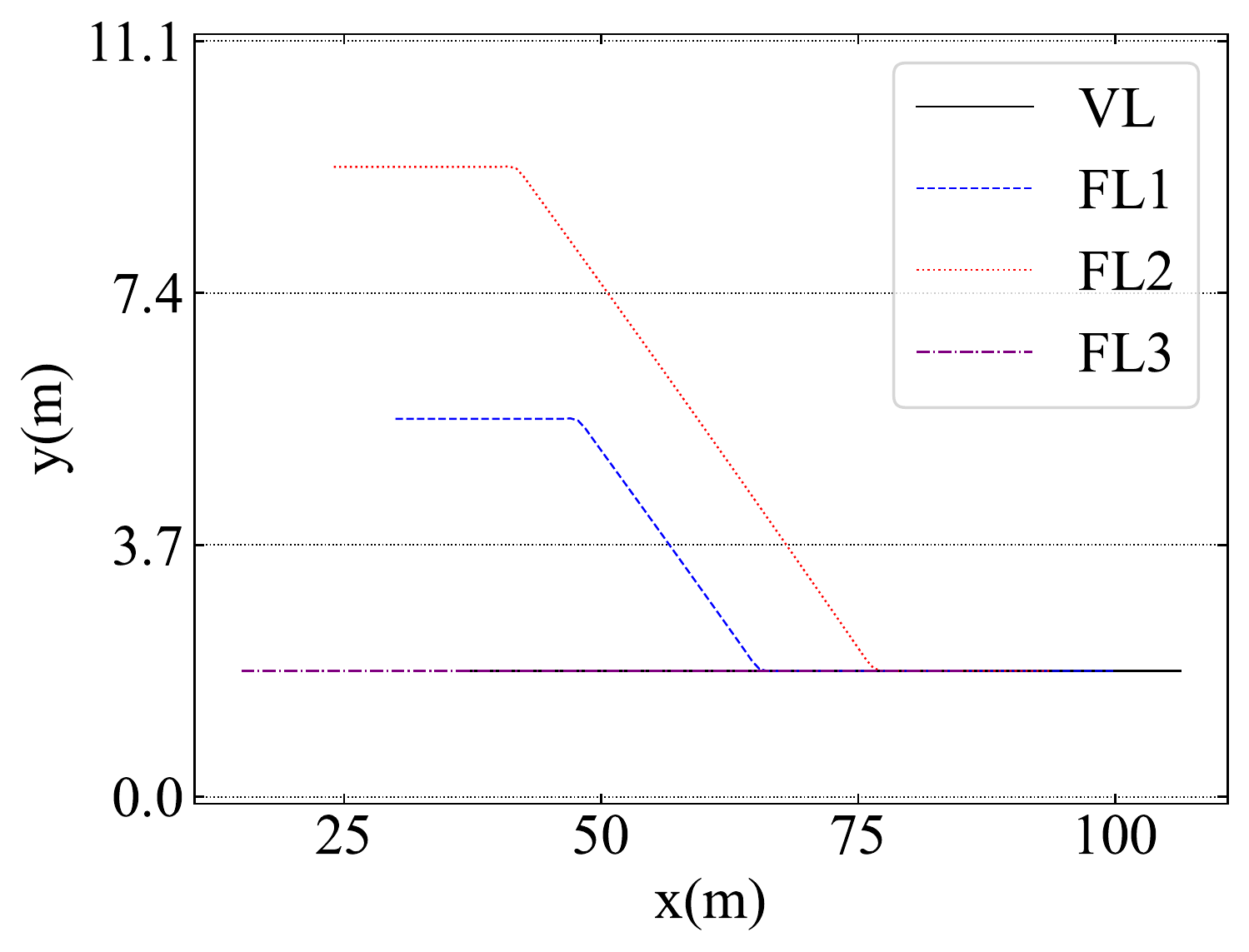}}
  \subfloat[]{
      \label{figure_8b}
      \includegraphics[width=0.3\linewidth]{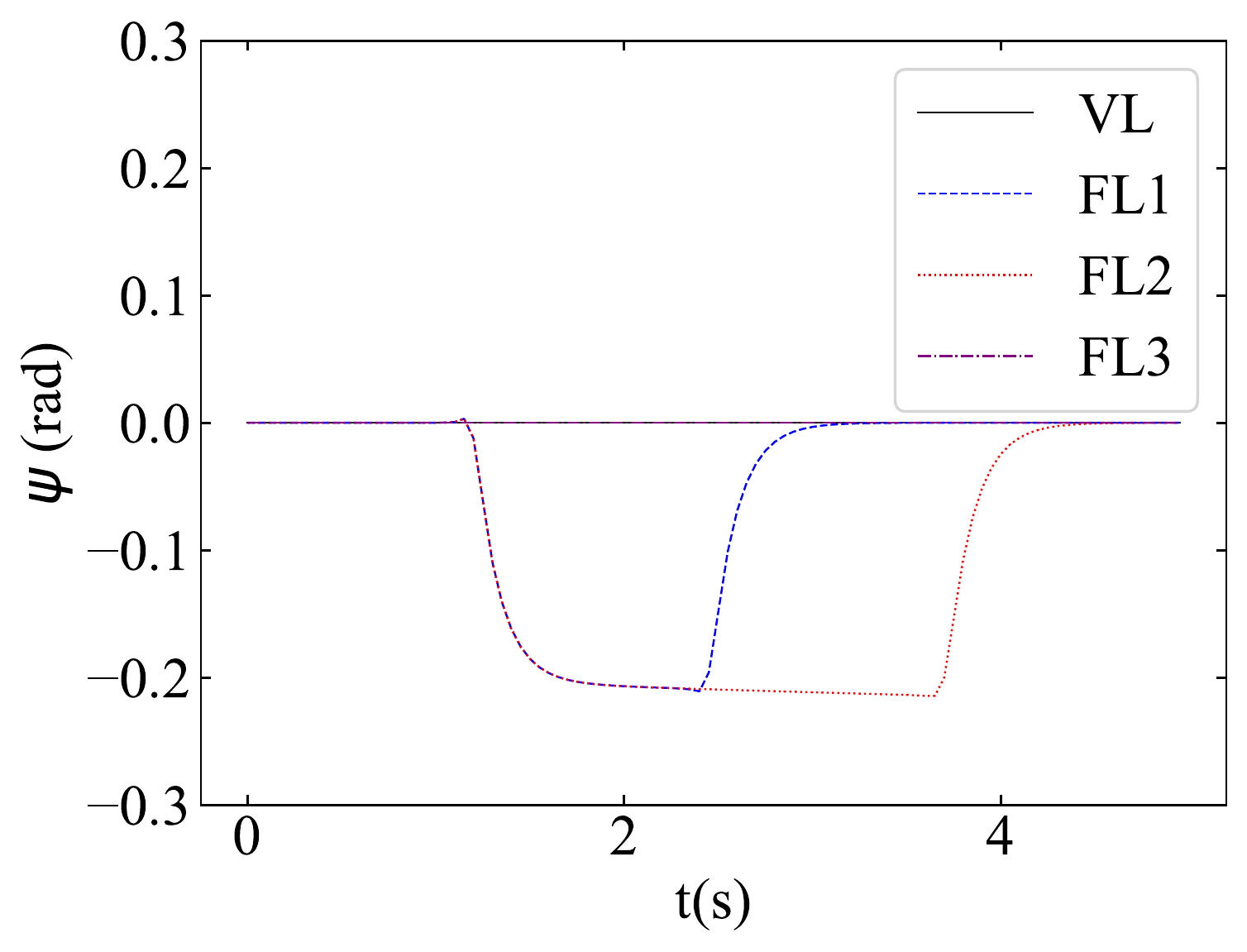}}
  \subfloat[]{
      \label{figure_8c}
      \includegraphics[width=0.3\linewidth]{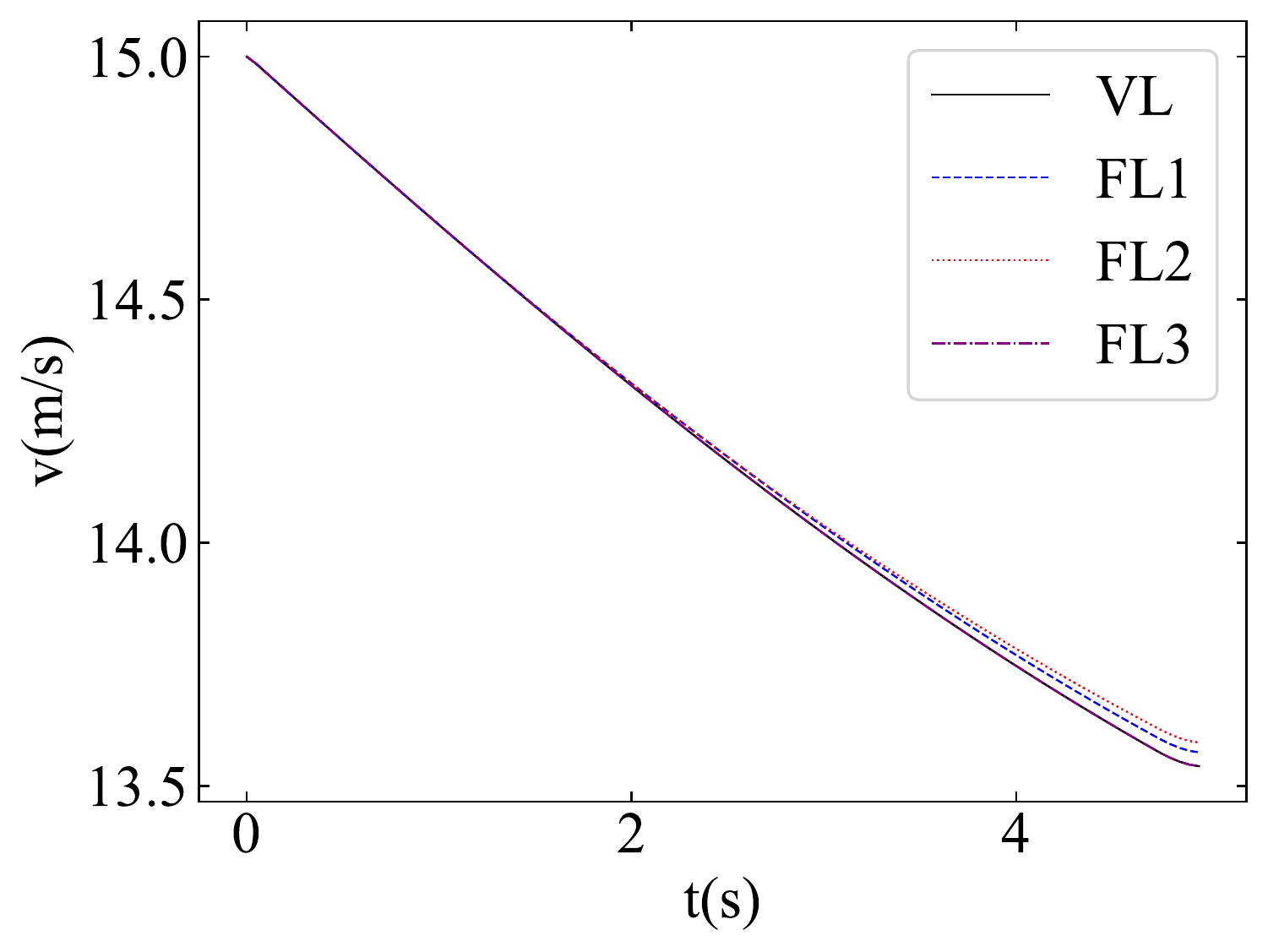}}
      
  \subfloat[]{
      \label{figure_8d}
      \includegraphics[width=0.3\linewidth]{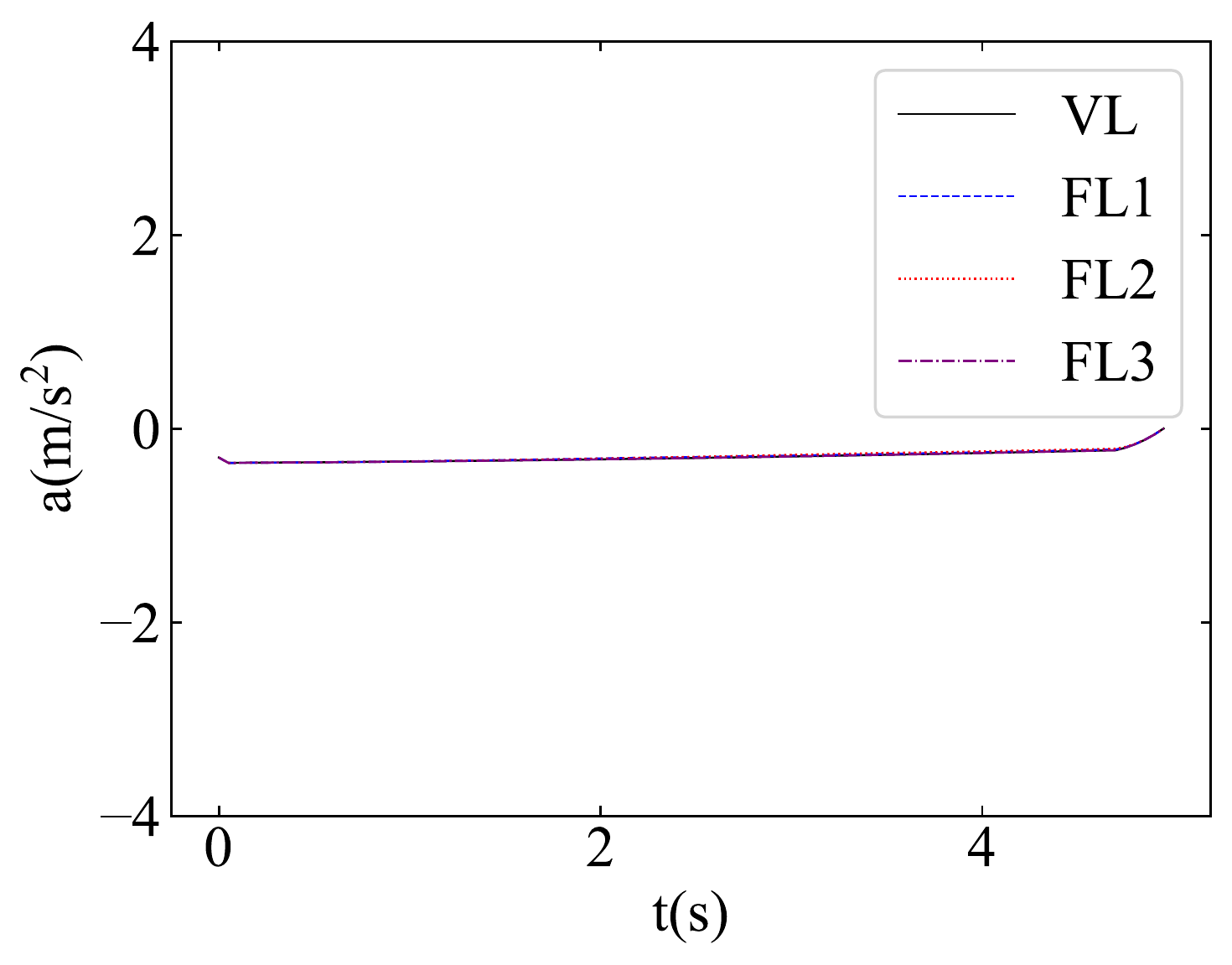}}
  \subfloat[]{
      \label{figure_8e}
      \includegraphics[width=0.3\linewidth]{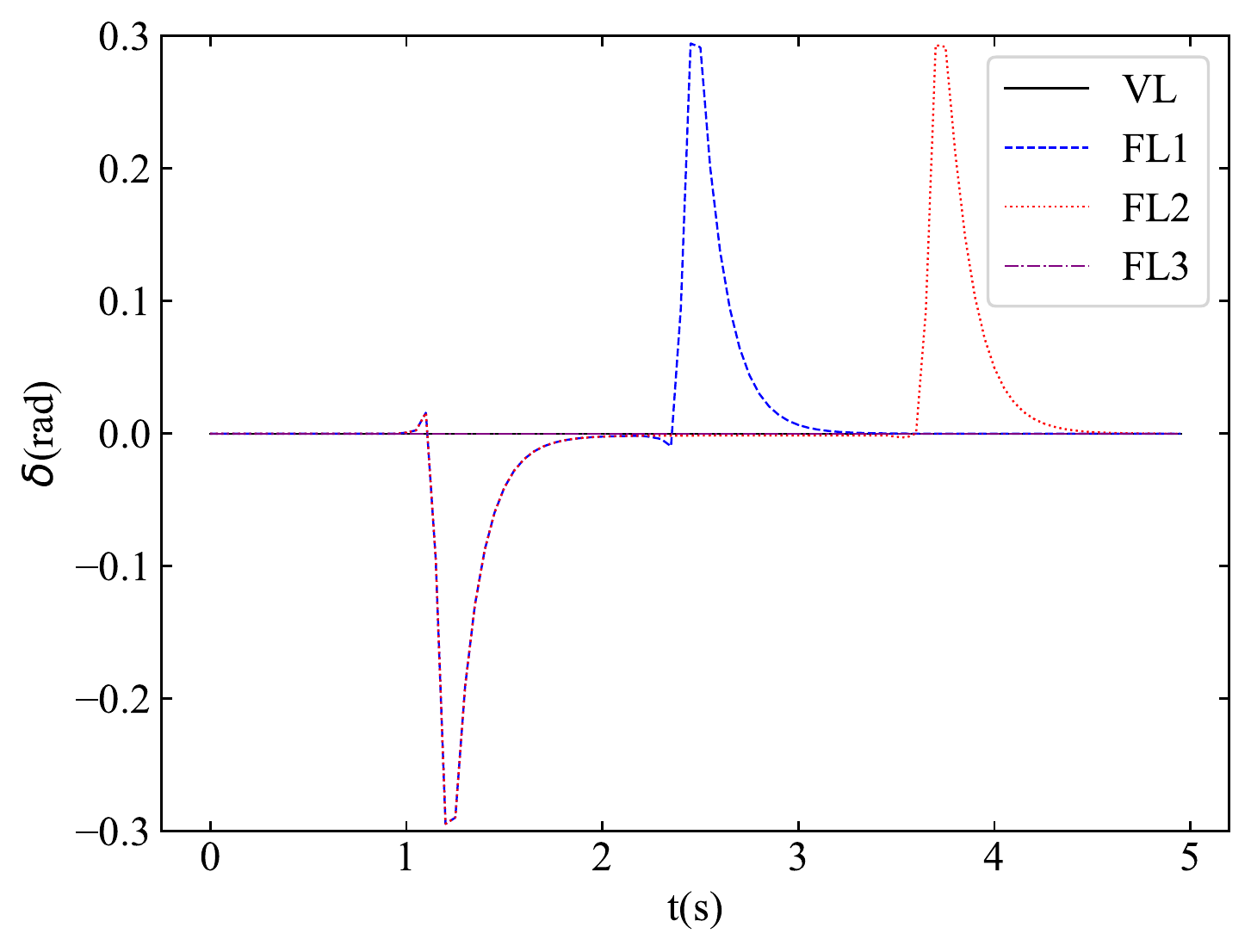}}
  \subfloat[]{
      \label{figure_8f}
      \includegraphics[width=0.3\linewidth]{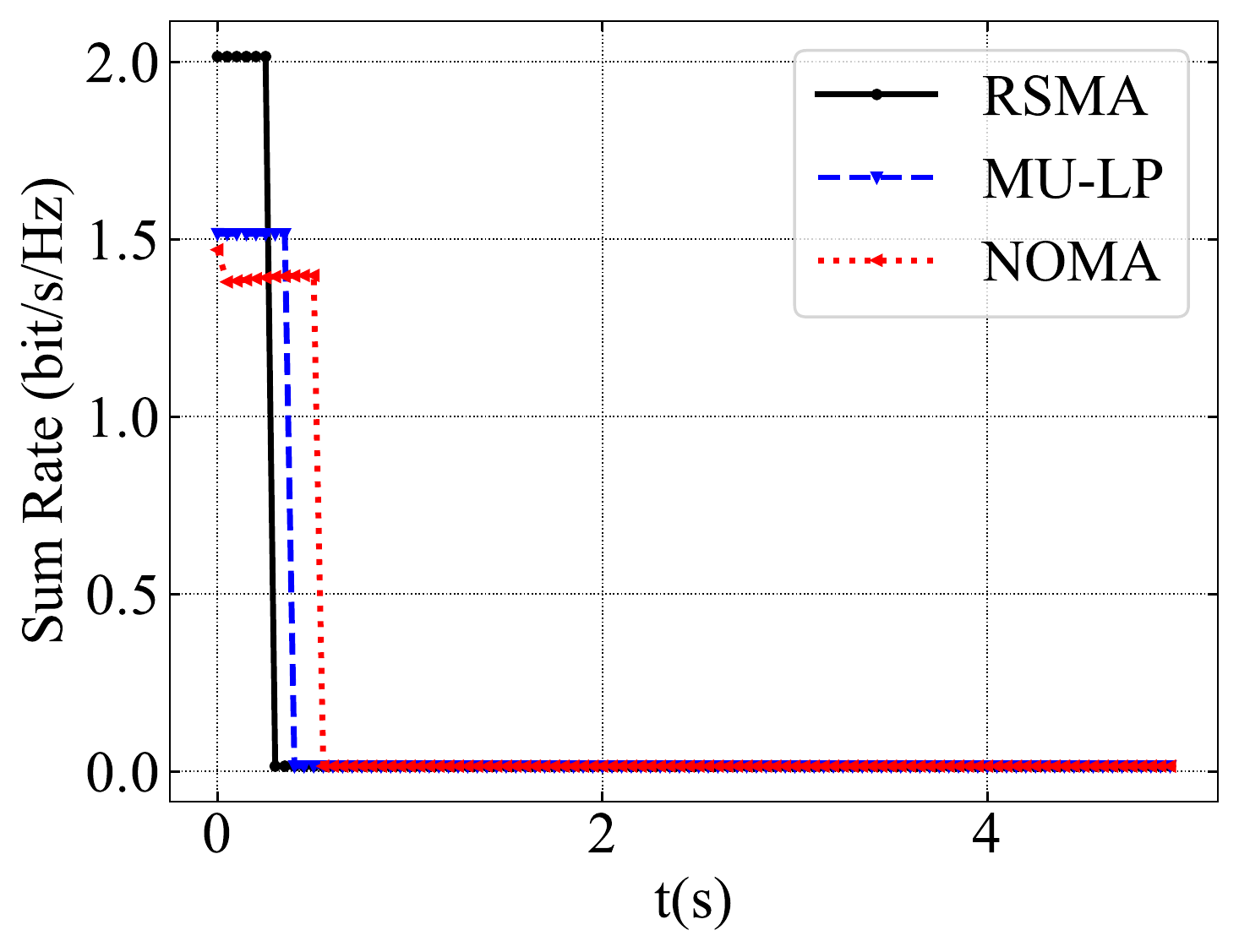}}
\vspace{-0.2cm} 
  \caption{Vehicles’ states and actions in S1 at rainy weather with RSMA: (a) motion trajectories, (b) heading angle, (c) velocity, (d) acceleration, (e) steering angle, (f) sum rate.}
  \label{figure_8}
\end{figure*}

Fig. \ref{figure_8} shows the vehicles' states and control inputs: (a) motion trajectories, (b) heading angle, (c) velocity, (d) acceleration, (e) steering angle, and (f) sum rate in S1 at rainy weather. As shown in the results, the IoV platoon performs the lane-change in a smooth manner. Furthermore, the RSMA still outperforms the MU-LP and NOMA scheme in terms of its sum rate. Thus, we investigate the robustness of our proposed framework in different road conditions.

\subsection{Convergence of the Algorithms Devised}
\begin{figure}[h]
\centering  
  \subfloat[]{
      \label{figure_9a}
      \includegraphics[width=0.45\linewidth]{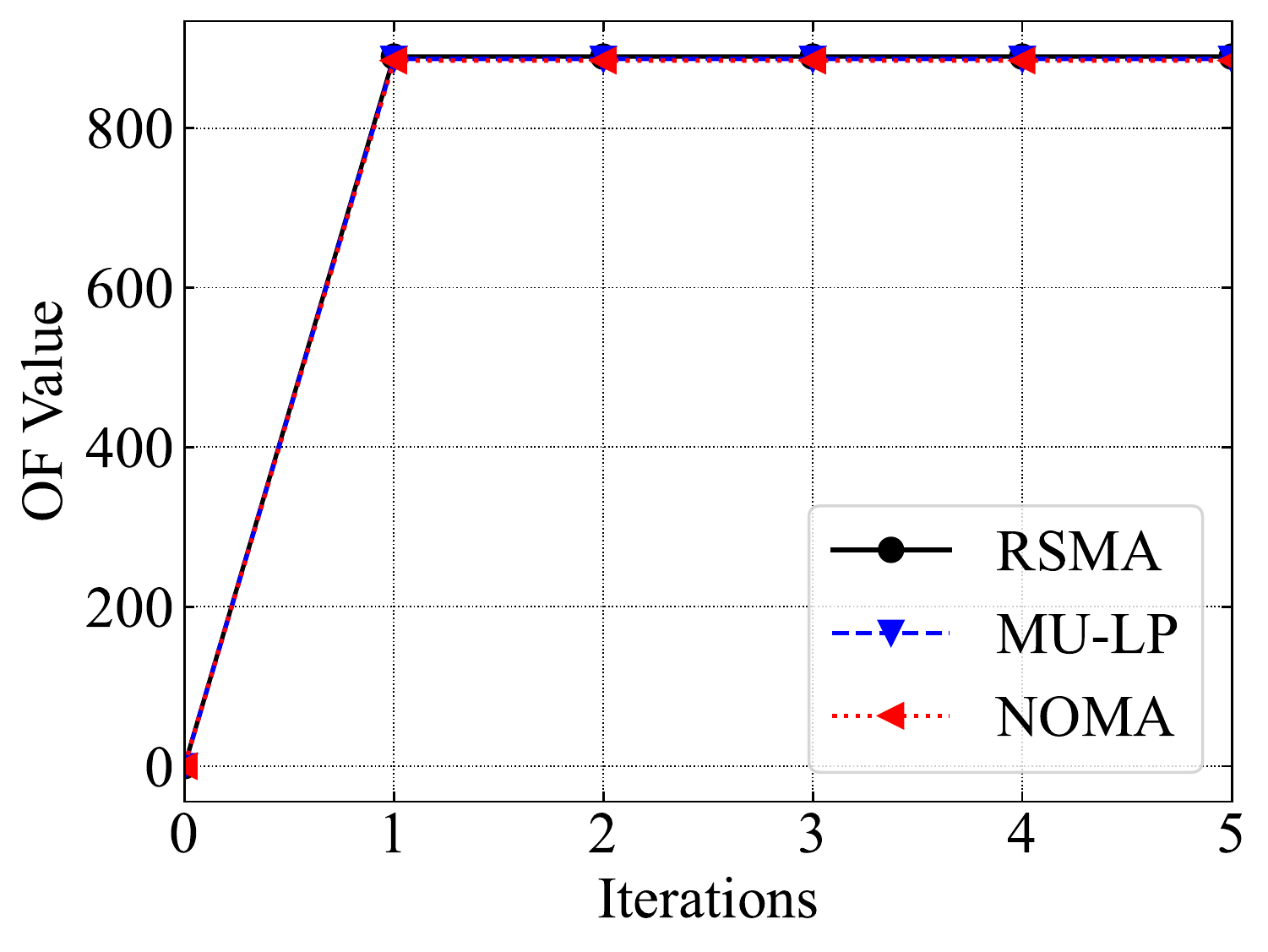}}
  \subfloat[]{
      \label{figure_9b}
      \includegraphics[width=0.45\linewidth]{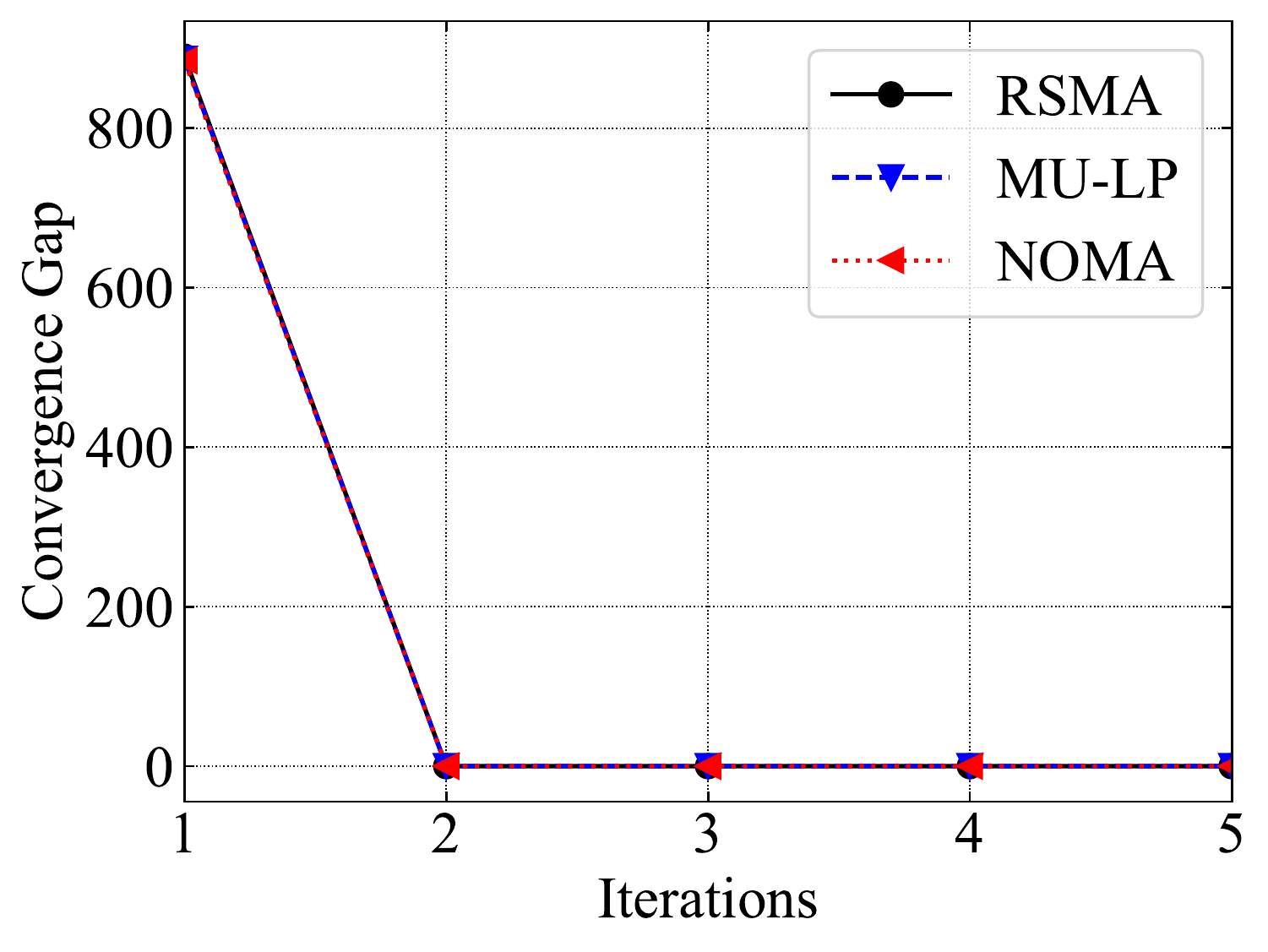}}

\vspace{-0.2cm} 
  \caption{Convergence of the BCD framework: (a) OF value, (b) convergence gap.}
  \label{figure_9}
\end{figure}

In this section, we investigate the convergence of the proposed BCD framework, which is related to Algorithm \ref{Algorithm1}. Here, we study the proposed IoV platoon model under S1 supporting four cars, and mainly focus our attention on the Objective Function (OF) value of the optimization problem formulated in \eqref{eqn_19} and on the corresponding convergence gap. Explicitly, the convergence gap is derived based on the difference of the OF values obtained in every iteration, as shown in Fig. \ref{figure_9}.  Specifically, we can observe that the OF value saturates after approximately 3 iterations. At this stage, the convergence gap falls within the tolerance threshold of the stopping criterion after approximately 3 iterations. This confirms that the proposed BCD framework is efficient in generating the near-optimal solutions of the RSMA-based communication problem formulated. 

\begin{figure}[h]
\centering  
  \subfloat[]{
      \label{figure_10a}
      \includegraphics[width=0.45\linewidth]{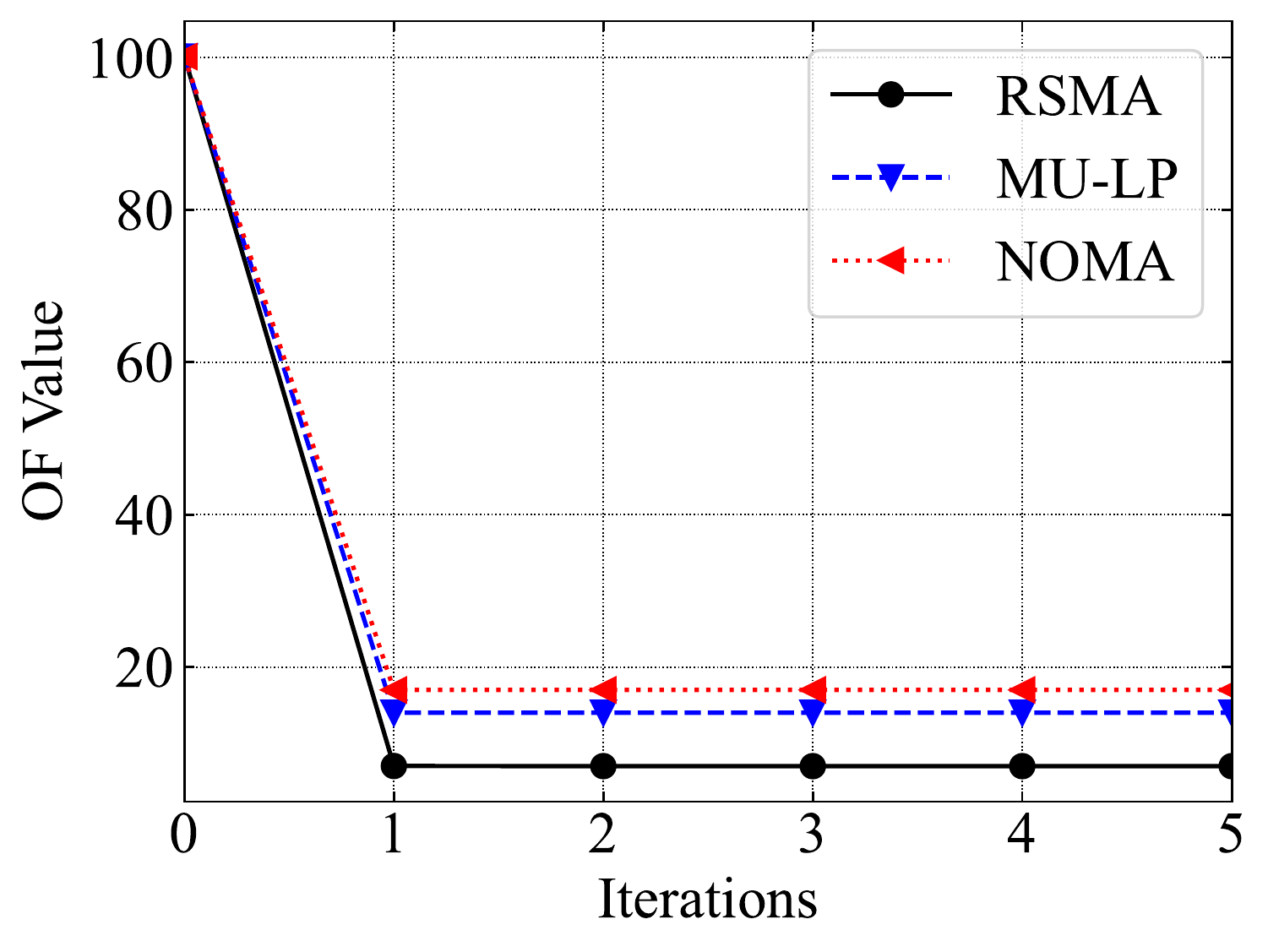}}
  \subfloat[]{
      \label{figure_10b}
      \includegraphics[width=0.45\linewidth]{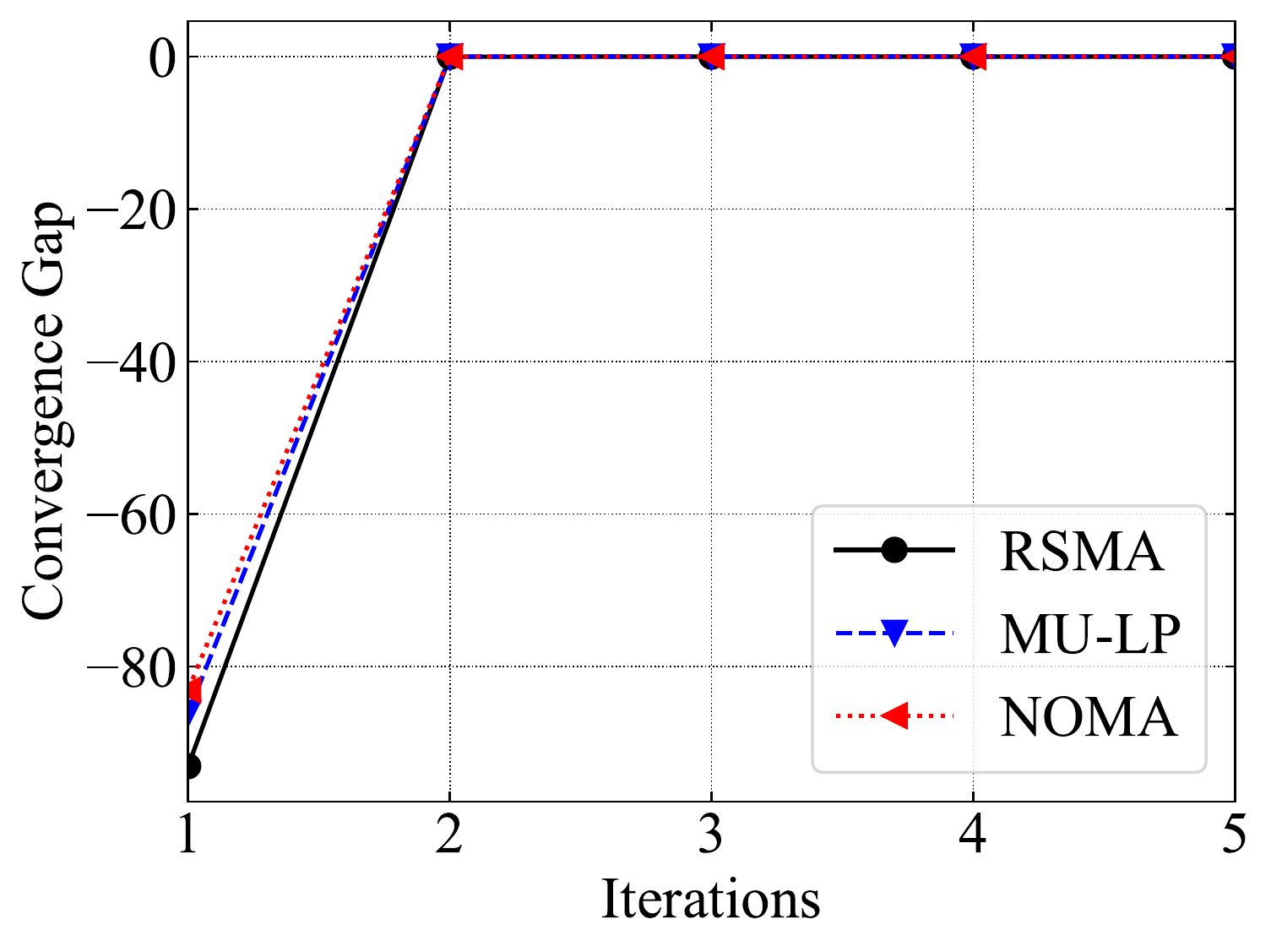}}

\vspace{-0.2cm} 
  \caption{Convergence of the SCA framework: (a) OF value, (b) convergence gap.}
  \label{figure_10}
\end{figure}

\begin{figure}[h]
\centering  
  \subfloat[]{
      \label{figure_11a}
      \includegraphics[width=0.45\linewidth]{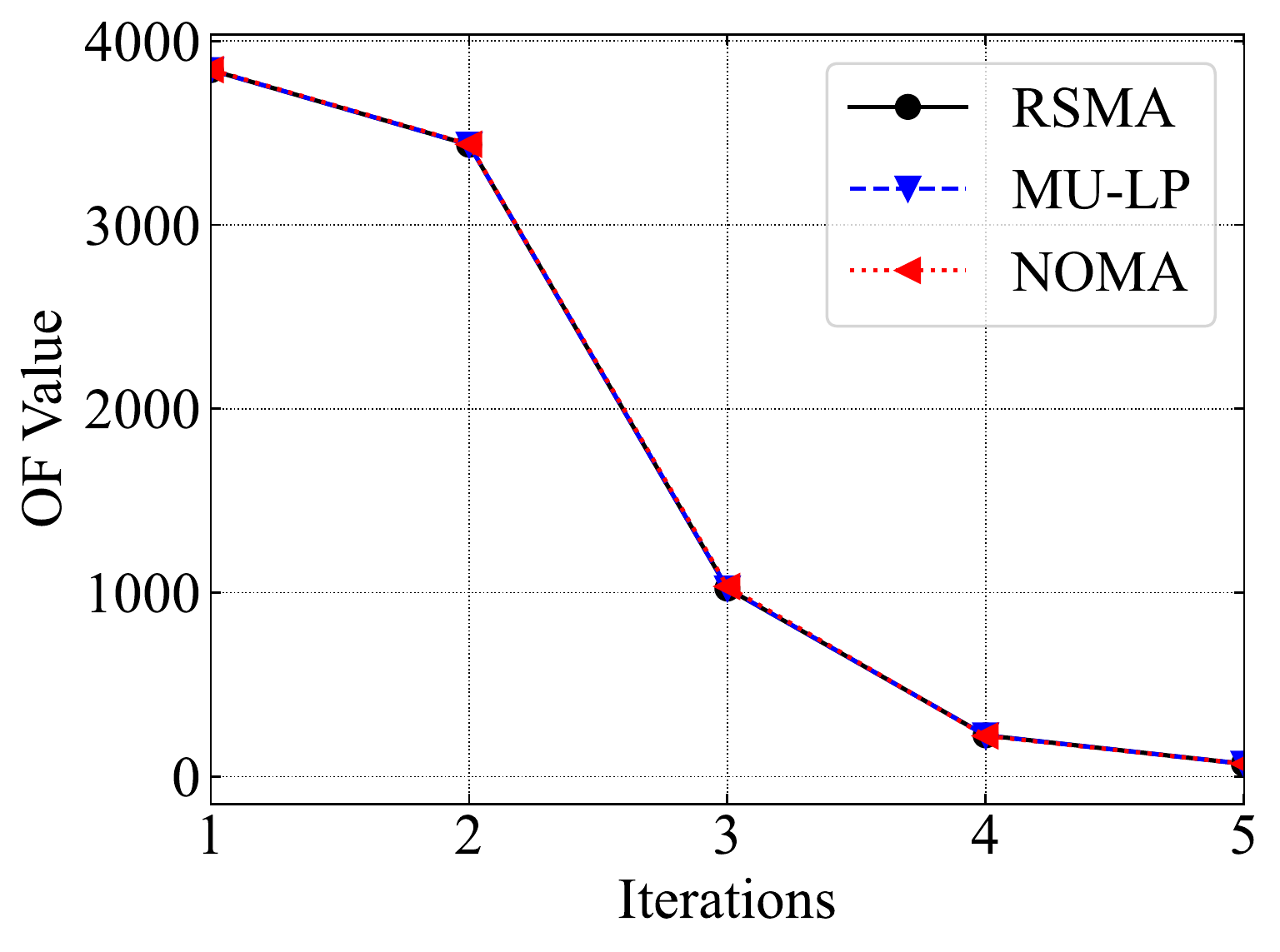}}
  \subfloat[]{
      \label{figure_11b}
      \includegraphics[width=0.45\linewidth]{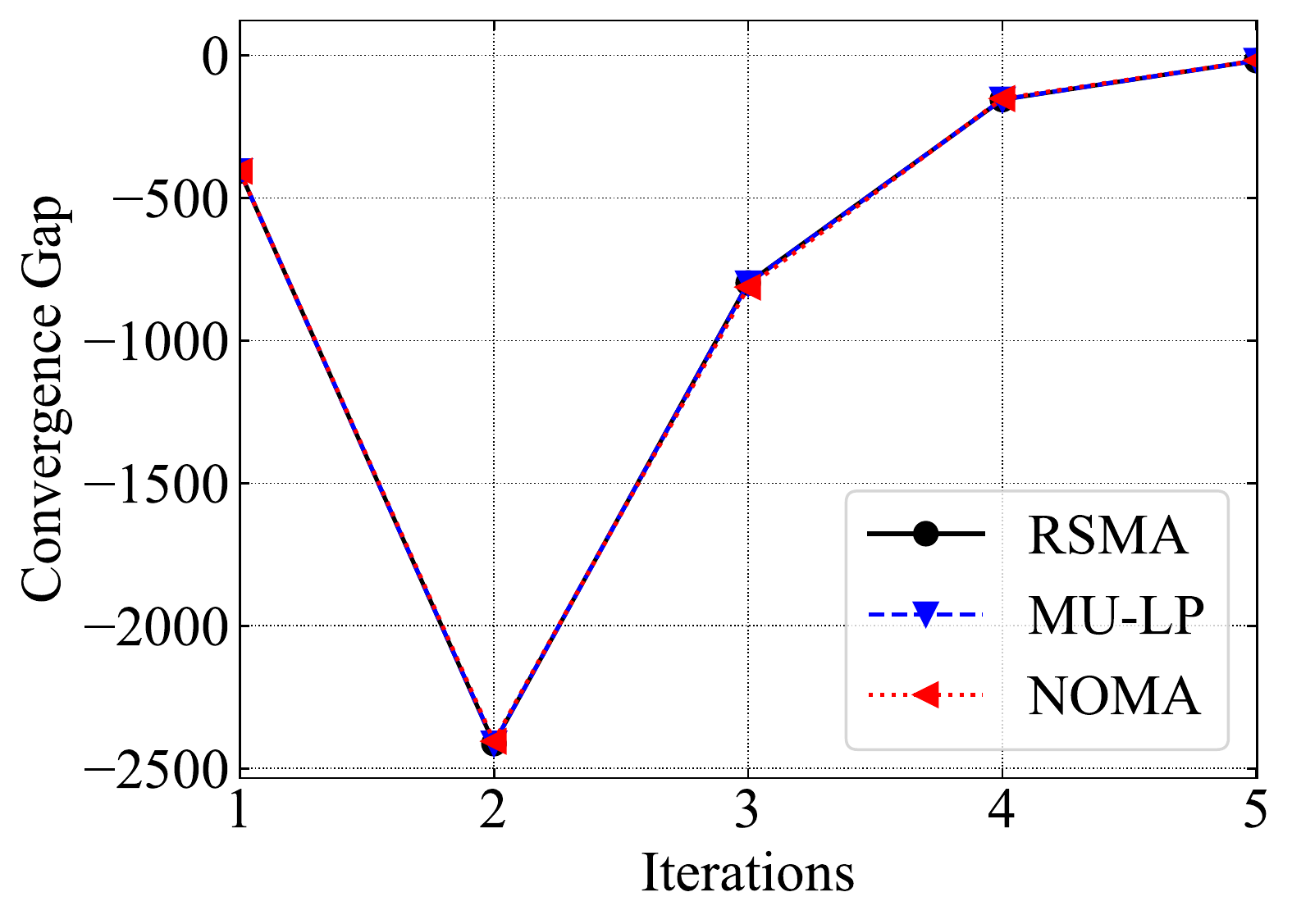}}

\vspace{-0.2cm} 
  \caption{Convergence of the MPC framework: (a) OF value, (b) convergence gap.}
  \label{figure_11}
\end{figure}

Additionally, we also investigate the convergence of the SCA-based FEEL downlink problem and the MPC-based platoon control problem, which are related to Algorithm \ref{Algorithm2} and Algorithm \ref{Algorithm3}, respectively. These two problems are studied in the same settings as previously. For the SCA-based FEEL downlink problem, we focus on the OF value in \eqref{eqn_30} and on the related convergence gap in Fig. \ref{figure_10}. Specifically, we can observe that the OF value saturates after 4 iterations, while the convergence gap falls within the tolerance threshold after 4 iterations. This confirms that the proposed SCA-based FEEL downlink problem is efficiently solved, and near-optimal solutions can be found. Similarly, for the platoon control problem in Fig. \ref{figure_11}, we study the convergence of the MPC approach by assessing the OF value of \eqref{eqn_31} and the convergence gap. Specifically, we can observe that the OF value saturates after 5 iterations, and the convergence gap satisfies the stopping criterion after 5 iterations. This demonstrates the efficiency of the algorithm devised for our platoon control problem.

\section{Conclusions}

\par  This paper investigated an RSMA-aided V2X downlink scenario in a vehicular platoon application. Specifically, an RSMA-based IoV solution was proposed for jointly considering two problems, namely, the IoV platoon control and FEEL downlink operations. The proposed framework was designed for minimizing both the latency of the FEEL downlink and the deviation of the vehicles' trajectories within the platoon system. Given this sophisticated framework, a multi-objective optimization problem was formulated. To efficiently solve this problem, a BCD framework was developed for decoupling the main multi-objective problem into two sub-problems. Additionally, for solving these non-convex sub-problems, the classic SCA and MPC approaches were deployed for solving the FEEL-based downlink and platoon control problems, respectively. Our numerical results indicated that the proposed RSMA-based IoV system outperforms both the popular MU–LP and the conventional NOMA system in terms of the FEEL downlink latency and the control input variance. Furthermore, the BCD framework was shown to generate near-optimal solutions at reduced complexity. 

\par In our future work, we will consider adopting the deep learning approach to estimate the channel information, since this work only considers the channel estimation through the known vehicle’s states. We will also explore how to incorporate perception-based concepts into the proposed system model in support of both the platoon control and the RSMA-based V2X system.

\bibliographystyle{IEEEtran}
\bibliography{reference.bib}

\begin{IEEEbiography}[{\includegraphics[width=1in,height=1.25in,clip,keepaspectratio]{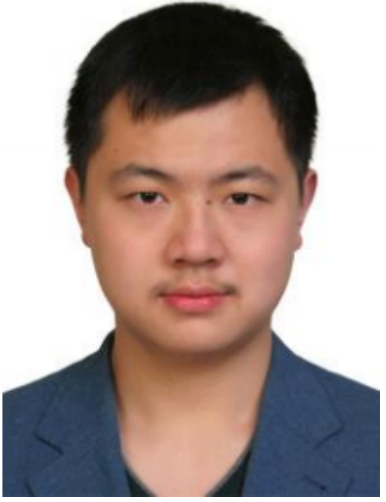}}]{Shengyu Zhang} received the B. Eng. and M. Eng. degrees in communication engineering from Southeast University, Nanjing, China, in 2016 and 2019, respectively. He is currently pursuing a Ph.D. in the Department of Electrical and Electronic Engineering, The University of Hong Kong. His research includes optical networks, satellite networks, quantum networks, and vehicular networks.
\end{IEEEbiography}
 \vspace{-30 mm}
\begin{IEEEbiography}[{\includegraphics[width=1in,height=1.25in,clip,keepaspectratio]{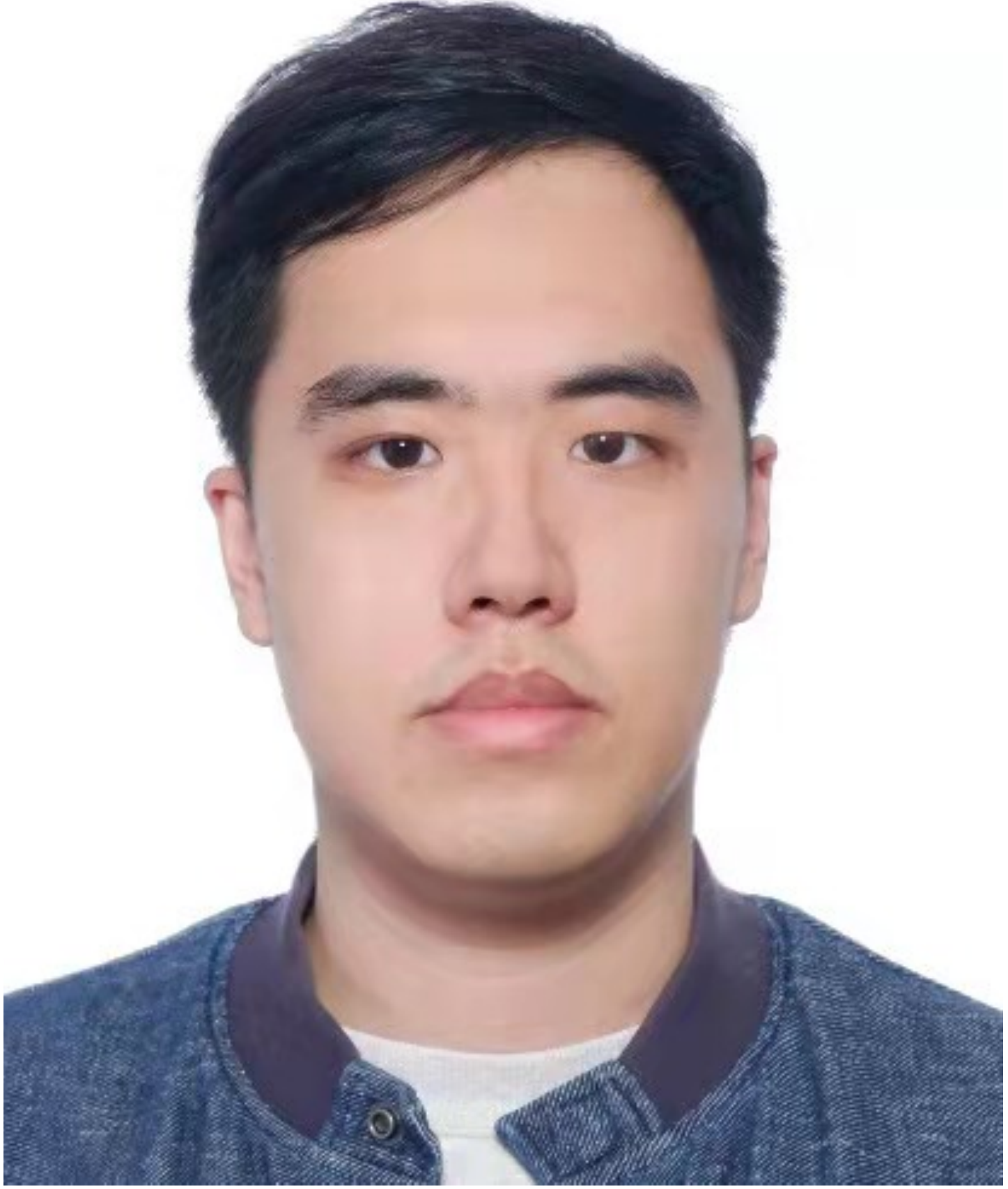}}]{Shiyao Zhang} (Member, IEEE) received the B.S. degree (Hons.) in Electrical and Computer Engineering from Purdue University, West Lafayette, IN, USA, in 2014, the M. S. degree in Electrical Engineering (Electric Power) from University of Southern California, Los Angeles, CA, USA, in 2016, and the Ph.D. degree from the University of Hong Kong, Hong Kong, China. He was a Post-Doctoral Research Fellow with the Academy for Advanced Interdisciplinary Studies, Southern University of Science and Technology from 2020 to 2022. He is currently a Research Assistant Professor with the Research Institute for Trustworthy Autonomous Systems, Southern University of Science and Technology. His research interests include smart cities, intelligent transportation systems, smart energy systems, optimization theory and algorithms, and deep learning applications.
\end{IEEEbiography}
 \vspace{-30 mm} 
\begin{IEEEbiography}[{\includegraphics[width=1in,height=1.25in,clip,keepaspectratio]{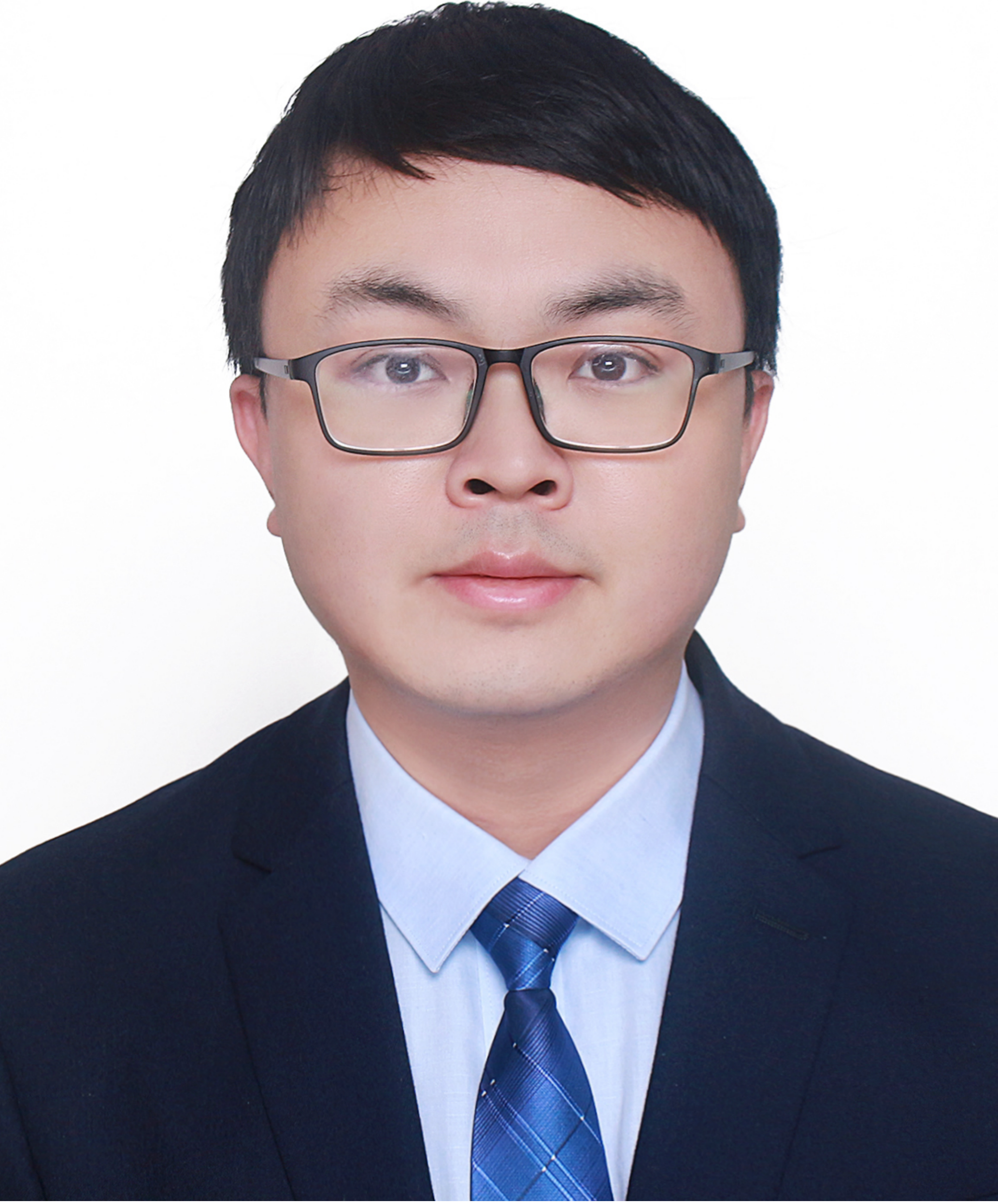}}]{Weijie Yuan} (Member, IEEE) received the B.E. degree from the Beijing Institute of Technology, China, in 2013, and the Ph.D. degree from the University of Technology Sydney, Australia, in 2019. In 2016, he was a Visiting Ph.D. Student with the Institute of Telecommunications, Vienna University of Technology, Austria. He was a Research Assistant with the University of Sydney, a Visiting Associate Fellow with the University of Wollongong, and a Visiting Fellow with the University of Southampton, from 2017 to 2019. From 2019 to 2021, he was a Research Associate with the University of New South Wales. He is currently an Assistant Professor with the Department of Electrical and Electronic Engineering, Southern University of Science and Technology, Shenzhen, China. He was a recipient of the Best Ph.D. Thesis Award from the Chinese Institute of Electronics and an Exemplary Reviewer from IEEE TCOM/WCL. He currently serves as an Associate Editor for the IEEE Communications Letters, an Associate Editor and an Award Committee Member for the EURASIP Journal on Advances in Signal Processing. He has led the guest editorial teams for three special issues in IEEE Communications Magazine, IEEE Transactions on Green Communications and Networking, and China Communications. He was an Organizer/the Chair of several workshops and special sessions on orthogonal time frequency space (OTFS) and integrated sensing and communication (ISAC) in flagship IEEE and ACM conferences, including IEEE ICC, IEEE/CIC ICCC, IEEE SPAWC, IEEE VTC, IEEE WCNC, IEEE ICASSP, and ACM MobiCom. He is the Founding Chair of the IEEE ComSoc Special Interest Group on Orthogonal Time Frequency Space (OTFS-SIG). 
\end{IEEEbiography}
 \vspace{-30 mm} 
\begin{IEEEbiography}[{\includegraphics[width=1in,height=1.25in,clip,keepaspectratio]{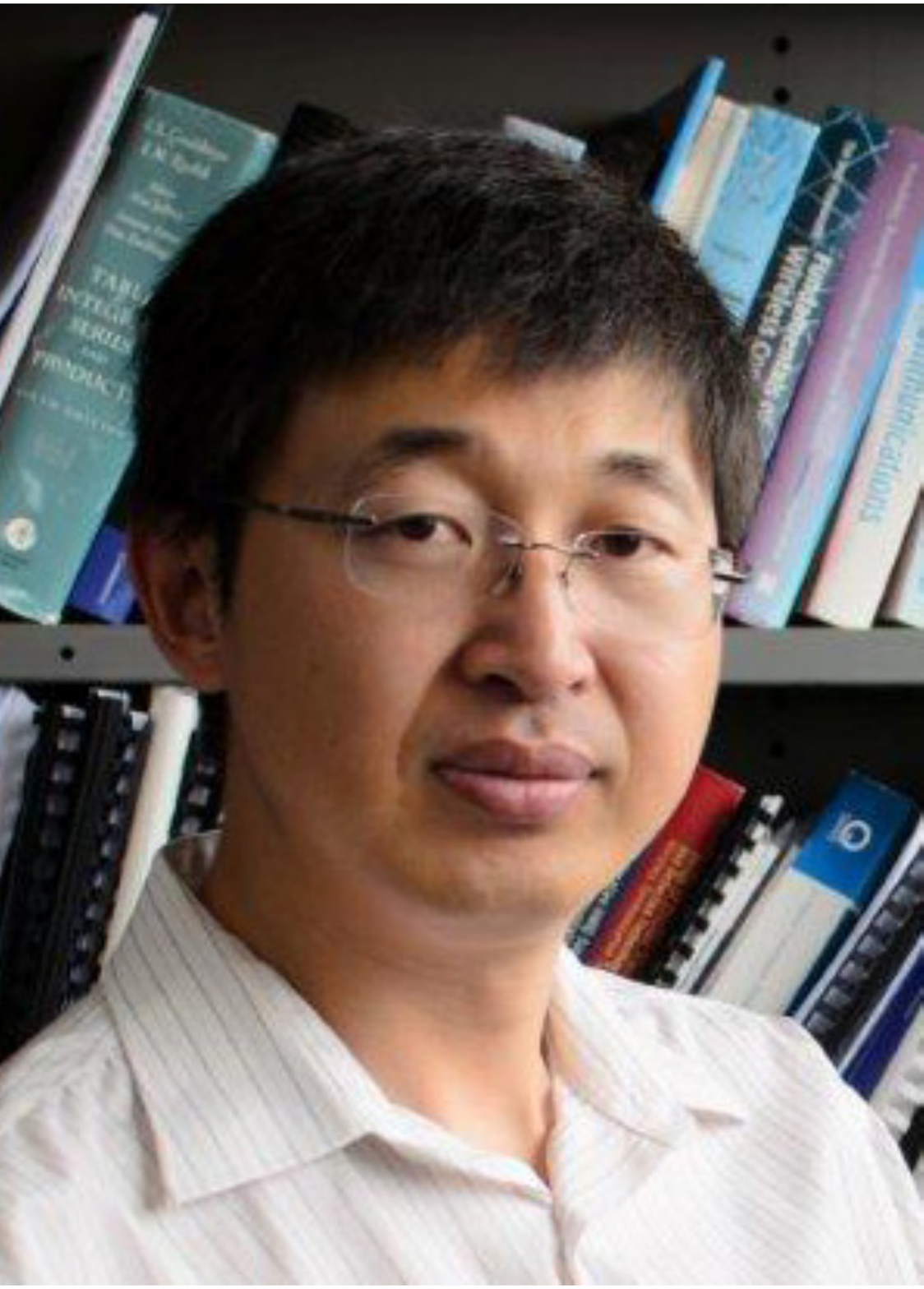}}]{Yonghui Li} (M’04-SM’09-F19) received his PhD degree in November 2002 from Beijing University of Aeronautics and Astronautics. Since 2003, he has been with the Centre of Excellence in Telecommunications, the University of Sydney, Australia. He is now a Professor and Director of Wireless Engineering Laboratory in School of Electrical and Information Engineering, University of Sydney. He is the recipient of the Australian Queen Elizabeth II Fellowship in 2008 and the Australian Future Fellowship in 2012. He is a Fellow of IEEE.

His current research interests are in the area of wireless communications, with a particular focus on MIMO, millimeter wave communications, machine to machine communications, coding techniques and cooperative communications. He holds a number of patents granted and pending in these fields. He is now an editor for IEEE transactions on communications, IEEE transactions on vehicular technology. He also served as the guest editor for several IEEE journals, such as IEEE JSAC, IEEE Communications Magazine, IEEE IoT journal, IEEE Access. He received the best paper awards from IEEE International Conference on Communications (ICC) 2014, IEEE PIRMC 2017 and IEEE Wireless Days Conferences (WD) 2014.
\end{IEEEbiography}

\begin{IEEEbiography}[{\includegraphics[width=1in,height=1.25in,clip,keepaspectratio]{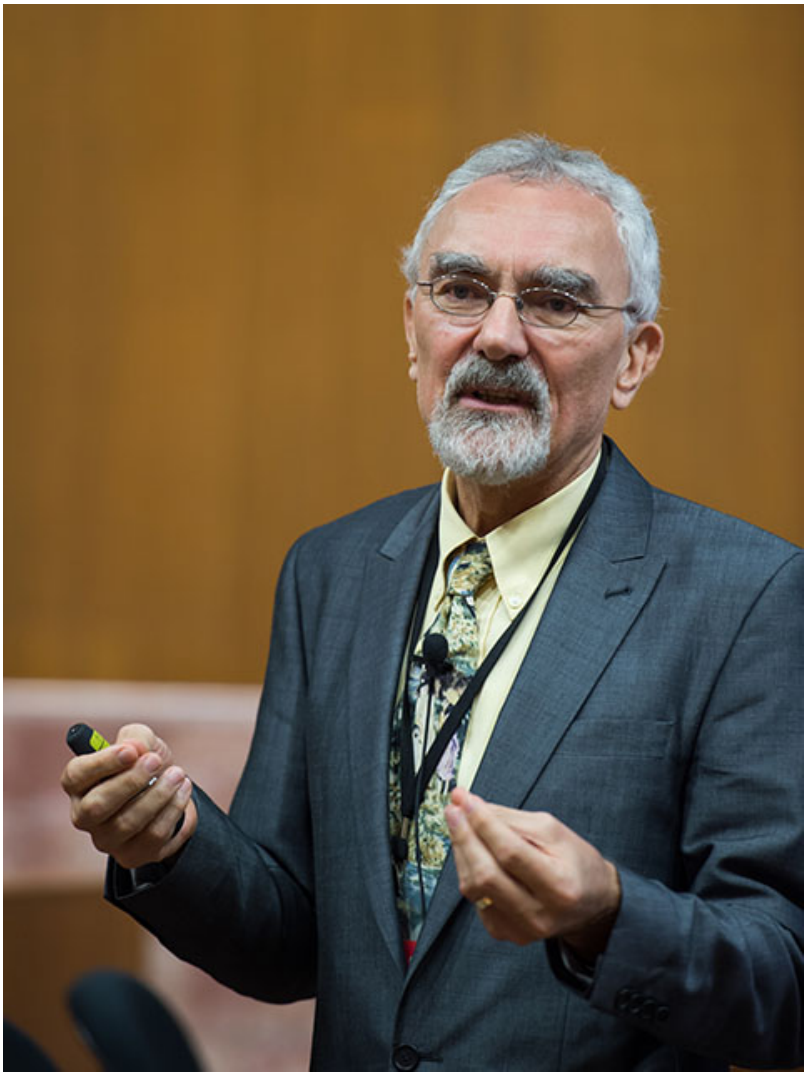}}]{Lajos Hanzo} (http://www-mobile.ecs.soton.ac.uk, https://en.wikipedia.org/wiki/Lajos\_Hanzo) (FIEEE'04) received his Master degree and Doctorate in 1976 and 1983, respectively from the Technical University (TU) of Budapest. He was also awarded the Doctor of Sciences (DSc) degree by the University of Southampton (2004) and Honorary Doctorates by the TU of Budapest (2009) and by the University of Edinburgh (2015).  He is a Foreign Member of the Hungarian Academy of Sciences and a former Editor-in-Chief of the IEEE Press.  He has served several terms as Governor of both IEEE ComSoc and of VTS.  He is also a Fellow of the Royal Academy of Engineering (FREng), of the IET and of EURASIP. He is the recipient of the 2022 Eric Sumner Field Award. He has published 19 John Wiley research monographs and 2000+ contributions at IEEE Xplore.
\end{IEEEbiography}
\end{document}